\documentclass[aps,prb,superscriptaddress,showpacs,floatfix,10pt,twocolumn]{revtex4-2}
\usepackage{graphicx,graphics}
\usepackage{appendix}
\usepackage{dcolumn}
\usepackage{amsmath,amssymb,amsfonts}
\usepackage{dsfont}
\usepackage{latexsym,verbatim}
\usepackage{bm}
\usepackage{braket}
\usepackage{color}
\usepackage{multirow}
\usepackage[abs]{overpic}
\usepackage{overpic}
\usepackage{array}
\usepackage{nicefrac}
\usepackage{float}
\usepackage[normalem]{ulem}
\usepackage[breaklinks=true,colorlinks,citecolor=blue,linkcolor=blue,urlcolor=blue]{hyperref}
\usepackage{cleveref}
\newcommand{\Tr}{\operatorname{Tr}}
\renewcommand{\Re}{\operatorname{Re}}
\renewcommand{\Im}{\operatorname{Im}}

\crefname{equation}{Eq.}{Eqs.}
\crefname{figure}{Fig.}{Figs.}
\crefname{appendix}{App.}{Apps.}

\begin{document}
\title{Phonon spectra, quantum geometry, and the Goldstone theorem}
\author{Guglielmo Pellitteri}
\email{guglielmo.pellitteri@sns.it}
\affiliation{Scuola Normale Superiore, Piazza dei Cavalieri 7, I-56126 Pisa,~Italy}
\author{Zenan Dai}
\affiliation{Zhiyuan College, Shanghai Jiao Tong University, Shanghai 200240, China}
\author{Haoyu Hu}
\affiliation{Department of Physics, Princeton University, Princeton, New Jersey 08544,~USA}
\author{Yi Jiang}
\affiliation{Donostia International Physics Center, P. Manuel de Lardizabal 4, 20018 Donostia-San Sebastian,~Spain}
\author{Guido Menichetti}
\affiliation{Dipartimento di Fisica dell'Università di Pisa, Largo Bruno Pontecorvo 3, I-56127 Pisa,~Italy}
\affiliation{Istituto Italiano di Tecnologia, Graphene Labs, Via Morego 30, I-16163 Genova, Italy}
\author{Andrea Tomadin}
\affiliation{Dipartimento di Fisica dell'Università di Pisa, Largo Bruno Pontecorvo 3, I-56127 Pisa,~Italy}
\author{B. Andrei Bernevig}
\affiliation{Department of Physics, Princeton University, Princeton, New Jersey 08544,~USA}
\affiliation{Donostia International Physics Center, P. Manuel de Lardizabal 4, 20018 Donostia-San Sebastian,~Spain}
\affiliation{IKERBASQUE, Basque Foundation for Science, Maria Diaz de Haro 3, 48013 Bilbao,~Spain}
\author{Marco Polini}
\affiliation{Dipartimento di Fisica dell'Università di Pisa, Largo Bruno Pontecorvo 3, I-56127 Pisa,~Italy}
\affiliation{ICFO-Institut de Ci\`{e}ncies Fot\`{o}niques, The Barcelona Institute of Science and Technology, Av. Carl Friedrich Gauss 3, 08860 Castelldefels (Barcelona),~Spain}
\begin{abstract}
Phonons are essential quasi-particles of all crystals and play a key role in fundamental properties such as thermal transport and superconductivity. In particular, acoustic phonons can be interpreted as Goldstone modes that emerge due to the spontaneous breaking of translational symmetry.
In this Article, we investigate the quantum geometric contribution to the phonon spectrum in the absence of Holstein phonons. 
Using graphene as a case study, we decompose the dynamical matrix into distinct terms that exhibit different dependencies on the electron energy and wavefunction. We then examine the role of quantum geometry in shaping the material's phonon spectrum, and we find that removing the nontrivial quantum geometric contribution from the dynamical matrix causes the acoustic phonon modes to behave in a non-analytic fashion.
\end{abstract}

\maketitle

\section{Introduction}
\begin{figure}[t]
    \centering
    \includegraphics[width=1.\columnwidth]{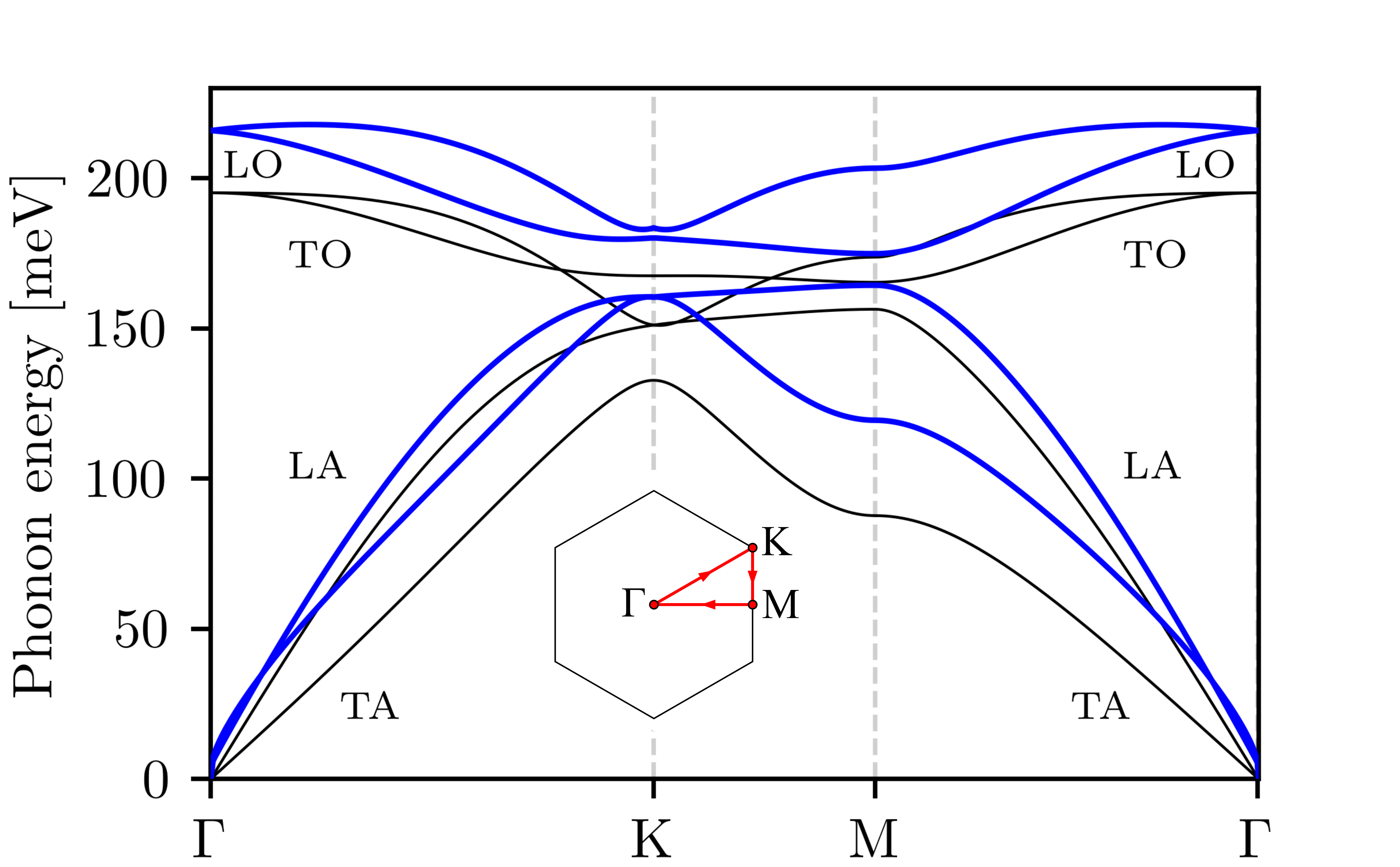}
    \caption{(Color online) Dispersion of the four in-plane phonon branches (LA, TA, LO, and TO) in graphene. Results in this figure are plotted along the $\Gamma$KM$\Gamma$ high-symmetry path in the first Brillouin zone illustrated in the inset. Black lines: Phonon dispersion $\omega_{\ell}(\bm q)$ calculated analytically within a next-nearest neighbor Born-von Karman framework~\cite{Falkovsky2007}. The relevant parameters needed to produce these plots have been determined by fitting {\it ab initio} density functional theory results. Blue lines: Phonon dispersion $\Tilde{\omega}_{\ell}(\bm q)$ as calculated after removing from the full dynamical matrix ${\cal D}({\bm q})$ the quantum geometric contribution ${\cal D}_{\rm g}({\bm q})$---see Eq.~\eqref{eq:dm-decomposition2}.  A zoom near $\Gamma$ is reported in Fig.~\ref{fig:2}, showing that both the LA and TA blue curves exhibit a non-analytic dependence on the wavevector $\bm q$ in a neighborhood of  $\Gamma$.}
    \label{fig:1}
\end{figure}

Every crystal supports elastic waves, which are traveling quanta of vibrational energy~\cite{Grosso2014,Marder2010} that Frenkel dubbed ``{\it phonons}'' in 1932. Phonons are quasiparticles that appear in a wide variety of condensed matter phenomena. Most notably, they serve as the bosonic glue that binds electrons into Cooper pairs in many conventional BCS superconductors~\cite{Tinkham2004}. They also play crucial roles in thermal and electrical transport, acting as an intrinsic and unavoidable source of scattering for electrons moving through a crystal. Additionally, phonons are important in atomically thin two-dimensional (2D) materials~\cite{Geim2013}: the authors of Ref.~\cite{wang_science_2013} demonstrated that graphene samples encapsulated in hexagonal Boron Nitride can display very large mobilities also at room temperature, which are solely limited by scattering of electrons against graphene's acoustic phonons~\cite{Marzari2005,Ferrari2006,Ferrari2007,Nika2012,Ferrari2013}. These are gapless modes near the center of the first Brillouin zone (FBZ)---see black lines labelled by ``LA'' and ``TA'' in Fig.~\ref{fig:1}---and play an important role in limiting plasmon lifetimes~\cite{principi_prb_2014,Woessner_naturemater_2015,Lundeberg_Science_2017,Ni_Nature_2018} in the same devices. From a theoretical point of view, acoustic phonons can be thought of as the Goldstones associated with the spontaneous breaking of continuous translation symmetry~\cite{Leutwyler_HPA_1997,Esposito_PRD_2020,Nicolis_JHEP_2015,Endlich_JCAP_2013}.

The physical properties listed above depend critically on the electron-phonon interaction (EPI)~\cite{Giustino_RMP_2017, engel2020electron}. The coexistence of electrons and phonons in a single host crystal has recently prompted various authors to search for topological contributions to phonons that originate microscopically from the EPI (see, for example, Refs.~\cite{Li_PRL_2023} and~\cite{Xu_Science_2024}). After two decades of intense research on applying topology to material science properties~\cite{Resta2000,vanderbilt_book,bernevig_book_2013,Xiao2010}, a more general mathematical framework---now universally known as ``quantum geometry'' (QG)~\cite{Provost1980}---is attracting significant attention in the community~\cite{Torma2023, Yu_review_2024, Bouhon2023}. Mathematically, QG is quantified by the gauge-invariant Quantum Geometric Tensor (QGT)~\cite{Provost1980, Page1987, Anandan_PRL_1990, Berry1984}, which is defined in Appendix~\ref{appendixA} and further discussed in Sect.~I in the Supplemental Material~\cite{SM}. QG plays a fundamental role in a vast number of physical phenomena. Its importance in the theory of the insulating state, for example, has been known for more than two decades~\cite{Resta1999, Souza2000, Resta2011}, and has been recently revisited~\cite{Lapa2019, Kwon2024, Jankowski2025, Onishi2024, Komissarov2024, Verma2024}. Interest in QG has been recently revitalized in the context of strongly correlated electron systems with flat bands such as flat-band superconductors~\cite{Peotta2015,Hofmann2020, Rossi2021,Torma2022,herzogarbeitman2022manybodysuperconductivitytopologicalflat, Peotta2023,Hofmann2023}. Quantum transport anomalies in systems with flat bands~\cite{Kruchkov2023} as well as nonlinear optical response functions~\cite{Ahn2021} have also been linked to the QG of states. Direct measurements of the QGT have been reported in optical lattices~\cite{Yi2023}, qubits~\cite{Yu2019, Zheng2022}, microcavities~\cite{Gianfrate2020}, and, more recently, in moiré materials~\cite{Kumar2024}; however, further experimental replication is necessary. 

A recent theoretical work~\cite{Yu_NaturePhys_2024} has incorporated the quantum geometry of electron bands into the theory of EPIs, demonstrating the crucial contributions of the Fubini–Study metric~\cite{Page1987} or its orbital selective version~\cite{Yu_NaturePhys_2024} to the dimensionless electron-phonon coupling constant. Concrete estimates have been provided for two materials, i.e.~graphene and ${\rm MgB}_2$, where the geometric contributions account
for approximately $50\%$ and $90\%$ of the total electron-phonon coupling constant, respectively. In this work, we study the role of the QGT in the dispersion of phonons {\it directly}, using the same Gaussian approximation introduced in Ref.~\cite{Yu_NaturePhys_2024}. We do not consider Holstein-type EPI \cite{holstein1959studiesI, holstein1959studiesII, Sous2021}, for which the approximation used fails. The ions in the lattice experience both their mutual repulsion and the attractive force of the electron cloud. The latter is represented, within the Born-Oppenheimer approximation~\cite{Grosso2014}, by the variation of the electronic ground-state energy induced by the displacement of an ion from its equilibrium position. We show that this electronic term in the ionic equation of motion contains a contribution which is entirely due to the electronic QGT, thus affecting the phonon dispersion of the crystal. We then choose graphene as a case study. The choice of this material is well motivated: it exhibits a non-zero QGT due to its two-site unit cell, the electron dispersion relation in graphene is very well-described by a simple tight-binding model, and the EPI has already been benchmarked in \cite{Yu_NaturePhys_2024} to be well-described by the approximation used here. We find that removing the non-trivial quantum geometric contribution deeply affects the vibrational modes of this material, causing the acoustic phonons to behave in a non-analytic fashion for large wavelengths.

\section{Harmonic theory of phonons and Goldstone sum rules}
Consider a generic multipartite lattice in $D$ dimensions, with $N_{\bm \tau}$ atoms in the unit cell. In the following, we will denote with
$
\bm R_{p\nu} \equiv \bm R_p + \bm \tau_{\nu\vphantom{'}} + \bm u_{p\nu\vphantom{'}} \in {\mathbb R}^D
$
the position of the ions in the lattice, where $\bm R_p$ is a Bravais lattice vector labelled by an integer $p = 1,\dots, N$ in a finite $N$-unit-cell Born-von Karman (BvK) supercell~\cite{Giustino_RMP_2017}, $\bm \tau_{\nu\vphantom{'}}$ is a basis vector in the $\bm R_p = \bm 0$ unit cell labelled by a sublattice number $\nu = 1,\dots,N_{\bm \tau}$, and $\bm u_{p\nu}$ is the displacement of an ion from its equilibrium position $\bm R^0_{p\nu} \equiv \bm R_{p} + \bm \tau_{\nu}$.

Choosing a basis $\{\varphi_{\nu\alpha}(\bm r)\}$ of sufficiently localized wavefunctions, the second-quantized one-electron Hamiltonian in the modified~\cite{Yu_NaturePhys_2024} tight-binding model can be expressed as
\begin{equation}\label{eq:electron-hamiltonian}
\hat{{\cal H}}_{{\rm e}}(\{\bm u_{p\nu}\}) = \sum_{p \nu \alpha \vphantom{'}} \sum_{p'\nu' \alpha'} t_{\nu\nu'}^{\alpha\alpha'} \left( \bm R_{p\nu} - \bm R_{p'\nu'}\right) \hat{c}^{\dagger}_{p\nu \alpha \vphantom{'}}\hat{c}^{\vphantom{\dagger}}_{p'\nu' \alpha'}~,
\end{equation}
where $ \hat{c}^{\dagger}_{p\nu\alpha \vphantom{'}}$ creates an electron in a generic orbital state identified by an additional set of quantum numbers $\alpha$ (including e.g. spin or internal atomic degrees of freedom) on the lattice site $\bm R^0_{p\nu}$, and $t^{\alpha\alpha'}_{\nu\nu'}(\bm r) = [t(\bm r)]_{\nu\nu'}^ {\alpha\alpha'}$ denotes the matrix elements of the hopping matrix $t({\bm r})$, which carries both sublattice and orbital indices. Each matrix element is a smooth hopping function satisfying $[t(-\bm r)]_{\nu\nu'}^ {\alpha\alpha'} = [t^*(\bm r)]_{\nu'\nu}^ {\alpha'\alpha}$ in order to guarantee the Hermiticity of $\hat{\cal H}_{\rm e}(\{\bm u_{p\nu}\})$. 

The Hamiltonian~\eqref{eq:electron-hamiltonian}, its eigenvalues, and eigenstates depend parametrically on the set of displacements $\left\{\bm u_{p\nu}\right\}$ of the ions from their equilibrium positions. All of these quantities are evaluated within the one-electron approximation. Generalizations to include electron-electron interactions are of course possible in the spirit of density functional perturbation theory~\cite{Giustino_RMP_2017} but are beyond the scope of the present work.

The strength of the interaction between neighboring atoms is determined by the interatomic force constants $\bigl[ {\cal C}\bigr]_{p\nu i}^ {p'\nu'j}  =   \bigl[ {\cal C}^{(\text{ion})}\bigr]_{p\nu i}^ {p'\nu'j} +  \bigl[ {\cal C}^{(\text{el})}\bigr]_{p\nu i}^ {p'\nu'j}$, where we highlighted the natural separation of the interaction into an ionic and an electronic contribution, and $i,j$ are Cartesian indices. The key quantity of this Article is the {\it dynamical matrix} (DM) ${\cal D}(\bm q) = {\cal D}^ {\rm (ion)}(\bm q) +  {\cal D}^ {\rm (el)}(\bm q)$, which is the Fourier transform of the force constant matrix and can be decomposed in the same fashion. The dynamical matrix yields the phonon dispersion lines $\omega_\ell(\bm q)$, where $\ell$ is a branch label, via the eigenvalue equation
\begin{equation}
\label{eq:full_eigenvalue_problem}
    \sum_{\nu'j}   \bigl[{\cal D}(\bm q)\bigr]_{\nu i}^ {\nu'j}\, w^{(\ell)}_{\nu'j}(\bm q) = \omega_\ell^2(\bm q) \,w^{(\ell)}_{\nu i}(\bm q)~.
\end{equation}
As shown in Sect.~II of Ref.~\cite{SM}, due to the invariance under global translations of the entire crystal, both the full DM and, separately, its ionic and electronic parts, obey the so-called acoustic sum rules~\cite{Gonze1997}:
\begin{align}
\label{eq:acoustic-sum-rule}
    \sum_{\nu'} \sqrt{M_{\nu'}}\bigl[{\cal D} ({\bm 0})\bigr]_{\nu i}^ {\nu'j}& =0\quad\forall~\nu, i, j~,\\
\label{eq:acoustic-sum-rules}
    \sum_{\nu'} \sqrt{M_{\nu'}}\bigl[{\cal D}^ {\rm (ion/el)}({\bm 0})\bigr]_{\nu i}^ {\nu'j}& =0\quad\forall~\nu, i, j~.
\end{align}
These sum rules imply that the acoustic phonon modes, i.e. those eigenmodes $\bm w_\nu^ {\ell}(\bm q)$ in which atoms in the unit cell vibrate in phase with each other, have to be {\it Goldstones}, i.e.~their frequency must go to zero in the long-wavelength limit $\bm q \to \bm 0$. 
Indeed, when $\bm q = \bm 0$,  acoustic modes correspond to global translations of the lattice, which are part of the symmetry group of the crystal.

In order to obtain formulas for ${\cal D}^ {\rm (el)}(\bm q)$, we must first evaluate its real-space counterpart ${\cal C}^ {(\rm el)}$. This purely electronic contribution to the force-constant matrix is traditionally calculated directly from Eq.~\eqref{eq:electron-hamiltonian}---see Sect.~III in Ref.~\cite{SM}---by exploiting the Hellmann-Feynman theorem~\cite{GiulianiVignale} and its Epstein generalization~\cite{Grosso2014}. Formulas for ${\cal C}^ {(\rm el)}$ obtained in such a way are given in Eqs.~\eqref{eq:cI}-\eqref{eq:cII} in App.~\ref{appendixB}. Here, we rewrite them in a form that has a clear and appealing physical interpretation. In calculating ${\cal C}^ {(\rm el)}$, one studies the response of the electronic Hamiltonian $\hat{\cal H}_{\rm e} (\{\bm u_{p\nu}\})$ to a static displacement field $\bm u_{p\nu}$, for small displacements. The latter quantity plays the role of a time-independent vector potential ${\bm A}({\bm r})$~\cite{Vozmediano2010} in the usual linear response theory for electronic systems coupled to the electromagnetic field~\cite{GiulianiVignale}. In such an analogy, it is easy to recognize that two key operators entering
Eqs.~\eqref{eq:cI}-\eqref{eq:cII} are: i) the ``paramagnetic'' current operator
\begin{equation}
    \label{eq:paramagnetic-current}
    \hat{j}_{p\nu i} \equiv \left.\frac{\partial \hat{\cal H}_{\rm e}}{\partial u_{p\nu i}}\right|_0~,
\end{equation}
which controls the first-order (i.e.~linear) coupling between the electronic degrees of freedom and $\bm u_{p\nu}$, and ii) the ``diamagnetic'' tensor
\begin{equation}
    \label{eq:diamagnetic-operator}
    \hat{\cal T}_{p\nu i}^ {p'\nu'j} \equiv \left.\frac{\partial^ 2 \hat{\cal H}_{\rm e}}{\partial u_{p\nu i}\partial u_{p'\nu'j}}\right|_0~,
\end{equation}
which controls the second-order coupling between the electronic degrees of freedom and $\bm u_{p\nu}$. Recognizing these operators in Eqs.~\eqref{eq:cI}-\eqref{eq:cII} leads to the following identities:
\begin{align}
\label{eq:cI-analogue}
    \bigl[ {\cal C}^{(\text{el,\,1})}\bigr]_{p\nu i}^ {p'\nu'j} & \equiv \chi_{j_{p\nu i}\, j_{p'\nu'j}}^ {\vphantom{\dagger}}~,\\[3pt]
\label{eq:cII-analogue}
    \bigl[ {\cal C}^{(\text{el,\,2})}\bigr]_{p\nu i}^ {p'\nu'j} & \equiv \left\langle \phi^ {(0)}_{\{\bm 0\}}\middle|  \hat{\cal T}_{p\nu i}^ {p'\nu'j}\middle|\phi^ {(0)}_{\{\bm 0\}}\right\rangle~,
\end{align}
where $\chi_{j_{p\nu i}\, j_{p'\nu'j}}^ {\vphantom{\dagger}}$ is the ``paramagnetic'' contribution to the $T=0$ current-current response function~\cite{GiulianiVignale}, in the static (i.e.~$\omega=0$) limit. We therefore conclude that Eq.~\eqref{eq:cI} is simply $\chi_{j_{p\nu i}\, j_{p'\nu'j}}^ {\vphantom{\dagger}}$ in the Lehmann representation~\cite{GiulianiVignale}, while Eq.~\eqref{eq:cII} is the ``diamagnetic'' contribution. The sum of these two contributions appearing in Eq.~(\ref{eq:force-constants-I-II}), which {\it is} the force constant matrix, actually coincides with the {\it physical} current-current response function of the electronic system.

In light of this formal analogy, we can rederive the acoustic sum rule~\eqref{eq:acoustic-sum-rules} in a manner that emphasizes the Goldstone nature of the acoustic branches. Indeed, it is clear that a finite, physical electronic current cannot flow in response to a static and {\it uniform} displacement ${\bm u}_{p\nu} \equiv {\bm u}~(\forall~p,\nu)$ of all the atoms, which is simply a global translation of the lattice. Mathematically, this statement translates into the so-called TRK sum rule~\cite{Sakurai2017}:
\begin{equation}
\label{eq:TRK-sumrule}
    \sum_{p'\nu'}  \left[\chi_{j_{p\nu i}\, j_{p'\nu'j}}^ {\vphantom{\dagger}} + \left\langle \phi^ {(0)}_{\{\bm 0\}}\middle|  \hat{\cal T}_{p\nu i}^ {p'\nu'j}\middle|\phi^ {(0)}_{\{\bm 0\}}\right\rangle\right] = 0~,
\end{equation}
which must hold  $\forall~p,\nu,i,j$. Eq.~(\ref{eq:TRK-sumrule}) is just the real-space version of the acoustic sum rule~\eqref{eq:acoustic-sum-rules} for the electronic contribution to the force constants. In the linear response theory to a static gauge field ${\bm A}({\bm r})$, the TRK sum rule expresses the fact that a finite physical current cannot flow in response to a static and spatially-uniform vector potential ${\bm A}({\bm r}) ={\bm A}$ since this can be gauged away~\cite{Andolina2019}. We refer the reader to Sects..~III and~IV in Ref.~\cite{SM} for all the necessary technical details on the calculation of the electronic contribution to the DM,  Sect.~V for the construction of the formal analogy with linear response theory, and  Sect.~VI for the explicit proof of the TRK sum rule in a simple 1D toy model. Furthermore, in Sect.~X we calculate the geometric group velocity renormalization for the acoustic mode in the case of the EFK model. 

\section{Quantum geometric terms in the dynamical matrix}
We now proceed to single out the quantum geometric contribution to ${\cal D}^{({\rm el})}({\bm q})$. As mentioned above, $\hbar \bm k$ has a sharp physical meaning in a crystal: it is the quasi-momentum of a Bloch electron and spans the FBZ. Each band, labeled by a discrete index $n$, can be interpreted as a separate manifold of states and is thus characterized by its own QGT, i.e.~$\mathcal{Q}^{(n)}_{ij}(\bm k)$. As shown in Sect.~I in Ref.~\cite{SM}, the QGT can be calculated from the following expression:
\begin{equation}
    \label{eq:qgt-projectors}
    \mathcal{Q}^{(n)}_{ij}(\bm k) = \Tr \left\{\partial_{k_i} P_n(\bm k) \bigl[1- P_n(\bm k)\bigr]\partial_{k_j} P_n(\bm k)\right\}~,
\end{equation}
where $P_n(\bm k)$ is the projection matrix onto a Bloch eigenvector~\cite{Yu_NaturePhys_2024, note_on_indices}. As the group velocity $\partial_{k_i} E_n(\bm k)$ characterizes the dispersion properties of a band~\cite{Grosso2014}, the derivatives $\partial_{k_i} P_{n}(\bm k)$ of the projector fully characterize the quantum geometric properties. For single-band systems, $P(\bm k) = 1$ is the only non-zero projector, leading to a trivial QG. 
\begin{figure}[t]
    \centering
    \begin{tabular}{c}
        \begin{overpic}[width=1\columnwidth]{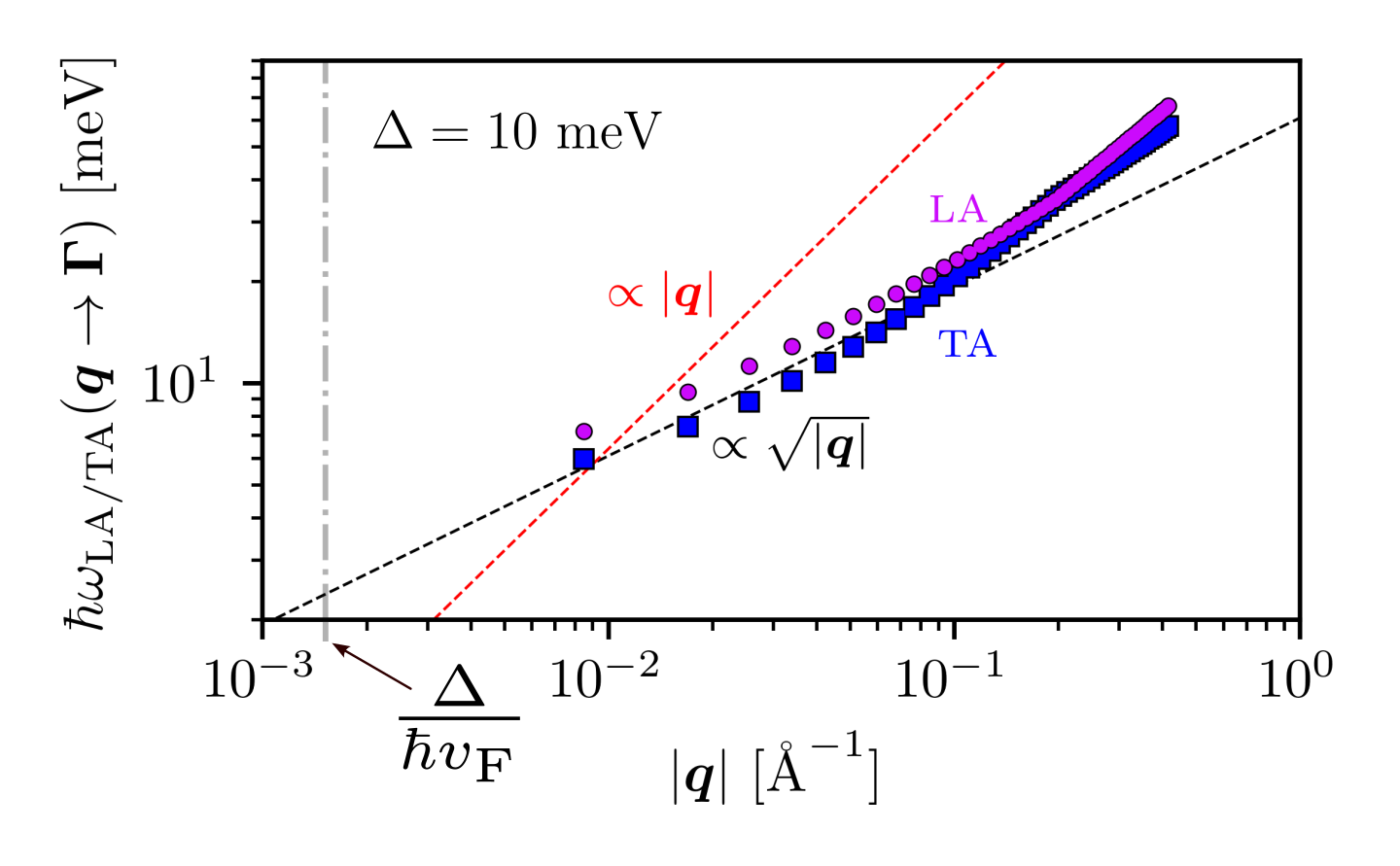}
            \put(25,150){\rm{(a)}}
        \end{overpic} \\   
        \begin{overpic}[width=1\columnwidth]{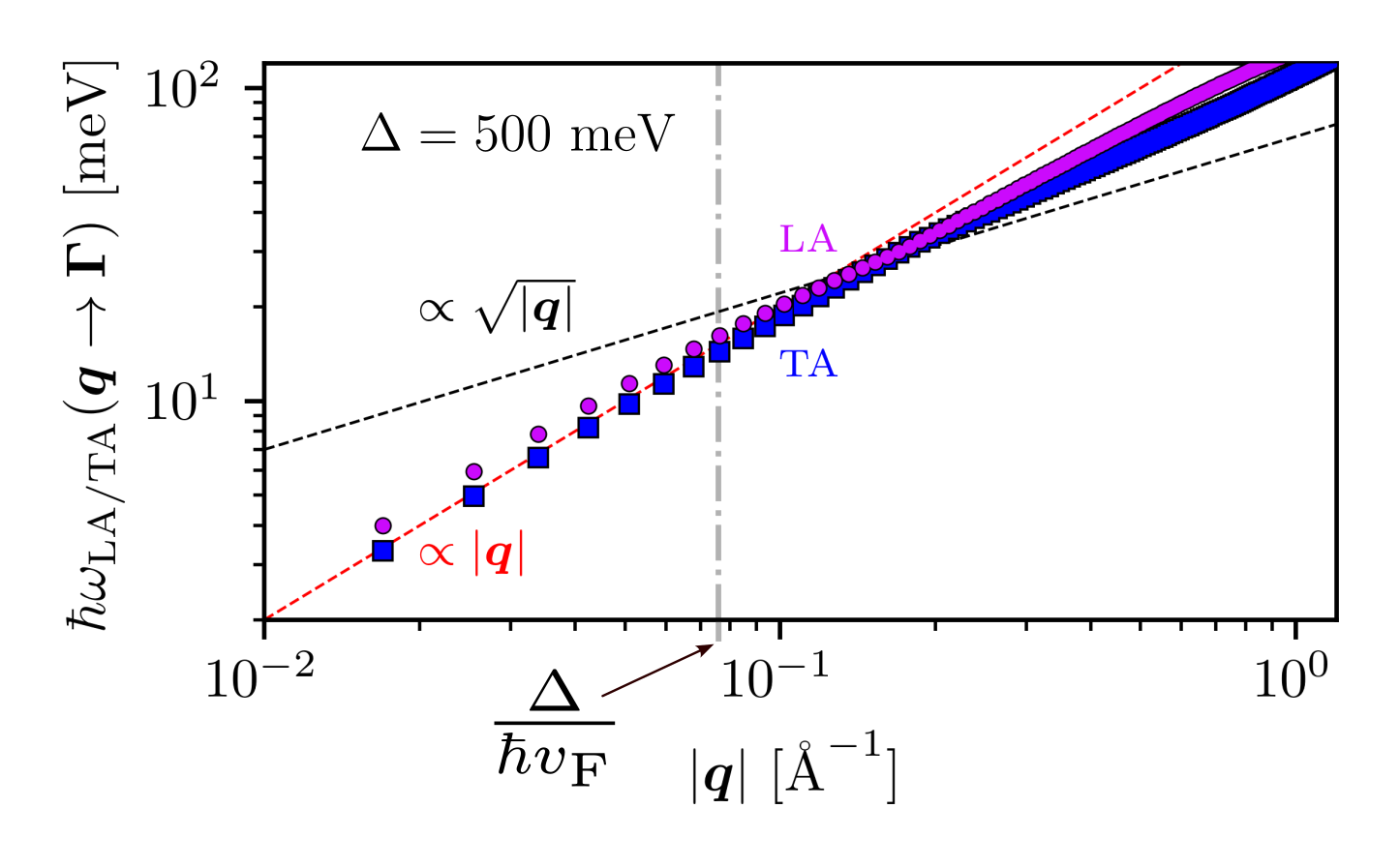}
            \put(25,150){\rm{(b)}}
        \end{overpic}
    \end{tabular}
    \caption{(Color online) { (a) Zoom-in of Fig.~\ref{fig:1} ($\Delta=10~{\rm meV}$) near the $\Gamma$ point. Data for the LA (TA) branch are denoted by magenta (blue) filled circles (squares). The two lowest-lying eigenfrequencies of ${\cal D}_{\rm ng}(\bm q)$, i.e. $\tilde{\omega}_{\rm LA/TA}(\bm q)$, follow a $\propto\sqrt{|\bm q|}$ trend for $\bm q$ near $\Gamma$ as long as $q$ is larger than the threshold $q_{\rm thr} = \Delta/\hbar v_{\rm F}$ (vertical dash-dotted line). This behavior is well-described by a massless Dirac fermion model. Dashed lines serve as guides to the eye, illustrating linear and $\sqrt{|\bm q|}$ trends. (b) For a relatively large fermion gap $\Delta = 500$ meV, the eigenfrequencies $\tilde{\omega}_{\rm LA/TA}(\bm q)$ behave linearly in $\bm q$ for $|\bm q| < q_{\rm thr}$, as the quantum metric is regularized by $\Delta$. This is explained by a massive Dirac fermion model.}}
    \label{fig:2}
\end{figure}
We now have a simple prescription to identify the non-trivial geometric terms in the DM. In order to do so, we exploit the same Gaussian approximation (GA) for the hopping amplitude $t_{\nu\nu'}^{\alpha\alpha'}(\bm r)$ that was proposed and extensively discussed in Ref.~\cite{Yu_NaturePhys_2024}, i.e.~we take
\begin{equation}\label{eq:GA_hopping_matrix}
t_{\nu\nu'}^{\alpha\alpha'}(\bm r) = t_{\nu\nu'}^{\alpha\alpha'}(\bm 0) \exp\left(\gamma_{\nu\nu'}^{\alpha\alpha'} r^ 2/2\right)~, 
\end{equation}
with $\gamma_{\nu\nu'}^{\alpha\alpha'} \equiv [\gamma]_{\nu\nu'}^{\alpha\alpha'} < 0$. The theory described in this Article relies on the GA for the hopping matrix elements and is therefore appropriate for materials whose hopping functions are well described by a spatial profile with a Gaussian shape.

The electronic term ${\cal D}^{\rm (el)}(\bm q)$ in the DM is built from the Bloch Hamiltonian $h(\bm k) = \sum_n E_n(\bm k)P_n(\bm k)$ and its gradient and Hessian $f_i(\bm k)$ and $M_{ij}(\bm k)$, defined in App.~\ref{appendixC}. As shown in the latter, the GA allows us to decompose these quantities into dispersive (or energetic) and geometric parts. For $f_i(\bm k)$, we have $f_i(\bm k) =  f_i^{\rm E}(\bm k) + f_i^{\rm g}(\bm k)$, with
\begin{align}
        \label{eq:f-e}
             f_i^{\rm E}(\bm k) & =  i \gamma \sum_n \partial_{k_i} E_n(\bm k) P_n(\bm k)~,\\
        \label{eq:f-g}
         f_i^{\rm g}(\bm k) & = i\gamma \sum_n  E_n(\bm k) \partial_{k_i} P_n(\bm k)~.
\end{align}
The quantity $f_i^{\rm g}(\bm k)$ is the geometric contribution to $f_i(\bm k)$~\cite{Yu_NaturePhys_2024}, since it arises due to the momentum dependence of $P_n(\bm k)$ and is zero in the case of trivial geometry. The term $f_i^{\rm E}(\bm k)$, on the other hand, is a {\it dispersive} term, as it scales with the group velocity and is zero in the case of a perfectly flat band. The Hessian tensor $M_{ij}(\bm k)$ is decomposed in the same fashion, as shown in detail in Sect.~VII of Ref.~\cite{SM}. While the geometric nature of the gradient $f_i(\bm k)$ of the Bloch Hamiltonian has been discussed at length in Ref.~\cite{Yu_NaturePhys_2024}---where it was shown that $f_i(\bm k)$ controls the strength of the EPI---in this Article we deal with higher-order contributions contained in the Hessian tensor $M_{ij}(\bm k)$, which scale as the second-order derivatives $\partial_{k_i}\partial_{k_j} P_n(\bm k)$ of the projection matrix. While these are obviously recognizable as non-trivial geometric terms, they do not enter the QGT. This suggests the intriguing possibility that higher-order geometric structures have a role in defining the properties of crystals~\cite{Hetenyi2023}.

Inserting the different contributions \eqref{eq:f-e}-\eqref{eq:f-g} to $f_i(\bm k)$ into Eqs.~\eqref{eq:cI}-\eqref{eq:cII} and doing the same for $M_{ij}(\bm k)$, we can decompose the purely electronic part of the DM, ${\cal D}^ {\rm (el)}(\bm q)$, which has been introduced immediately after Eq.~(\ref{eq:dynamical-matrix}) into the sum of non-geometric and geometric contributions: ${\cal D}^ {\rm (el)}(\bm q) =  {\cal D}^ {\rm (el)}_{\rm ng}(\bm q) + {\cal D}_{\rm g}(\bm q)$. Here, ${\cal D}_{\rm g}(\bm q)$ is the non-trivial geometric contribution to the DM, which is solely of electronic nature and therefore has no ionic contribution, while ${\cal D}^ {\rm (el)}_{\rm ng}(\bm q)$ is that part of the electronic term in the DM which would be non-zero also in the case of a trivial QG.  The entire DM can thus be decomposed as
\begin{equation}
    \label{eq:dm-decomposition2}
        {\cal D}(\bm q) =  {\cal D}_{\rm ng}(\bm q) +  {\cal D}_{\rm g}(\bm q)~,
\end{equation}
where~${\cal D}_{\rm ng}(\bm q) \equiv  {\cal D}^ {\rm (ion)}(\bm q ) +{\cal D}_{\rm ng}^ {\rm (el)}(\bm q)$. Therefore, we can calculate the phonon dispersions after the removal of nontrivial geometric terms by solving the eigenvalue equation
\begin{equation}
\label{eq:full_eigenvalue_problem-ng}
    \sum_{\nu'j}   \bigl[{\cal D}_{\rm ng}(\bm q)\bigr]_{\nu i}^ {\nu'j}\, \Tilde w^{(\ell)}_{\nu'j}(\bm q) = \Tilde \omega_\ell^2(\bm q) \,\Tilde w^{(\ell)}_{\nu i}(\bm q)~.
\end{equation}
Explicit expressions for the geometric contribution ${\cal D}_{\rm g}(\bm q)$ to the DM are provided in Sect.~VII in Ref.~\cite{SM}. 

\section{Graphene as a case study}

The calculation of the geometric contribution to the phonons in graphene proceeds as schematized in App.~\ref{appendixD}. Details on the numerics and {\it ab initio} methods can be found in Sect.~VIII of Ref.~\cite{SM} and Refs.~\cite{Reich2002, Falkovsky2007, DFPT_Baroni, QE1, QE2, prandini2018precision,DalCorso, PBE, Grimme2006, MP_smearing, MP, Sohier2017}. Our main numerical results for $\omega_\ell(\bm q)$ and $\Tilde{\omega}_\ell(\bm q)$ for the case of in-plane phonons in graphene are reported in Figs.~\ref{fig:1}-\ref{fig:2}. A finite fermion gap $\Delta = 10$~meV has been introduced to perform this calculation~\cite{Katsnelson2012, Giovannetti2007}. 

{
The quantitative impact of quantum geometry can be estimated by looking at the normalized difference $\delta_{\ell}(\bm q) \equiv [\tilde{\omega}_{\ell}(\bm q) - \omega_{\ell}(\bm q)]/\omega_{\ell}(\bm q)$. Regarding optical modes, we have $\delta_{\rm LO/TO}(\bm q) \simeq 10\%$ at the $\Gamma$ point while values spanning the range $5\%$-$15\%$ are achieved near the edge of the FBZ. As for the acoustic modes, we have $\delta_{\rm LA/TA}( \Gamma) = 0$ at the zone center, while at the edges of the FBZ we find $\delta_{\rm LA}( K) \simeq \delta_{\rm LA}( M) \simeq 6\%$ and $\delta_{\rm TA}( K) \simeq 20\%$, $\delta_{\rm TA}( M) \simeq 27\%$. Although these quantitative effects are significant, the most striking feature is that $\delta_{\rm LA/TA}(\bm q)$ diverges as $\bm q \to \Gamma$, while remaining exactly zero at $\bm q \equiv \Gamma$. Indeed, by zooming in near the $\Gamma$ point, we find that $\tilde{\omega}_{\rm LA/TA}(\bm q)$ follows a non-analytic $\propto\sqrt{|\bm q|}$ trend, as emphasized in Fig.~\ref{fig:2}(a).

This rather singular behavior is explained by analytical calculations based on a massless Dirac fermion model, as in Sect.~IX of Ref.~\cite{SM}. The non-analytic $\propto \sqrt{|\bm q|}$ behavior arises from the singularity of electronic eigenstate derivatives at the Dirac point, which cause a divergence of the quantum metric. Indeed, by performing analytical calculations for massive Dirac fermions with a gap $\Delta$, we find that the eigenstate derivatives are properly regularized, and the acoustic phonon dispersions recover their linear dependence on $\bm q$ at long wavelength. This can be seen numerically by choosing a value of $\Delta$ that is large enough to resolve the long-wavelength behavior below the threshold $q_{\rm thr} = \Delta/\hbar v_{\rm F}$, as shown in Fig.~\ref{fig:2}(b).

We emphasize that this effect is purely geometric, as the removal of the entire electronic contribution of the $\pi$ bands} ${\cal D}^{(\rm el)}(\bm q)$ does not introduce any singular behavior to the large-wavelength phonon dispersions. In fact, the removal of ${\cal D}^{(\rm el)}(\bm q)$ is shown to introduce a group velocity renormalization of $\sim 20\%$ for a complete tight-binding model, and to be exactly vanishing for the case of massless Dirac fermions (see Sect.~VIII in Ref.~\cite{SM}). 

We note that the symmetry of the phonon spectrum~\cite{Manes2007} is preserved when either the electronic or geometric contribution is removed, as evidenced by the retention of the twofold degenerate bands at $\Gamma$ and K (see Fig.~\ref{fig:1} and Ref.~\cite{SM}). While a finite $\Delta$ breaks inversion symmetry and can, in principle, lift the twofold degeneracy at K by opening a gap, the effect remains negligible for small $\Delta$.

\section{Discussion}
In summary, we developed a framework to study the electronic contribution to the dynamical matrix of a generic crystal. We demonstrated that this contribution can be neatly separated into geometric and non-geometric terms within the paradigm of the Gaussian approximation, and that both kinds of terms satisfy acoustic sum rules. Using graphene as a case study, we examined and quantified the behavior of these contributions to the acoustic modes through both analytical and numerical methods. We found that the geometric contribution behaves in a highly non-analytic $\propto \sqrt{|\bm q|}$ fashion in the long-wavelength limit $\bm q \to \Gamma$, while the total electronic contributions remains regular at every $\bm q$. Additionally, we find that the total electronic contribution to the acoustic mode from a gapless Dirac node with linear dispersion vanishes. However, in a realistic tight-binding model, high-energy electronic states---beyond those captured by the simple Dirac Hamiltonian---can yield nonzero contributions, resulting in a weak renormalization of the acoustic mode velocity.

\section*{Acknowledgments}
G.~P. and M.~P. wish to thank Angelo Esposito for useful and inspiring discussions. This work was supported by the MUR - Italian Ministry of University and Research under the ``Research projects of relevant national interest  - PRIN 2020''  - Project No.~2020JLZ52N (``Light-matter interactions and the collective behavior of quantum 2D materials, q-LIMA'') and by the European Union under grant agreements No. 101131579 - Exqiral and No.~873028 - Hydrotronics. Views and opinions expressed are however those of the author(s) only and do not necessarily reflect those of the European Union or the European Commission. Neither the European Union nor the granting authority can be held responsible for them. 
B.~A.~B. and H.~H. were supported by the Gordon and Betty Moore Foundation through Grant No. GBMF8685 towards the Princeton theory program, the Gordon and Betty Moore Foundation’s EPiQS Initiative (Grant No. GBMF11070), the Office of Naval Research (ONR Grant No. N00014-20-1-2303), the Global Collaborative Network Grant at Princeton University, the Simons Investigator Grant No. 404513, the BSF Israel US foundation No. 2018226, the NSF-MERSEC (Grant No. MERSEC DMR 2011750), the Simons Collaboration on New Frontiers in Superconductivity (Grant No. SFI-MPS-NFS-00006741-01), and the Schmidt Foundation at the Princeton University. Y.~J. was supported by the European Research Council (ERC) under the European Union’s Horizon 2020 research and innovation program (Grant Agreement No. 101020833), as well as by the IKUR Strategy under the collaboration agreement between Ikerbasque Foundation and DIPC on behalf of the Department of Education of the Basque Government. 

\appendix
\section{The quantum geometric tensor}
\label{appendixA}
In 1980, Provost and Vallee~\cite{Provost1980} understood that the usual Hermitian product on the projective Hilbert space induces a meaningful metric tensor on any manifold of quantum states. This intuition led them to introduce the quantum geometric tensor (QGT):
\begin{equation}\label{eq:QGT_explicit}
\mathcal{Q}_{ij}(\bm k) = g_{ij}({\bm k}) - \frac{i}{2} {\cal F}_{ij}({\bm k})~.
\end{equation}
Here, $g_{ij}({\bm k})$ is a real, symmetric, and gauge-invariant tensor usually dubbed quantum metric (or Fubini-Study metric~\cite{Page1987,Anandan_PRL_1990}) and defined by
\begin{equation}\label{eq:quantum_metric_tensor_explicit}
g_{ij}({\bm k}) = \Re \bigl[\left\langle\partial_{k_i}\psi_{\bm k}\middle\vert\partial_{k_j}\psi_{\bm k}\right\rangle \bigr]-  \left\langle\partial_{k_i}\psi_{\bm k}\middle\vert\psi_{\bm k}\right\rangle\left\langle\psi_{\bm k}\middle\vert \partial_{k_j}\psi_{\bm k}\right\rangle~.
\end{equation}
In Eq.~(\ref{eq:quantum_metric_tensor_explicit}), $\{\vert\psi_{\bm k}\rangle\}$ is a family of normalized vectors of some Hilbert space which smoothly depend on an $n$-dimensional parameter ${\bm k}\in {\mathbb R}^n$ (this is usually the Bloch quasi-momentum in solid-state physics). The quantum metric tensor $g_{ij}({\bm k})$ is a measure of the distance in amplitude between infinitesimally close wavefunctions in ${\bm k}$-space.

The second term in Eq.~(\ref{eq:QGT_explicit}) is instead the more familiar Berry curvature~\cite{Berry1984,vanderbilt_book,bernevig_book_2013}:
\begin{equation}
    \label{eq:berry-curvature}
    {\cal F}_{ij}({\bm k})  =\left\langle\partial_{k_i}\psi_{\bm k}\middle\vert\partial_{k_j}\psi_{\bm k}\right\rangle- \left\langle\partial_{k_j}\psi_{\bm k}\middle\vert\partial_{k_i}\psi_{\bm k}\right\rangle~,
\end{equation}
which, of course, is also gauge invariant. In Sect.~I of the Supplemental Material~\cite{SM}, we discuss the QGT for electrons in a periodic potential and prove the equivalence of the two formulas~\eqref{eq:qgt-projectors} and~\eqref{eq:QGT_explicit}. 
\section{Force constants and dynamical matrix}
\label{appendixB}
In the Born-Oppenheimer approximation, the classical ionic Hamiltonian can be expanded up to second order in the ionic displacements (harmonic approximation) as follows:
\begin{equation}
    \label{eq:harmonic-hamiltonian}
    \hat{\cal H}_\text{ion} = \sum_{p\nu} \frac{\bm P^2_{p\nu}}{2 M_{\nu}}  
    +\frac{1}{2}\sum_{pp'}\sum_{\nu \nu'} \sum_{ij}\bigl[ {\cal C}\bigr]_{p \nu i}^ {p'\nu'j} u_{p\nu i} u_{p'\nu'j}~,
\end{equation}
The interatomic force constant can be naturally decomposed into the sum of two terms, i.e.
\begin{equation}
    \begin{split}
    \bigl[{\cal C}\bigr]_{p\nu i}^ {p'\nu'j} & = \bigl[{\cal C}^ {\rm (ion)}\bigr]_{p\nu i}^ {p'\nu'j} +  \bigl[{\cal C}^ {\rm (el)}\bigr]_{p\nu i}^ {p'\nu'j}\\
    & = \frac{\partial^ 2 {\cal V}}{\partial u_{p\nu i} \partial u_{p'\nu' j}} + \frac{\partial^ 2 {\cal E}_0}{\partial u_{p\nu i} \partial u_{p'\nu' j}}~,
    \end{split}
\end{equation}
obtained respectively by differentiating the inter-ionic Coulomb potential ${\cal V}(\{\bm u_{p\nu}\})$ and the electronic ground state energy ${\cal E}_{0}(\{\bm u_{p\nu}\})$.

We now formally introduce the dynamical matrix, whose diagonalization results in the phonon dispersion. Indeed, the dispersion $\omega_\ell(\bm q)$ of the $\ell$-th phononic branch, where~$\ell = 1, \dots, D \times N_{\bm \tau}$ and $\bm q$ is the phonon wavevector, is determined by the eigenvalue equation~\eqref{eq:full_eigenvalue_problem} in the main text, where the DM $\bigl[{\cal D}(\bm q)\bigr]_{\nu i}^ {\nu'j}$ is defined as
\begin{equation}
    \label{eq:dynamical-matrix}
\bigl[{\cal D}(\bm q)\bigr]_{\nu i}^ {\nu'j} \equiv \frac{1}{\sqrt{M_{\nu\vphantom{'}} M_{\nu'}}}\sum_{p} e^{-i\bm q\cdot(\bm R_p + \bm \tau_{\nu\vphantom{'}} -\bm \tau_{\nu'})} \,\bigl[ {\cal C}\bigr]_{p\nu i}^ {p'= 0,\,\nu'j}~.
\end{equation}
The decomposition~${\cal D}(\bm q) = {\cal D}^{\rm (ion)}(\bm q) + {\cal D}^{\rm (el)}(\bm q)$, with the latter term depending exclusively on electronic degrees of freedom, comes naturally from Eq.~\eqref{eq:force-constants-I-II} and linearity of the Fourier transform. 

The general formulas for the electronic force constants are derived from the ground state energy
\begin{equation}
    {\cal E}_0 = \left\langle \phi_{\{\bm u_{p\nu}\}}^{(0)} \middle| \hat{\cal H}_{\rm e}(\{ \bm u_{p\nu}\})\middle|\phi_{\{\bm u_{p\nu}\}}^{(0)}\right\rangle
\end{equation}
by differentiating twice with respect to the atomic displacements, and applying the Hellmann-Feynman theorem~\cite{GiulianiVignale} and its Epstein generalization~\cite{Grosso2014}. We obtained
\begin{equation}
\label{eq:force-constants-I-II}
        \bigl[ {\cal C}^{(\text{el})}\bigr]_{p\nu i}^ {p'\nu'j} = 
        \bigl[ {\cal C}^{(\text{el,\,1})}\bigr]_{p\nu i}^ {p'\nu'j}  +
        \bigl[ {\cal C}^{(\text{el,\,2})}\bigr]_{p\nu i}^ {p'\nu'j}~,
\end{equation}
with
\begin{align}
\label{eq:cI}
    \begin{split}
    \bigl[ {\cal C}^{(\text{el,\,1})}\bigr]_{p\nu i}^ {p'\nu'j}  \equiv & \sum_{m\neq 0} \frac{1}{{\cal E}_0-{\cal E}_m} \left \langle \phi^{(m )}_{\{\bm 0\}} \middle| \partial_{p\nu i} \hat{\cal H}_{\rm e}\middle| \phi^ {(0)}_{\{\bm 0\}}\right\rangle \times\\&\times \left \langle \phi^{(0 )}_{\{\bm 0\}} \middle| \partial_{p'\nu' j} \hat{\cal H}_{\rm e}\middle| \phi^ {(m)}_{\{\bm 0\}}\right\rangle+ {\rm H.c.}~,
    \end{split}\\[3pt]
\label{eq:cII}
    \bigl[ {\cal C}^{(\text{el,\,2})}\bigr]_{p\nu i}^ {p'\nu'j} & \equiv \left\langle \phi^ {(0)}_{\{\bm 0\}}\middle| \partial_{p\nu i}\partial_{p'\nu'j} \hat{\cal H}_{\rm e}\middle|\phi^ {(0)}_{\{\bm 0\}}\right\rangle~,
\end{align}
where $\partial_{p\nu i}\equiv \partial/\partial u_{p\nu i}$, $\vert \phi^{(m)}_{\{\bm 0\}} \rangle$ with $m\neq 0$ is an excited state of $\hat{\cal H}_{\rm e}(\{\bm u_{p\nu} =\bm 0\})$ and $\mathcal{E}_m$ is the associated eigenenergy. 
\begin{figure}[t]
    \centering
    \includegraphics[width=0.9\linewidth]{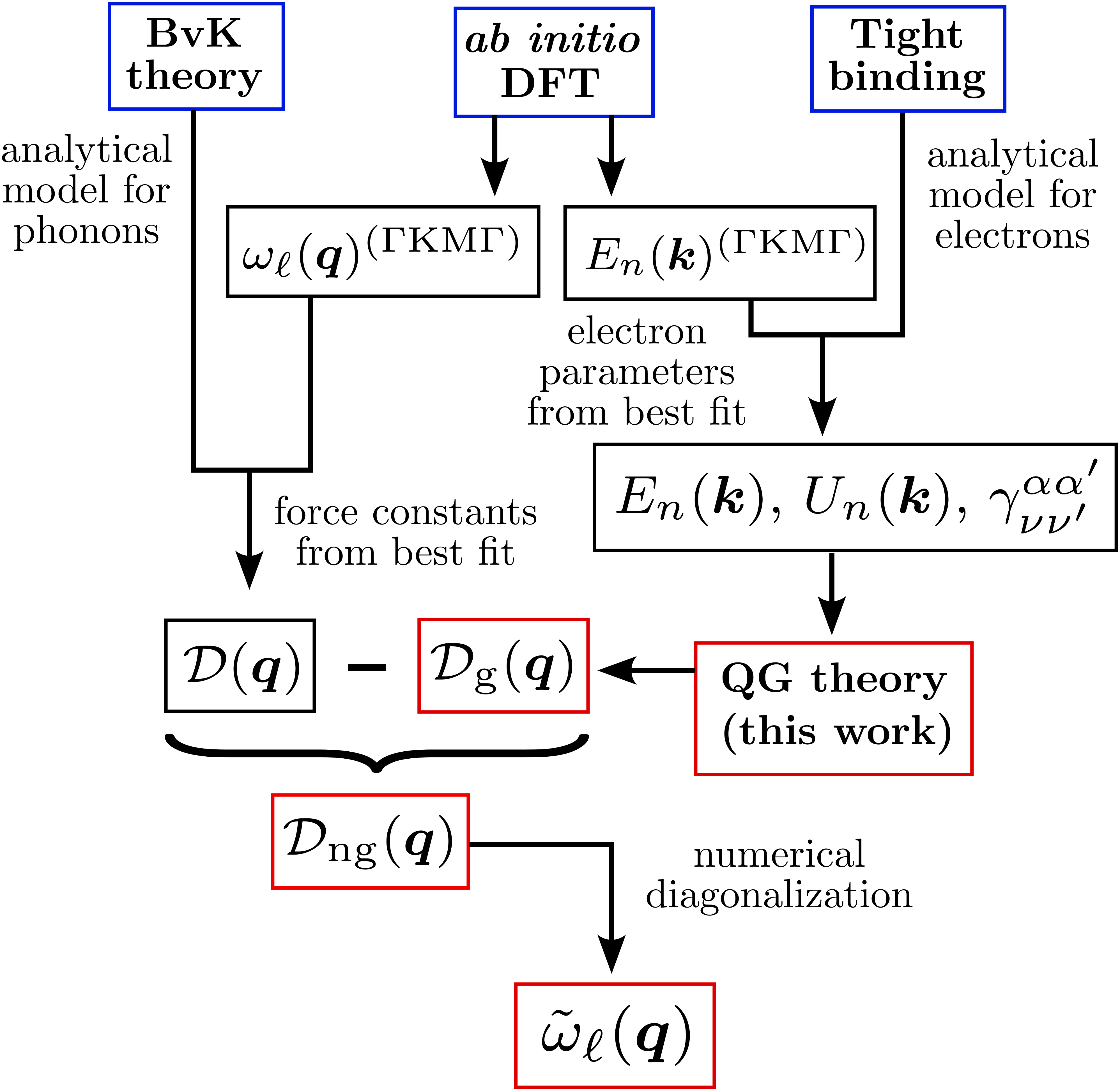}
    \caption{(Color online) Workflow followed for the case of graphene. The same method can be applied to any material for which the relevant hopping integrals are well-described by the GA. Fitting the BvK model to {\it ab initio} data on the high-symmetry line $\Gamma$KM$\Gamma$ allows us to extract the force constants and thus calculate the DM on the entire FBZ. The same fitting procedure is followed in order to obtain all the relevant information on electrons, thus allowing us to calculate the geometric contribution to the DM ${\cal D}_{\rm g}(\bm q)$, from which we obtained the phonon dispersion lines $\Tilde{\omega}_{\ell}(\bm q)$ illustrated in Figs.~\ref{fig:1} and~\ref{fig:2} in the main text.}
    \label{fig:4}
\end{figure}
\section{Nontrivial geometric terms in the DM}
\label{appendixC}
In order to isolate nontrivial terms as defined in the main text, we start from the Fourier transforms of the hopping matrix and of its derivatives, i.e.~$h(\bm k) = \sum_{p} e^{-i\bm k \cdot (\bm R_{p} + \bm \tau_{\nu} - \bm \tau_{\nu'})}\left. t(\bm r)\right|_{\bm R_{p} + \bm \tau_{\nu} - \bm \tau_{\nu'}}$, $f_i(\bm k) = \sum_{p} e^{-i\bm k \cdot (\bm R_{p} + \bm \tau_{\nu} - \bm \tau_{\nu'})} \left.\partial_{r_i} t(\bm r)\right|_{\bm R_{p} + \bm \tau_{\nu} - \bm \tau_{\nu'}}$ and $M_{ij}(\bm k) = \sum_{p} e^{-i\bm k \cdot (\bm R_{p} + \bm \tau_{\nu} - \bm \tau_{\nu'})} \partial_{r_i r_j}^ 2 \left. t(\bm r)\right|_{\bm R_{p} + \bm \tau_{\nu} - \bm \tau_{\nu'}}$. The key property of the GA is that the following identities hold true~\cite{note_on_indices}: $\partial_{r_i} t(\bm r) = \gamma r_i \,t(\bm r)$ and $\partial^2_{r_i r_j} t(\bm r) = (\gamma \delta_{ij} + \gamma^2 r_i r_j) \,t(\bm r)$. We therefore find the following results, which are valid in the framework of the GA:
\begin{align}\label{eq:f-function-ga}
f_i(\bm k) & =  i \gamma\partial_{k_i} h(\bm k)~,\\
\label{eq:m-function-ga}
M_{ij}(\bm k) & =  \big(\gamma\delta_{ij} - \gamma^2 \partial_{k_i k_j}^2\big) h(\bm k)~.
\end{align}
The Bloch Hamiltonian $h(\bm k)$ defines both the energy bands $E_n(\bm k)$ and the Bloch eigenvectors $U_n(\bm k)$ via the eigenvalue equation~$h(\bm k) U_n(\bm k) = E_n(\bm k)U_n(\bm k)$, thus providing the following explicit form $P_{n}(\bm k) = U_n(\bm k)U_n^ \dagger(\bm k)$ for the projection matrices. Identities~\eqref{eq:f-function-ga} and \eqref{eq:m-function-ga}, together with the spectral decomposition $h(\bm k) = \sum_n E_n(\bm k) P_n(\bm k)$, therefore allow us to get to Eqs.~\eqref{eq:f-e}-\eqref{eq:f-g}. The analogous procedure to follow for the Hessian $M_{ij}(\bm k)$ is shown in Sect.~VII of the SM~\cite{SM}.

\section{Details on the case of graphene}
\label{appendixD}
The calculation for the case of graphene proceeds as schematized in Fig.~\ref{fig:4}. The DM is calculated analytically within a next-nearest-neighbor (NNN) BvK model~\cite{Falkovsky2007}, leaving the force constants as free parameters. The force constants are then determined by fitting the analytical results to {\it ab initio} density functional perturbation theory (DFPT)~\cite{DFPT_Baroni} calculations along the high-symmetry $\Gamma$KM$\Gamma$ path.

The Hamiltonian for graphene electrons is calculated analytically by using a linear combination of atomic orbitals (LCAO) within the next-next-nearest-neighbor (NNNN) tight-binding approximation, leaving hopping and overlap integrals as free parameters~\cite{Reich2002}. These are in turn determined by fitting the analytical model to {\it ab initio} density functional theory (DFT) data on the $\Gamma$KM$\Gamma$ high-symmetry path. This allows us to reconstruct the Hamiltonian on the full FBZ, thus obtaining, by means of numerical diagonalization, all the information on the $\pi$ bands and their geometric structure, i.e. the energy bands $E_{\pm}(\bm k)$ and the Bloch eigenvectors $U_{\pm}(\bm k)$, where the subscript $+$ ($-$) denotes the conduction (valence) band. Only these two bands are considered in this calculation, as the hopping integrals that generate them are very well-described by the GA, as shown in Sect.~VIII of Ref.~\cite{SM}. The spatial dependence of $t^{p_z p_z}_{AA}(\bm r)$ and $t^{p_z p_z}_{AB}(\bm r)$, where $A$ and $B$ are sublattice labels, is obtained by repeating the best-fit procedure to determine the hopping integrals on a strained lattice, i.e.~a graphene lattice on which a lattice constant different from that of the relaxed structure has been imposed.

For what regards QG, the relevant quantum number in graphene is the sublattice label, usually referred to as {\it pseudospin}~\cite{Katsnelson2012}. Bloch states corresponding to the $\pi$ bands at a fixed wavevector $\bm k$ are obtained by producing a linear combination of states localized on the $A$ and $B$ sublattices. Thanks to this superposition, QG in graphene is non-trivial~\cite{Torma2023}. However, the singularity in the energy gradient at the location of the Dirac points---i.e.~at the ${\bm K}$ and ${\bm K}^\prime$ points in the FBZ---translates into an infrared singularity in the $\bm k$-derivatives of the projection matrices, namely $\bm \nabla_{\bm k}P_n(\bm k) \sim 1/k$, so the QGT~\eqref{eq:qgt-projectors} is ill-defined. In order to correct this behavior, we propose to use a natural regularization by shifting the energy of $p_z$ orbitals localized on $A$-type sublattice sites by a small amount $\Delta$ with respect to that of the $B$-type sublattice sites. This staggered potential breaks inversion symmetry (as in the case of graphene aligned to hBN~\cite{Giovannetti2007}) and opens a gap $\Delta$ between the two $\pi$ bands, generating a finite band gap~\cite{Katsnelson2012}. The trace of the Fubini-Study metric and the Berry curvature for a gapped graphene are shown in Sect.~VIII of Ref.~\cite{SM}. 

Remarkably, both the first-order and second-order contributions to ${\cal D}_{\rm g}(\bm q)$---namely, the ``paramagnetic'' and ``diamagnetic'' terms as defined above---are found to satisfy acoustic sum rules {\it independently}, i.e.
\begin{equation}
    \sum_{\nu'} \sqrt{M_{\nu'}} \,\bigl[{\cal D}_{\rm g}^{(1)}(\bm 0)\bigr]_{\nu i}^{\nu' j} = \sum_{\nu'} \sqrt{M_{\nu'}} \,\bigl[{\cal D}_{\rm g}^{(2)}(\bm 0)\bigr]_{\nu i}^{\nu' j} = 0~,
\end{equation}
as proven in Sect.~VII. This is a consequence of continuous translational invariance, and it is indeed a stronger constraint compared to the TRK sum rule above. As a direct consequence of this sum rule, the energies of the two lowest-lying eigenmodes of ${\cal D}_{\rm ng}(\bm q)$ will be vanishing in the long-wavelength limit, i.e.
\begin{equation}
    \lim_{\bm q \to \bm 0}\tilde{\omega}_{\rm LA/TA}(\bm q) = 0~.
\end{equation}

\clearpage 
\setcounter{section}{0}
\setcounter{equation}{0}%
\setcounter{figure}{0}%
\setcounter{table}{0}%

\setcounter{page}{1}

\renewcommand{\thetable}{S\arabic{table}}
\renewcommand{\bibnumfmt}[1]{[S#1]}
\renewcommand{\citenumfont}[1]{S#1}

\onecolumngrid
\setcounter{secnumdepth}{3}

\clearpage 
\setcounter{section}{0}
\setcounter{equation}{0}%
\setcounter{figure}{0}%
\setcounter{table}{0}%

\setcounter{page}{1}

\renewcommand{\thetable}{S\arabic{table}}
\renewcommand{\bibnumfmt}[1]{[S#1]}
\renewcommand{\citenumfont}[1]{S#1}

\onecolumngrid
\setcounter{secnumdepth}{3}

\begin{center}

\textbf{\Large Supplemental Material for:\\ ``Phonon spectra, quantum geometry, and the Goldstone theorem''}

\bigskip

Guglielmo Pellitteri,$^{1,\,*}$
Zenan Dai,$^{2}$
Haoyu Hu,$^{3}$
Yi Jiang,$^{4}$
Guido Menichetti,$^{5,\,6}$
Andrea Tomadin,$^{5}$
B. Andrei Bernevig$^{3,\,4,\,7}$ and
Marco Polini$^{5,\,8}$

\bigskip

$^1$\!{\it Scuola Normale Superiore, Piazza dei Cavalieri 7, I-56126 Pisa, Italy}

$^2$\!{\it Zhiyuan College, Shanghai Jiao Tong University, Shanghai 200240, China}

$^3$\!{\it Department of Physics, Princeton University, Princeton, New Jersey 08544,~USA}

$^4$\!{\it Donostia International Physics Center, P. Manuel de Lardizabal 4, 20018 Donostia-San Sebastian,~Spain}

$^5$\!{\it Dipartimento di Fisica dell'Universit\`a di Pisa, Largo Bruno Pontecorvo 3, I-56127 Pisa,~Italy}

$^6$\!{\it Istituto Italiano di Tecnologia, Graphene Labs, Via Morego 30, I-16163 Genova,~Italy}

$^7$ \!{\it IKERBASQUE, Basque Foundation for Science, Maria Diaz de Haro 3, 48013 Bilbao,~Spain}

$^8$\!{\it ICFO-Institut de Ci\`{e}ncies Fot\`{o}niques, The Barcelona Institute of Science and Technology, Av. Carl Friedrich Gauss 3, 08860 Castelldefels (Barcelona),~Spain}

\end{center}

\bigskip

This Supplemental Material is organized into ten Sections. In Sect.~\ref{section1}, we briefly present some fundamental identities regarding the QGT in a crystal. In Sect.~\ref{section2}, we give a proof of the acoustic sum rules for the interatomic force constants. In Sect.~\ref{section3}, we show how to write the electronic part of the force-constant matrix in terms of derivatives of the electronic Hamiltonian only. In Sect.~\ref{section4} we derive explicit, general analytical formulas for the electronic part of the dynamical matrix of a crystal. Sect.~\ref{section5} contains details on the construction of the formal analogy between phonon theory and linear response theory. In Sect.~\ref{section6}, we verify the TRK sum rule for a simple 1D toy model. Sect.~\ref{section7} is devoted to the isolation of nontrivial geometric terms in the dynamical matrix. In Sect.~\ref{section8}, we further discuss the application of the theory to graphene. In Sect.~\ref{section9}, we calculate ${\cal D}_{\rm g}(\bm q)$ in the long-wavelength limit for the case of massive and massless Dirac fermions. In Sect.~\ref{section10}, we derive the geometric dynamical matrix and the group velocity renormalization analytically for the EFK Hamiltonian.

\onecolumngrid

\renewcommand{\thesection}{\Roman{section}}  % Change to Roman numerals

\section{The Quantum Geometric Tensor in solid-state systems}
\label{section1}

The problem of electrons in a periodic potential is described by the Hamiltonian
\begin{equation}
\label{sm-eq:electron-hamiltonan}
    \hat{{\cal H}}_{\rm e} = \frac{\hat{\bm p}^2}{2m_{\rm e}} + \hat{V}(\bm r) \qquad {\rm with} \qquad \hat{V}(\bm r + \bm R^ 0_{p\nu}) = \hat{V}(\bm r)~,
\end{equation}
where we have used the same notation for the lattice structure as in the main text. The eigenstates $\psi_{n \bm k}(\bm r)$ of the Hamiltonian~\eqref{sm-eq:electron-hamiltonan} are of the Bloch type~\cite{Grosso2014_SM}, i.e.~$\psi_{n \bm k}(\bm r) = e^{i\bm k \cdot \hat{\bm r}} u_{n\bm k}(\bm r)$, with $u_{n \bm k}(\bm r) = u_{n \bm k}(\bm r + \bm R^0_{p\nu})$. They are labeled by a discrete band number $n$ and depend parametrically on the Bloch quasi-momentum $\bm k \in {\rm FBZ}$. The crystal is therefore described by the  Hamiltonian $\hat{{\cal H}}_{\rm e}$ which does not explicitly depend on the Bloch quasi-momentum, with boundary conditions that do depend on it. To transition to a formulation where the boundary conditions remain $\bm k$-independent while the Hamiltonian retains its $\bm k$-dependence, it is useful to apply a momentum shift~\cite{bernevig_book_2013_SM}:
\begin{equation}
\label{sm-eq:electron-hamiltonian-kdep}
    \hat{\cal H}_{\rm e} \to \hat{{\cal H}}_{\rm e}(\bm k) \equiv e^{-i\bm k \cdot \hat{\bm r}}\,\hat{{\cal H}}_{\rm e}\, e^{i\bm k \cdot \hat{\bm r}} = \frac{{(\hat{\bm p} + \hbar \bm k)}^2}{2m_{\rm e}} + \hat{V}(\bm r)~.
\end{equation}
The eigenstates of the Hamiltonian~\eqref{sm-eq:electron-hamiltonian-kdep} reduce from being of the Bloch type to being only the periodic part of the Bloch wavefunction, i.e.~they coincide with the quantity $u_{n \bm k}(\bm r)$. This representation is well-suited to discuss topology and, more in general, quantum geometry. 

As stated in the main text, in the case in which bands do {\it not} touch, the analysis of the geometry of wavefunctions requires the definition of a different QGT for each band~\cite{Resta2011_SM}:
\begin{equation}
\label{sm-eq:qgt}
    \mathcal{Q}^{(n)}_{ij}(\bm k) = \Tr \left\{\partial_{k_i} \hat{P}_n(\bm k) \left[1- \hat{P}_n(\bm k)\right]\partial_{k_j} \hat{P}_n(\bm k)\right\}~,
\end{equation}
where $\hat{P}_n(\bm k) \equiv \vert u_{n\bm k}\rangle\langle u_{n\bm k}\vert$ is the projection operator on the state $\vert u_{n\bm k}\rangle$. The expression (\ref{sm-eq:qgt}) of the QGT is appropriate for numerical calculations, as it does not require any type of gauge smoothing. Indeed, the projectors $\hat{P}_n(\bm k)$ are manifestly gauge-invariant. We note that, due to the presence of a band gap $\Delta$ acting as a regularizer in our description of the quantum geometry of graphene, we do not need to deal with band crossings. 

We now show that the definition~\eqref{sm-eq:qgt} is equivalent to the original definition provided by Provost and Vallee~\cite{Provost1980_SM} and given in Eqs.~(16)-(18) of the main text. We begin by transforming the projector-based expression~\eqref{sm-eq:qgt} into a formulation that explicitly incorporates an exact eigenstate representation: 
\begin{align}
\label{sm-eq:resta-identity}
 \mathcal{Q}^{(n)}_{ij}(\bm k) 
& = \Tr \{\partial_{k_i}\Bigl[\vert u_{n\bm k}\rangle\langle u_{n\bm k} \vert \Bigr]   \Bigl[1- \vert u_{n\bm k}\rangle\langle u_{n\bm k}\vert\Bigr] \partial_{k_j}\Bigl[\vert u_{n\bm k}\rangle\langle u_{n\bm k} \vert \Bigr] \}\nonumber\\
& = \sum_{m} \langle u_{m \bm k} \vert \Bigl[\vert\partial_{k_i} u_{n\bm k}\rangle\langle u_{n\bm k}\vert + \vert u_{n\bm k}\rangle\langle \partial_{k_i} u_{n\bm k}\vert\Bigr]   \Bigl[\sum_{l\neq n} \vert u_{l\bm k}\rangle\langle u_{l\bm k} \vert \Bigr] \Bigl[\vert\partial_{k_j} u_{n\bm k}\rangle\langle u_{n\bm k}\vert + \vert u_{n\bm k}\rangle\langle \partial_{k_j} u_{n\bm k}\vert\Bigr] \vert u_{m\bm k} \rangle \nonumber\\
& = \sum_m \langle u_{m \bm k} \vert u_{n\bm k}\rangle\langle \partial_{k_i} u_{n\bm k}\vert  \Bigl[\sum_{l\neq n} \vert u_{l\bm k}\rangle\langle u_{l\bm k} \vert \Bigl] \vert \partial_{k_j} u_{n\bm k}\rangle\langle u_{n\bm k}\vert u_{m \bm k} \rangle \nonumber\\
& = \sum_m \delta_{mn} \langle \partial_{k_i} u_{n\bm k}\vert  \Bigl[\sum_{l\neq n} \vert u_{l\bm k}\rangle\langle u_{l\bm k} \vert \Bigr] \vert \partial_{k_j} u_{n\bm k}\rangle \delta_{nm} \nonumber\\
& = \bigl\langle \partial_{k_i} u_{n \bm k}|\partial_{k_j} u_{n \bm k} \bigr\rangle  -\bigl\langle \partial_{k_i} u_{n \bm k}| u_{ n\bm k}\bigr\rangle \bigl\langle u_{ n \bm k}|\partial_{k_j} u_{n \bm k} \bigr\rangle~.
\end{align}
Here, we used the completeness relation $\sum_{n} \vert u_{n\bm k}\rangle\langle u_{n\bm k} \vert =\openone$ and orthogonality of the eigenstates, i.e.~$\langle u_{n\bm k}\vert u_{m\bm k} \rangle = \delta_{nm}$, which are both valid at each fixed value of $\bm k$, as well as the definition of the trace on the parametric Hilbert space, $\Tr[\cdots] = \sum_m \langle u_{m\bm k} \vert\cdots \vert u_{m\bm k}\rangle$, again valid at each fixed value of ${\bm k}$. 

We now note that the second term in the last line of Eq.~\eqref{sm-eq:resta-identity} is real. We therefore have:
\begin{subequations}
    \begin{eqnarray}
    \label{sm-eq:reQ}
    \Re \mathcal{Q}^{(n)}_{ij}(\bm k) & = & \Re \Bigl[\bigl\langle \partial_{k_i} u_{n \bm k}|\partial_{k_j} u_{n \bm k} \bigr\rangle\Bigr]  -\bigl\langle \partial_{k_i} u_{n \bm k}| u_{ n\bm k}\bigr\rangle \bigl\langle u_{ n \bm k}|\partial_{k_j} u_{n \bm k} \bigr\rangle~, \\
    \label{sm-eq:imQ}
    2i \Im \mathcal{Q}^{(n)}_{ij}(\bm k) & = & \mathcal{Q}^{(n)}_{ij}(\bm k) - \bigl[\mathcal{Q}^{(n)}_{ij}(\bm k)\bigr]^* = \bigl\langle \partial_{k_i} u_{n \bm k}|\partial_{k_j} u_{n \bm k} \bigr\rangle - \bigl\langle \partial_{k_j} u_{n \bm k}|\partial_{k_i} u_{n \bm k} \bigr\rangle~.
    \end{eqnarray}
\end{subequations}
By substituting $\vert u_{n\bm k} \rangle \leftrightarrow \vert \psi_{\bm k}\rangle $, we see that Eqs.~\eqref{sm-eq:reQ} and \eqref{sm-eq:imQ} coincide with the definitions of $g_{ij}(\bm k)$ and ${\cal F}_{ij}(\bm k)$ given in Eqs.~(17) and~(18) of the main text, respectively.

The form given in Eq.~\eqref{sm-eq:qgt}, however, is not quite equivalent to that provided in Eq.~(10) of the main text: the formal correspondence $\hat{P}_n(\bm k) \to P_n(\bm k) \equiv U_n(\bm k)U_n^ \dagger(\bm k)$ between the projection operator and the projection matrix is only valid under the tight-binding approximation, within which all the information on the $\bm k$-dependence of the Bloch eigenstates $\vert u_{n\bm k}\rangle$ is contained in the Bloch eigenvector $U_n(\bm k)$. A proof of this statement can be found in the Supplementary information of Ref.~\cite{Yu_NaturePhys_2024_SM}.

\section{Proof of the acoustic sum rules}\label{section2}

We start by proving the acoustic sum rule for the complete dynamical matrix and for the ionic and electronic contributions to it. It is sufficient to impose invariance of the interatomic potential energy $V(\{\bm u_{p\nu}\})$ felt by each atom under a global translation of the entire crystal by a constant vector $\bm b$. Let us calculate the potential energy felt by an atom in the $p = 0$ unit cell, with basis vector $\bm \tau_\nu$:
\begin{equation}
    V(\bm u_{0\nu}) = \sum_{p'} \sum_{\nu'} u_{0\nu i} \,\bigl[{\cal C}\bigr]_{0\nu i}^ {p'\nu'j} u_{p'\nu'j} =  \sum_{p'} \sum_{\nu'} \bigl(u_{0\nu i}+b_i\bigr) \,\bigl[{\cal C}\bigr]_{0\nu i}^ {p'\nu'j}\bigl( u_{p'\nu'j}+b_j\bigr)~,
\end{equation}
implying 
\begin{equation}
    \sum_{p'} \sum_{\nu'}b_i \,\bigl[{\cal C}\bigr]_{0\nu i}^ {p'\nu'j}b_j=0 \qquad \forall~\bm b \in \mathds{R}^ D~,
\end{equation}
and therefore 
\begin{equation}
\label{sm-eq:acoustic-sum-rule}
    \sum_{\nu'}  \sqrt{M_{\nu'}}\bigl[{\cal D}(\bm q = \Gamma)\bigr]_{\nu i}^ {\nu' j}  = \frac{1}{\sqrt{M_{\nu}}}\sum_{p',\nu'}\bigl[{\cal C}\bigr]_{0\nu i}^ {p'\nu'j} = 0\qquad \forall~\nu,i,j ~.
\end{equation}
The same line of reasoning can be applied to ${\cal D}^ {(\rm ion)}(\bm q = \Gamma)$, substituting the interatomic potential with the bare Coulomb repulsive potential, and thus to the electronic contribution ${\cal D}^ {(\rm el)}(\bm q = \Gamma)$. 

With the two-center approximation of the tight-binding model, we can also derive the sum rule for the electronic contribution. To demonstrate this, we examine the sum of the derivatives of the Hamiltonian.

\begin{equation}
\begin{aligned}
\sum_{p\nu}\partial_{p\nu i} \hat{\cal H}_{\text{e}} & = 
\sum_{p\nu}
 \sum_{p' \nu' \alpha' \vphantom{'}}
 \sum_{p''\nu'' \alpha''} \partial_{p\nu i} t_{\nu'\nu''}^{\alpha'\alpha''} \left( \bm R_{p'\nu'} - \bm R_{p''\nu''}\right) \hat{c}^{\dagger}_{p'\nu' \alpha' \vphantom{'}}\hat{c}^{\vphantom{\dagger}}_{p''\nu'' \alpha''}~\\
 & = \sum_{p\nu}
 \sum_{p' \nu' \alpha' \vphantom{'}}
 \sum_{p''\nu'' \alpha''}
 \delta_{p,p'}\delta_{\nu,\nu'} \partial_{p'\nu' i} t_{\nu'\nu''}^{\alpha'\alpha''} \left( \bm R_{p'\nu'} - \bm R_{p''\nu''}\right) \hat{c}^{\dagger}_{p'\nu' \alpha' \vphantom{'}}\hat{c}^{\vphantom{\dagger}}_{p''\nu'' \alpha''}\\
 &+
 \delta_{p,p''}\delta_{\nu,\nu''} \partial_{p''\nu'' i} t_{\nu'\nu''}^{\alpha'\alpha''} \left( \bm R_{p'\nu'} - \bm R_{p''\nu''}\right) \hat{c}^{\dagger}_{p'\nu' \alpha' \vphantom{'}}\hat{c}^{\vphantom{\dagger}}_{p''\nu'' \alpha''}\\
 & = 
 \sum_{p' \nu' \alpha' \vphantom{'}}
 \sum_{p''\nu'' \alpha''}
 \partial_{p'\nu' i} t_{\nu'\nu''}^{\alpha'\alpha''} \left( \bm R_{p'\nu'} - \bm R_{p''\nu''}\right)+ \partial_{p''\nu'' i} t_{\nu'\nu''}^{\alpha'\alpha''} \left( \bm R_{p'\nu'} - \bm R_{p''\nu''}\right)
 \hat{c}^{\dagger}_{p'\nu' \alpha' \vphantom{'}}\hat{c}^{\vphantom{\dagger}}_{p''\nu'' \alpha''}\\
& = 0
\end{aligned}
\label{sumpH}
\end{equation}
where we have used the fact that for any function $f(x-y)$
\begin{equation}
\begin{aligned}
 \partial_{x}f(x-y)+\partial_{y}f(x-y) =  \partial_{(x-y)}f(x-y)-\partial_{(x-y)}f(x-y)  = 0
\end{aligned}
\label{}
\end{equation}
Substituting \ref{sumpH} into Eqs.~(23)-(25) in the main text, we arrive to
\begin{equation}
\begin{aligned}
 \sum_{p^{\prime},\nu^{\prime}} [\mathcal{C}^{(\text{el})}]_{p \nu i}^{p^{\prime}\nu^{\prime}}
 &= \sum_{p^{\prime},\nu^{\prime}} [\mathcal{C}^{(\text{el},1)}]_{p \nu i}^{p^{\prime}\nu^{\prime}}+
 \sum_{p^{\prime},\nu^{\prime}} [\mathcal{C}^{(\text{el},2)}]_{p \nu i}^{p^{\prime}\nu^{\prime}}\\
 &=\sum_{m\neq 0} \frac{1}{{\cal E}_0-{\cal E}_m} \left \langle \phi^{(m )}_{\{\bm 0\}} \middle| \partial_{p\nu i} \hat{\cal H}_{\rm e}\middle| \phi^ {(0)}_{\{\bm 0\}}\right\rangle \times \left \langle \phi^{(0 )}_{\{\bm 0\}} \middle| \sum_{p^{\prime},\nu^{\prime}}\partial_{p'\nu' j} \hat{\cal H}_{\rm e}\middle| \phi^ {(m)}_{\{\bm 0\}}\right\rangle\\
 &+\frac{1}{{\cal E}_0-{\cal E}_m} \left \langle \phi^{(m )}_{\{\bm 0\}} \middle| \sum_{p^{\prime},\nu^{\prime}}\partial_{p^{\prime}\nu^{\prime} j} \hat{\cal H}_{\rm e}\middle| \phi^ {(0)}_{\{\bm 0\}}\right\rangle \times \left \langle \phi^{(0 )}_{\{\bm 0\}} \middle|  \partial_{p\nu i} \hat{\cal H}_{\rm e}\middle| \phi^ {(m)}_{\{\bm 0\}}\right\rangle\\
 &+\left\langle \phi^ {(0)}_{\{\bm 0\}}\middle| \partial_{p\nu i} \left( \sum_{p^{\prime},\nu^{\prime}}\partial_{p'\nu'j} \hat{\cal H}_{\rm e}\right)\middle|\phi^ {(0)}_{\{\bm 0\}}\right\rangle~,\\
 & = 0
\end{aligned}
\label{smr}
\end{equation}
which is identical to Eq.~\ref{sm-eq:acoustic-sum-rule}. This proof does not require the Gaussian approximation.

Only continuous translational symmetry is required for the acoustic sum rules. We consider a general Hamiltonian parameterized by the set of displacements $\{\bm{u}_{p \nu}\}$. Translational symmetry implies that if all atoms are uniformly displaced by an arbitrary vector $ \Delta \bm{u}$, the Hamiltonian remains invariant.

\begin{equation}
\begin{aligned}
 \mathcal{\hat{H}}_{e}(\{\bm{u}_{p \nu}\}) = \mathcal{\hat{H}}_{e}(\{\bm{u}_{p \nu}+\Delta \bm{u}\}) 
\end{aligned}
\label{}
\end{equation}
We consider a infinitsimall $\Delta \bm{u}$ in $i$ direction, we obtained 

\begin{equation}
\begin{aligned}
 \partial_{pvi} \hat{\mathcal{H}}_{e}(\{\bm{u}_{p \nu}\}) \Delta u_{i} &= 0\\
 \partial_{pvi} \hat{\mathcal{H}}_{e}(\{\bm{u}_{p \nu}\}) &= 0\\
\end{aligned}
\label{pat0}
\end{equation}
Eq.~\ref{pat0} leads to sum rule as show in Eq.~\ref{smr}. This is nothing but the Goldstone theorem for translationally invariant systems.

\section{The force constants in terms of Hamiltonian derivatives}\label{section3}

To derive an explicit expression for the quantity ${\cal C}^ {\rm (el)}$ as defined in Eq.~(23) of the main text, we use the Hellmann-Feynmann theorem~\cite{GiulianiVignale_SM} carefully combined with the second-order derivative:
\begin{align}
\label{sm-eq:hellmann-feynman}
\bigl[ {\cal C}^{\text{(el)}}\bigr]^{p'\nu'j}_{p\nu i} & = \left.\frac{\partial^2}{\partial u_{p\nu i}\,\partial u_{p'\nu'j}} \left\langle\phi^{(0)}_{\{\bm u_{p\nu}\}}\right| \hat{\cal H}_{\text{e}}(\{\bm u_{p\nu}\}) \left|\phi^{(0)}_{\{\bm u_{p\nu}\}}\right\rangle \,\right|_0  \nonumber\\
& = \left.\partial_{p\nu i}\left\langle\phi^{(0)}\right| \partial_{p'\nu' j}\hat{\cal H}_{\text{e}} \left|\phi^{(0)}\right\rangle \,\right|_0  \nonumber\\
& = \left.\left\langle\partial_{p\nu i}\phi^{(0)}\right| \partial_{p'\nu' j}\hat{\cal H}_{\text{e}} \left|\phi^{(0)}\right\rangle \,\right|_0 +  \left.\left\langle\phi^{(0)}\right| \partial_{p'\nu' j}\hat{\cal H}_{\text{e}} \left|\partial_{p\nu i}\phi^{(0)}\right\rangle \,\right|_0  \nonumber\\
& \,+\left.\left\langle\phi^{(0)}\right| \partial_{p\nu i}\partial_{p'\nu' j}\hat{\cal H}_{\text{e}} \left|\phi^{(0)}\right\rangle \,\right|_0~.
\end{align}
Here, we used the shorthand notation $\partial_{p\nu i} \equiv \partial/\partial u_{p\nu i}$ and omitted, starting from the second equality, the explicit parametric dependence of the Hamiltonian and its ground state on $\{\bm u_{p\nu}\}$.

The parametric derivatives of the many-body ground state in the previous expression are formally well-defined but impractical to compute directly.  Consequently, we replace them with derivatives of the Hamiltonian. However, this substitution necessitates the introduction of the exact excited eigenstates of the Hamiltonian, which are determined by the Schr\"odinger equation:
\begin{equation}
    \hat{\cal H}_{\text{e}} \vert{\phi^{(m)}}\rangle = \mathcal{E}_m\vert{\phi^{(m)}}\rangle ~. 
\end{equation}
In the following, we simplify the parametric derivatives of the many-body ground state using the excited states. From 
\begin{equation}
\begin{aligned}
  &\bra{\phi^{(m)}} \mathcal{\hat{H}}_{e} \ket{\phi^{(n)}}
  = \mathcal{E}_{m} \delta_{m,n} \\
  &\bra{\partial_{p\nu i}\phi^{(m)}}\mathcal{\hat{H}}_{e} \ket{\phi^{(n)}}
  +
  \bra{\phi^{(m)}} \partial_{p\nu i}\mathcal{\hat{H}}_{e} \ket{\phi^{(n)}}
  +
  \bra{\phi^{(m)}} \mathcal{\hat{H}}_{e} \ket{\partial_{p\nu i}\phi^{(n)}}
  = \partial_{p\nu i}\mathcal{E}_{m} \delta_{m,n} \\
\end{aligned}
\label{}
\end{equation}
we arrive at the Epstein generalization~\cite{Grosso2014_SM} of the Hellmann-Feynman theorem for $m \neq n $:
\begin{equation}
\label{sm-eq:epstein}
\left\langle \phi^{(m)} \middle\vert \partial_{p\nu i} \phi^{(n)}\right\rangle =\frac{\left\langle\phi^{(m)} \left|\partial_{p\nu i} \hat{\cal H}_\text{e}\right|\phi^{(n)}\right\rangle}{\mathcal{E}_n - \mathcal{E}_m}~,\qquad (m \neq n)~.
\end{equation}
For $m = n$ we get the Hellmann-Feynman theorem 
\begin{equation}
\begin{aligned}
    \bra{\phi^{(n)}} \partial_{p\nu i}\mathcal{\hat{H}}_{e} \ket{\phi^{(n)}}
    = \partial_{p\nu i} \mathcal{E}_{n}
\end{aligned}
\label{}
\end{equation}
Using Eq.~\eqref{sm-eq:epstein} inside Eq.~\eqref{sm-eq:hellmann-feynman}, we get 
\begin{equation}
\begin{aligned}
    \bigl[ {\cal C}^{\text{(el)}}\bigr]^{p'\nu'j}_{p\nu i} 
    & = \left.\left\langle\partial_{p\nu i}\phi^{(0)}\right| \partial_{p'\nu' j}\hat{\cal H}_{\text{e}} \left|\phi^{(0)}\right\rangle \,\right|_0 +  \left.\left\langle\phi^{(0)}\right| \partial_{p'\nu' j}\hat{\cal H}_{\text{e}} \left|\partial_{p\nu i}\phi^{(0)}\right\rangle \,\right|_0 \\
    & \,+\left.\left\langle\phi^{(0)}\right| \partial_{p\nu i}\partial_{p'\nu' j}\hat{\cal H}_{\text{e}} \left|\phi^{(0)}\right\rangle \,\right|_0~.    \\
    & = \left.\left\langle\partial_{p\nu i}\phi^{(0)}\right| \sum_{m} \left\vert  \phi^{(m)} \right\rangle \left\langle \phi^{(m)}  \right\vert \partial_{p'\nu' j} \hat{\cal H}_{\text{e}} \left|\phi^{(0)}\right\rangle \,\right|_0 +  \left.\left\langle\phi^{(0)}\right| \partial_{p'\nu' j}\hat{\cal H}_{\text{e}}  \sum_{m} \left\vert  \phi^{(m)} \right\rangle \left\langle \phi^{(m)}  \right\vert \left.\partial_{p\nu i}\phi^{(0)}\right\rangle \,\right|_0 \\
    & \,+\left.\left\langle\phi^{(0)}\right| \partial_{p\nu i}\partial_{p'\nu' j}\hat{\cal H}_{\text{e}} \left|\phi^{(0)}\right\rangle \,\right|_0~.    \\
    & = \left.\left\langle\partial_{p\nu i}\phi^{(0)}\right| \sum_{m\neq 0} \left\vert  \phi^{(m)} \right\rangle \left\langle \phi^{(m)}  \right\vert  \partial_{p'\nu' j} \hat{\cal H}_{\text{e}} \left|\phi^{(0)}\right\rangle \,\right|_0 +  \left.\left\langle\phi^{(0)}\right| \partial_{p'\nu' j}\hat{\cal H}_{\text{e}}  \sum_{m\neq 0} \left\vert  \phi^{(m)} \right\rangle \left\langle \phi^{(m)}  \right\vert \left.\partial_{p\nu i}\phi^{(0)}\right\rangle \,\right|_0 \\
    & \,+\left.\left\langle\phi^{(0)}\right| \partial_{p\nu i}\partial_{p'\nu' j}\hat{\cal H}_{\text{e}} \left|\phi^{(0)}\right\rangle \,\right|_0~.    \\    
    & = \sum_{m\neq 0} \left. \frac{1}{{\cal E}_0-{\cal E}_m} \left \langle \phi^{(m )}_{\{\bm 0\}} \middle| \partial_{p\nu i} \hat{\cal H}_{\rm e}\middle| \phi^ {(0)}_{\{\bm 0\}}\right\rangle \times  \left \langle \phi^{(0 )}_{\{\bm 0\}} \middle| \partial_{p'\nu' j} \hat{\cal H}_{\rm e}\middle| \phi^ {(m)}_{\{\bm 0\}}\right\rangle\right|_0~ + {\rm H.c.} \\ 
    & \,+\left.\left\langle\phi^{(0)}\right| \partial_{p\nu i}\partial_{p'\nu' j}\hat{\cal H}_{\text{e}} \left|\phi^{(0)}\right\rangle \,\right|_0~. 
\end{aligned}
\label{app:eq:force_constant_el}
\end{equation}
In the derivation, we used the fact that 
\begin{equation}
\begin{aligned}
\left\langle\partial_{p \nu i}\phi^{(m)}\right. \left\vert \phi^{(n)}\right\rangle
 +
 \left\langle\phi^{(m)}\right. \left\vert \partial_{p\nu i}\phi^{(n)}\right\rangle = 0. 
\end{aligned}
\label{app:eq:prod-derivative}
\end{equation}
The two terms in \ref{app:eq:force_constant_el} are nothing but ${\cal C}^{\text{(el, 1)}}$ and ${\cal C}^{\text{(el,2)}}$ defined in Eqs.~(24)-(25) of the main text. 

We remark that from~\ref{sm-eq:epstein} one can also obtain
\begin{equation}
\begin{aligned}
\label{sm-eq:epstein-detail}
\left.\middle\vert \partial_{p\nu i} \phi^{(n)} \right\rangle 
&=\sum_{m} \left.\middle\vert \phi^{m} \right\rangle \left\langle \phi^{(m)} \middle\vert \partial_{p\nu i} \phi^{(n)} \right\rangle \\
&= \sum_{m\neq n} \left.\middle\vert \phi^{m} \right\rangle \left\langle \phi^{(m)} \middle\vert \partial_{p\nu i} \phi^{(n)} \right\rangle \\
&= \sum_{m\neq n} \left.\middle\vert \phi^{m} \right\rangle 
\frac{\left\langle\phi^{(m)} \left|\partial_{p\nu i} \hat{\cal H}_\text{e}\right|\phi^{(n)}\right\rangle}{\mathcal{E}_n - \mathcal{E}_m}~.
\end{aligned}
\end{equation}
In the second equation of~\ref{sm-eq:epstein-detail}, a parallel-transport gauge is assumed such that the Berry connection vanishes:
\begin{equation}
\left\langle \phi^{(n)} \middle\vert \partial_{p\nu i} \phi^{(n)} \right\rangle=0.    
\label{app:eq:prod-derivative2}
\end{equation}
However, this gauge condition cannot be imposed globally if the system exhibits a nonzero Berry phase. We emphasize that the derivation of~\ref{app:eq:force_constant_el} only requires the use of~\ref{app:eq:prod-derivative}, not the stricter condition in~\ref{app:eq:prod-derivative2}.

\section{Formulas for the dynamical matrix}\label{section4}

In this section, we derive the formulas for the dynamical matrix from the electronic contributions. We first present the main results and then show the detailed derivations. 

The amplitudes $\left \langle \phi^{(m )}_{\{\bm 0\}} \middle|\partial_{p\nu i}\hat{\cal H}_{\rm e}\middle| \phi^ {(0)}_{\{\bm 0\}}\right\rangle$ appearing in Eq.~(24) of the main text are nonzero only for the following class of states:
\begin{equation}
\label{sm-eq:nonzero-states}
    \vert{\phi^{(n \bm k,\, n'\bm k')}_{\{\bm 0\}}}\rangle = \hat{\gamma}_{n'\bm k'}^\dagger \hat{\gamma}_{n \bm k\vphantom{'}}^{\vphantom{\dagger}} \left\vert \phi^ {(0)}_{\{\bm 0\}} \right \rangle~, \qquad \text{with} \qquad 
    \begin{cases}
        E_{n}(\bm k) < \mu\\
        E_{n'}(\bm k') > \mu
    \end{cases}~,
\end{equation}
where $\mu$ is the chemical potential, and $\hat{\gamma}_{n\bm k}^\dagger$ creates an electron in a Bloch state labeled by $n, \bm k$: $\hat{\gamma}_{n\bm k}^\dagger|0\rangle = e^ {-i\bm k\cdot \hat{\bm r}}|u_{n\bm k}\rangle$.

The first derivatives of the Hamiltonian with respect to the displacements, which appear in Eq.~\eqref{app:eq:force_constant_el}, can be calculated starting from the modified tight-binding Hamiltonian---see Eq.~(1) of the main text. The evaluation of these derivatives on the states~\eqref{sm-eq:nonzero-states} allows us to get to an explicit form for the term ${\cal C}^{(\rm{el,\,1})}$ as defined in Eq.~(24) in the mian text. The expectation value on the Fermi sea that appears in the Eq.~(25) for $ \bigl[ {\cal C}^{(\rm{el,\,2})}_{ij}\bigr]^{p p'}_{\nu \nu'} $ is evaluated by explicit calculation of the second derivatives of the Hamiltonian with respect to the displacements. By taking the Fourier transform with respect to the ion positions, thereby defining
\begin{equation}
    \label{sm-eq:dynamical-matrix}
    \bigl[{\cal D}^{\rm (el,\,1/2)}(\bm q)\bigr]_{\nu i}^ {\nu'j} \equiv \frac{1}{\sqrt{M_{\nu\vphantom{'}} M_{\nu'}}}\sum_{p} e^{-i\bm q\cdot(\bm R_p + \bm \tau_{\nu\vphantom{'}} -\bm \tau_{\nu'})} \,\bigl[ {\cal C}^{\rm (el,\,1/2)}_{ij}\bigr]^{p, p'= 0}_{\nu\nu'}
\end{equation}
so that ${\cal D}^{\rm (el)}(\bm q)  = {\cal D}^{\rm (el,\,1)}(\bm q)  + {\cal D}^{\rm (el,\,2)}(\bm q)$, we obtain the following explicit formulas for the dynamical matrix:
\begin{equation}\label{sm-eq:dI-explicit}
\bigl[{\cal D}^{\rm (el,\,1)}(\bm q)\bigr]_{\nu i}^ {\nu'j} = \frac{1}{\sqrt{M_{\nu} M_{\nu'}}}\, \frac{2}{N} \sum_{n}^{\rm occ.} \sum_{n'}^{\rm unocc.} \sum_{\bm k}^{\text{FBZ}}
\left\{\frac{\bigl[F_i(n\bm k, n'\bm k+\bm q)\bigr]_{\nu} \bigl[F_j(n'\bm k+\bm q, n\bm k)\bigr]_{\nu'}}{E_{n}(\bm k) - E_{n'}(\bm k+\bm q)}  \right\} + {\rm H.c.}
\end{equation}
and
\begin{equation}
\label{sm-eq:dII-explicit}
\bigl[{\cal D}^{\rm (el,\,2)}(\bm q)\bigr]_{\nu i}^ {\nu'j} =\frac{1}{\sqrt{M_\nu M_{\nu'}}}\,\frac{2}{N} \sum_{\Tilde{\nu}} \sum_{n}^{\rm occ.} \sum_{\bm k \Tilde{\bm k}}^{\text{FBZ}} \sum_{\alpha\alpha'}\left[\delta_{\nu\nu'}\delta_{\Tilde{\bm k},\,\bm 0} - \delta_{\Tilde{\nu} \nu'} \delta_{\Tilde{\bm k},\,\bm q}\right]\left\{  \bigl[M_{ij}(\bm k + \Tilde{\bm k})\bigr]_{\nu \alpha}^ {\Tilde{\nu}\alpha'} \bigl[P_n(\bm k)\bigr]_{\Tilde{\nu}\alpha'}^{\nu\alpha}  \right\}+\rm{H.c.}~,
\end{equation}
Notice that here the hermitian conjugation accounts for taking the complex conjugate and exchanging the $\nu\nu'$ indices. The factor of two comes from the spin degrees of freedom. $P_n(\bm k) = U_n(\bm k)U^ \dagger_n(\bm k)$ is the projection matrix onto the eigenvector $U_n(\bm k)$ of the Bloch Hamiltonian $h(\bm k)$, while $N$ is the number of atoms in the Born-von Karman supercell, which is equal to the number of Bloch wave vectors involved in the sum over the first Brillouin zone (FBZ). In writing Eqs.~(\ref{sm-eq:dI-explicit})-(\ref{sm-eq:dII-explicit}) we employed the following shorthand notation for summations over the Fermi sea,
\begin{equation}
   \sum_{nn'}  \sum_{\bm{k}\bm{k}'}^{\text{FBZ}} \Theta\bigl(\mu-E_{\nu}(\bm k)\bigr) \,\Theta\bigl(E_{n'}(\bm k')-\mu\bigr)  = \sum_{\bm k \in \text{FBZ}}^{E_{n}(\bm k) < \mu} \sum_{\bm k' \in \text{FBZ}}^{E_{\nu'}(\bm k') > \mu} \equiv  \sum_{n\bm k}^{\rm occ.} \sum_{n'\bm k'}^{\rm unocc.}~,
\end{equation}
and defined the following quantity:
\begin{equation}
\label{sm-eq:ftensor}
    \bigl[F_i(n\bm k, n'\bm k')\bigr]_{\nu} \equiv  \sum_{\Tilde{\nu}} \sum_{\nu_1 \nu_2 } \sum_{\alpha\alpha'}\bigl[U^\dagger_{n}(\bm k)\bigr]_{\nu_1\alpha}\biggl\{ \delta_{\nu\nu_1} \delta_{\Tilde{\nu}\nu_2} \bigl[ f_i(\bm k')\bigr]_{\nu\alpha}^ {\Tilde{\nu}\alpha'}- \delta_{\nu \nu_2} \delta_{\Tilde{\nu}\nu_1} \bigl[f_i(\bm k)]_{\Tilde{\nu}\alpha}^ {\nu\alpha'} \biggr\}\bigl[U^{\vphantom{\dagger}}_{n'}(\bm k')\bigr]_{\nu_2\alpha'}~.
\end{equation}
We now turn to a detailed calculation of the Hamiltonian parametric derivatives. We start from the modified tight-binding (TB) form for the electron Hamiltonian:
\begin{equation}
\label{He-simplified}
    \hat{\mathcal{H}_{\text{e}}} = \sum_{p,\, \nu,\,\alpha} \sum_{p',\,\nu',\alpha'} t_{\nu\nu'}^{\alpha\alpha'}\left( \bm R_{p\nu} - \bm R_{p'\nu'}\right)\hat{c}^{\dagger}_{p\nu\alpha}\hat{c}^{\vphantom{\dagger}}_{p'\nu'\alpha'}\,.
\end{equation}
where $\bm R_{p \nu} = \bm R_p + \bm \tau_\nu + \bm u_{p\nu}$, and $t_{\nu\nu'}^{\alpha\alpha'}(\bm r)$ is supposed to be a real function of $\bm r$.

The parametric gradient of the TB Hamiltonian can be calculated component-wise by directly differentiating the form in Eq.~\eqref{He-simplified}:
\begin{equation}
    \begin{split}
        \frac{\partial \hat{\mathcal{H}_\text{e}}}{\partial u_{p\nu i}} & = \sum_{\alpha'\alpha''}\sum_{p' \nu'} \sum_{p''\nu''} \left\{ \left.\partial_{p'\nu'i} t_{\nu'\nu''}^ {\alpha'\alpha''}\right|_0 \delta_{pp'}\delta_{\nu\nu'}  
 + \left.\partial_{p''\nu''i} t_{\nu'\nu''}^ {\alpha'\alpha''}\right|_0 \delta_{pp''}\delta_{\nu\nu''}\right\}\,\hat{c}^{\dagger}_{p'\nu'\alpha'}\hat{c}^{\vphantom{\dagger}}_{p''\nu''\alpha''}\\
         & = \sum_{\alpha\alpha'}\sum_{p'\nu'} \left\{ \left.\bm[\partial_{i} t_{\nu\nu'}^ {\alpha\alpha'}(\bm r)\right|_{\bm R^0_{p\nu}-\bm  R^0_{p'\nu'}}
         c^{\dagger}_{p\nu\alpha}c^{\vphantom{\dagger}}_{p'\nu'\alpha'}
 - \left.\partial_{i} t_{\nu'\nu}^ {\alpha\alpha'}(\bm r)\right|_{\bm R^0_{p'\nu'}- \bm R^0_{p\nu}}  c^{\dagger}_{p'\nu'\alpha'}c^{\vphantom{\dagger}}_{p\nu\alpha} \right\}\\
        & =  \sum_{\alpha\alpha'}\sum_{p'\nu'} \left.\partial_i t_{\nu\nu'}^ {\alpha\alpha'}(\bm r)\right|_{\bm R^0_{p\nu}- \bm R^0_{p'\nu'}} \hat{c}^{\dagger}_{p\nu\alpha}\hat{c}^{\vphantom{\dagger}}_{p'\nu'\alpha'} + \text{H.c.} \,,
    \end{split}
\end{equation}
where $\partial_i \equiv \partial/\partial r_i$. 

Now, we Fourier-transform the electron operators
\begin{equation}
\label{cdaggerc-fourier}
    \hat{c}^{\dagger}_{p\nu\vphantom{'}\alpha} \hat{c}^{\vphantom{\dagger}}_{p'\nu'\alpha'} = \frac{1}{N}\sum_{\bm k \bm k'}^{\text{FBZ}} e^{-i\bm k \cdot \bm R^0_{p\nu}} e^{i\bm k' \cdot \bm R^0_{p'\nu'}}  \hat{c}^{\dagger}_{\bm k\nu\vphantom{'}\alpha}\hat{c}^{\vphantom{\dagger}}_{\bm k'\nu'\alpha'}\,.
\end{equation}
According to the same convention, we define the Fourier transform of the gradient of the hopping function:
\begin{equation}
\label{kspace-gradient}
    \bigl[ f_{i} (\bm \kappa) \bigr]_{\nu\nu'}^ {\alpha\alpha'} \equiv \sum_{p} e^{-i\bm \kappa \cdot\left( \bm R_{p} + \bm \tau_{\nu\vphantom{'}} - \bm \tau_{\nu'}\right)} \left.\partial_i t_{\nu\nu'}^ {\alpha\alpha'}(\bm r)\right|_{\bm R_{p} + \bm \tau_{\nu\vphantom{'}} - \bm \tau_{\nu'}} \,.
\end{equation}
It is easy to prove that this object is anti-hermitian with respect to the exchange of sublattice indices:
\begin{equation}
\label{kspace-gradient-skew-hermitian}
    \begin{split}
       [ f_i^*(\bm \kappa)]_{\nu\nu'}^ {\alpha\alpha'} & = \sum_p e^{i\bm \kappa \cdot\left( \bm R_{p} + \bm \tau_{\nu\vphantom{'}} - \bm \tau_{\nu'}\right)} \left.\partial_i t_{\nu\nu'}^ {\alpha\alpha'}(\bm r)\right|_{\bm R_{p} + \bm \tau_{\nu\vphantom{'}} - \bm \tau_{\nu'}} \\
       & = \sum_p e^{-i\bm \kappa \cdot\left( -\bm R_{p} + \bm \tau_{\nu'} - \bm \tau_{\nu\vphantom{'}}\right)}\left( -\left.\partial_i t_{\nu'\nu}^ {\alpha\alpha'}(\bm r)\right|_{\bm -R_{p} + \bm \tau_{\nu'} - \bm \tau_{\nu\vphantom{'}}}\right) \\
        & = -\sum_p e^{-i\bm \kappa \cdot\left( \bm R_{p} + \bm \tau_{\nu'} - \bm \tau_{\nu\vphantom{'}}\right)}\left.\partial_i t_{\nu'\nu}^ {\alpha\alpha'}(\bm r)\right|_{\bm R_{p} + \bm \tau_{\nu'} - \bm \tau_{\nu\vphantom{'}}} \\
        & = -\bigl[f_i(\bm \kappa)\bigr]_{\nu'\nu}^ {\alpha\alpha'}\,.
    \end{split}
\end{equation}
Here, we exploit invariance under inversion of the Bravais vectors set $\{\bm R_p\}$. Moreover, by virtue of $t_{\nu\nu'}^ {\alpha\alpha'}(\bm r)$ being a real number, we have $\bigl[f_i(-\bm \kappa)\bigl]_{\nu\nu'}^{\alpha\alpha'} = \bigl[f_i^*(\bm \kappa)\bigl]_{\nu\nu'}^{\alpha\alpha'} $. Exploiting both these properties, we thus obtain
\begin{equation}
    \begin{split}
       \frac{\partial \hat{\mathcal{H}_\text{e}}}{\partial u_{p\nu i}} & =  \frac{1}{N^2}\sum_{\alpha\alpha'}\sum_{p'\nu'} \sum_{\bm \kappa \bm k \bm k'}^{\text{FBZ}} e^{i\bm\kappa \cdot \left(\bm R^0_{p\nu} - \bm R^0_{p'\nu'}\right)} \bigl[ f_{i} (\bm \kappa) \bigr]_{\nu\nu'}^ {\alpha\alpha'}  \left\{e^{-i\bm k \cdot \bm R^0_{p\nu}} e^{i\bm k' \cdot \bm R^0_{p'\nu'}} c^{\dagger}_{\bm k\nu\vphantom{'}\alpha}c^{\vphantom{\dagger}}_{\bm k'\nu'\alpha'}\right\}+ \text{H.c.}\\
        & = \frac{1}{N} \sum_{\alpha\alpha'}\sum_{\nu'}\sum_{\bm \kappa \bm k \bm k'}^{\text{FBZ}}  \bigl[ f_{i} (\bm \kappa) \bigr]_{\nu\nu'}^ {\alpha\alpha'}\,\left\{e^{i\left(\bm \kappa - \bm k\right) \cdot \bm R^0_{p\nu}} \,\delta_{\bm \kappa, \bm k'}\, c^{\dagger}_{\bm k\nu\vphantom{'}\alpha}c^{\vphantom{\dagger}}_{\bm k'\nu'\alpha'} +
        e^{i\left(\bm \kappa + \bm k\right) \cdot \bm R^0_{p\nu}} \,\delta_{\bm \kappa, -\bm k'}\,c^{\dagger}_{\bm k'\nu'\alpha'}c^{\vphantom{\dagger}}_{\bm k\nu\alpha\vphantom{'}} \right\} \\
        & = \frac{1}{N} \sum_{\alpha\alpha'}\sum_{\nu'}\sum_{\bm k \bm k'}^{\text{FBZ}} \left\{e^{i\left(\bm k' - \bm k\right) \cdot \bm R^0_{p\nu}}  c^{\dagger}_{\bm k\nu\vphantom{'}\alpha} \bigl[ f_{i} (\bm \kappa) \bigr]_{\nu\nu'}^ {\alpha\alpha'} c^{\vphantom{\dagger}}_{\bm k'\nu'\alpha'} -
        e^{i\left(\bm k - \bm k'\right) \cdot \bm R^0_{p\nu}}\,c^{\dagger}_{\bm k'\nu'\alpha'}  \bigl[ f_{i} (\bm \kappa) \bigr]_{\nu'\nu}^ {\alpha'\alpha} c^{\vphantom{\dagger}}_{\bm k\nu\vphantom{'}\alpha} \right\} \\
        & = \frac{1}{N}\sum_{\alpha\alpha'} \sum_{\nu'}\sum_{\bm k \bm k'}^{\text{FBZ}} \left\{e^{i\left(\bm k' - \bm k\right) \cdot \bm R^0_{p\nu}}  c^{\dagger}_{\bm k\nu\vphantom{'}\alpha}  \bigl[ f_{i} (\bm \kappa) \bigr]_{\nu\nu'}^ {\alpha\alpha'} c^{\vphantom{\dagger}}_{\bm k'\nu'\alpha'}  \right\} + \text{H.c.}\,.
    \end{split}
\end{equation}
Now, we rotate the creation and annihilation operators in the diagonal basis $\hat{\gamma}^{\vphantom{\dagger}}_{n\bm k\alpha}$, i.e.
\begin{equation}
    \hat{c}^{\dagger}_{\bm k\nu\vphantom{'}\alpha} \hat{c}^{\vphantom{\dagger}}_{\bm k'\nu'\alpha'} = \sum_{n n'} \bigl[U^\dagger_{n}(\bm k)\bigr]_{\nu\alpha} \bigl[U^{\vphantom{\dagger}}_{n'}(\bm k')\bigr]_{\nu'\alpha'} \hat{\gamma}^\dagger_{n\bm k} \hat{\gamma}^{\vphantom{\dagger}}_{n'\bm k'}\,.
\end{equation}
Our final expression for the Hamiltonian first-order parametric derivative is thus
\begin{equation}
    \label{kspace-first-order-derivative}
    \begin{split}
         \frac{\partial \hat{\mathcal{H}_\text{e}}}{\partial u_{p\nu i}}
        & = \frac{1}{N} \sum_{\alpha\alpha'}\sum_{\nu'} \sum_{nn'} \sum_{\bm k \bm k'}^{\text{FBZ}} \biggl\{e^{i\left(\bm k' - \bm k\right) \cdot \bm R^0_{p\nu}} \bigl[U^\dagger_{n}(\bm k)\bigr]_{\nu\alpha}\, \bigl[ f_{i} (\bm \kappa) \bigr]_{\nu\nu'}^ {\alpha\alpha'}  \bigl[U^{\vphantom{\dagger}}_{n'}(\bm k')\bigr]_{\nu'\alpha'} \hat{\gamma}^\dagger_{n\bm k} \hat{\gamma}^{\vphantom{\dagger}}_{n'\bm k'}
        \biggr\} + \text{H.c.}\,.
    \end{split}
\end{equation}
We shall now calculate the second-order parametric derivative of the Hamiltonian~\eqref{He-simplified}. We proceed in the exact same way as above, omitting orbital indices for the sake of simplicity:
\begin{equation}
    \begin{split}
        \left.\partial_{p\nu i}\partial_{p'\nu'j}\mathcal{H}_{\text{e}} \right|_0 & =\left. \partial_{p\nu i} \sum_{p_1 \nu_1} \sum_{p_2\nu_2} \Bigl\{ \partial_{p_1\nu_1 j} \,t_{\nu_1\nu_2} \delta_{p p_1}\delta_{\nu\nu_1}  
 + \partial_{p_2\nu_2j}\, t_{\nu_1\nu_2} \delta_{p p_2}\delta_{\nu\nu_2} \Bigr\}\,\hat{c}^{\dagger}_{p_1\nu_1}\hat{c}^{\vphantom{\dagger}}_{p_2\nu_2}\right|_0
        \\
        & = \sum_{p_1\nu_1} \sum_{p_2 \nu_2} \left\{\left.\partial_{p_1\nu_1i}\partial_{p_1\nu_1 j}\, t_{\nu_1\nu_2}  \right|_0\delta_{p p_1}\delta_{\nu\nu_1}   \delta_{p' p_1} \delta_{\nu'\nu_1} \right.\\
            & \phantom{\sum_{p_1\nu_1} \sum_{p_2 \nu_2} } \quad  +\left.\partial_{p_2\nu_2i}\partial_{p_2\nu_2 j} t_{\nu_1\nu_2}  \right|_0\delta_{p p_2}\delta_{\nu\nu_2}   \delta_{p' p_2} \delta_{\nu'\nu_2}  \\
            & \phantom{\sum_{p_1\nu_1} \sum_{p_2 \nu_2} } \quad +  \left. \partial_{p_1\nu_1i}\partial_{p_2\nu_2 j}  t_{\nu_1\nu_2}  \right|_0\delta_{p 'p_1}\delta_{\nu'\nu_1}   \delta_{p p_2} \delta_{\nu\nu_2} \\
            & \phantom{\sum_{p_1\nu_1} \sum_{p_2 \nu_2} } \quad + \left. \left.\partial_{p_2\nu_2i}\partial_{p_1\nu_1 j}  t_{\nu_1\nu_2}  \right|_0\delta_{p p_1}\delta_{\nu\nu_1}   \delta_{p' p_2} \delta_{\nu'\nu_2} \right\} \,\hat{c}^{\dagger}_{p_1\nu_1}\hat{c}^{\vphantom{\dagger}}_{p_2\nu_2} 
        \\
        & = \delta_{pp'}\delta_{\nu\nu'} \sum_{p_2 \nu_2} \left.   \partial_i\partial_j t_{\nu\nu_2}(\bm r)  \right|_{\bm R^0_{p\nu} - \bm R^0_{p_2\nu_2}} \,\hat{c}^{\dagger}_{p\nu}\hat{c}^{\vphantom{\dagger}}_{p_2\nu_2}\\
            & \,+ \delta_{pp'}\delta_{\nu\nu'} \sum_{p_1 \nu_1} \left.   \partial_i\partial_j t_{\nu_1\nu}(\bm r)  \right|_{\bm R^0_{p_1\nu_1} - \bm R^0_{p\nu}} \,\hat{c}^{\dagger}_{p_1\nu_1}\hat{c}^{\vphantom{\dagger}}_{p\nu}\\
            & \phantom{\delta_{pp'}\delta_{\nu\nu'}\sum_{p_1 \nu_1}  }\,\,- \left.  \partial_i\partial_j t_{\nu\nu'}(\bm r)  \right|_{\bm R^0_{p\nu} - \bm R^0_{p'\nu'}} \,\hat{c}^{\dagger}_{p\nu}\hat{c}^{\vphantom{\dagger}}_{p'\nu'}\\
            & \phantom{\delta_{pp'}\delta_{\nu\nu'}\sum_{p_1 \nu_1}  }\,\ - \left.  \partial_i\partial_j t_{\nu'\nu}(\bm r)  \right|_{\bm R^0_{p'\nu'} - \bm R^0_{p\nu}} \,\hat{c}^{\dagger}_{p'\nu'}\hat{c}^{\vphantom{\dagger}}_{p\nu}
        \\
        & = \sum_{p''\nu''} \left(\delta_{pp'}\delta_{\nu\nu'} 
 - \delta_{p'p''}\delta_{\nu'\nu''}\right)  \partial_i\partial_j\left. t_{\nu\nu''}(\bm r)  \right|_{\bm R^0_{p\nu} - \bm R^0_{p''\nu''}}\hat{c}^{\dagger}_{p\nu}\hat{c}^{\vphantom{\dagger}}_{p''\nu''}  + \text{H.c.} \,.
    \end{split}
\end{equation}
In calculating this form, we exploited the parity properties of the second derivatives of the hopping functions, and arbitrarily relabeled some dummy indices.
We define the Fourier transform of the Hessian of the hopping function as 
\begin{equation}
\label{kspace-hessian}
     \bigl[M_{ij}(\bm \kappa)\bigr]_{\nu\nu'} \equiv \sum_{p} e^{-i\bm \kappa \cdot\left( \bm R_{p} + \bm \tau_{\nu\vphantom{'}} - \bm \tau_{\nu'}\right)} \left.  \partial_i\partial_j t_{\nu\nu'}(\bm r)\right|_{\bm R_{p} + \bm \tau_{\nu\vphantom{'}} - \bm \tau_{\nu'}} \,.
\end{equation}
This object can be easily proven to be hermitian, following the same line of reasoning as above. The following constraints are easy to prove:
\begin{equation}
    \bigl[M_{ij}(\bm \kappa)\bigr]_{\nu\nu'}= \bigl[M_{ij}^*(\bm \kappa)\bigr]_{\nu'\nu} \qquad \text{and} \qquad  \bigl[M_{ij}(-\bm \kappa)\bigr]_{\nu\nu'} = \bigl[M_{ij}^ *(\bm \kappa)\bigr]_{\nu\nu'}\,.
\end{equation}
We obtain the following form:
\begin{equation}
\label{kspace-second-order-derivative}
    \begin{split}
         \left.\partial_{p\nu i}\partial_{p'\nu'j} \mathcal{H}_{\text{e}} \right|_0  & = \frac{1}{N^2}\sum_{p''\nu''} \sum_{\bm \kappa \bm k \bm k'}^{\text{FBZ}} \left(\delta_{pp'}\delta_{\nu\nu'} 
 - \delta_{p'p''}\delta_{\nu'\nu''}\right)  \bigl[M_{ij}(\bm \kappa)\bigr]_{\nu\nu''}\times\\
             &\phantom{\frac{1}{N^2}\sum_{p''\nu''}} \times e^{i\kappa \cdot \left(\bm R^0_{p\nu} - \bm R^0_{p''\nu''}\right)} \left\{e^{-i\bm k \cdot \bm R^0_{p\nu}} e^{i\bm k' \cdot \bm R^0_{p''\nu''}}  c^{\dagger}_{\bm k\nu\vphantom{'}}c^{\vphantom{\dagger}}_{\bm k'\nu''}\right\} + \text{H.c.}
    \end{split}
\end{equation}
\begin{equation*}
    \begin{split}
        &  = \frac{1}{N^2}\sum_{p''\nu''} \sum_{\bm \kappa \bm k \bm k'}^{\text{FBZ}} \left(\delta_{pp'}\delta_{\nu\nu'} 
 - \delta_{p'p''}\delta_{\nu'\nu''}\right)  \bigl[M_{ij}(\bm \kappa)\bigr]_{\nu\nu''}\times\\
             &\phantom{e^{i\kappa \cdot \left(\bm R^0_{p\nu} - \bm R^0_{p''\nu''}\right)}} \times \left\{e^{i\left(\kappa - \bm k\right) \cdot \bm R^0_{p\nu}} e^{-i\left(\bm \kappa - \bm k'\right)\cdot \bm R^0_{p''\nu''} }  c^{\dagger}_{\bm k\nu\vphantom{'}}c^{\vphantom{\dagger}}_{\bm k'\nu''}\right.\\
             & \phantom{e^{i\kappa \cdot \left(\bm R^0_{p\nu} - \bm R^0_{p''\nu''}\right)}}  \quad\left.+
            e^{i\left(\kappa + \bm k\right) \cdot \bm R^0_{p\nu}} e^{-i\left(\bm \kappa + \bm k'\right)\cdot \bm R^0_{p''\nu''} } c^{\dagger}_{\bm k'\nu''}c^{\vphantom{\dagger}}_{\bm k\nu\vphantom{'}} \right\} \\
        & = \delta_{pp'}\delta_{\nu\nu'}  \frac{1}{N}  \sum_{\nu''}\sum_{\bm k \bm k'}^{\text{FBZ}} \left\{e^{i\left(\bm k' - \bm k\right) \cdot \bm R^0_{p\nu}}  c^{\dagger}_{\bm k\nu\vphantom{'}}   \bigl[M_{ij}(\bm k')\bigr]_{\nu\nu''} c^{\vphantom{\dagger}}_{\bm k'\nu''} \right\} + \text{H.c.}\\
            & \phantom{\delta_{pp'}\delta_{\nu\nu'}}  - \frac{1}{N^2} \sum_{\bm \kappa \bm k \bm k'}^{\text{FBZ}}  e^{i\bm \kappa \cdot \left(\bm R^0_{p\nu} - \bm R^0_{p'\nu'}\right)} \left\{e^{-i\bm k \cdot \bm R^0_{p\nu}} e^{i\bm k' \cdot \bm R^0_{p'\nu'}}  c^{\dagger}_{\bm k\nu\vphantom{'}}   \bigl[M_{ij}(\bm \kappa)\bigr]_{\nu\nu'}c^{\vphantom{\dagger}}_{\bm k'\nu'} \right\}+ \text{H.c.}\,.
    \end{split}
\end{equation*}
\noindent Expectation values will be taken by once again rotating the annihilation and creation operators in the diagonal basis.

We now take the expectation values on states of the form~\eqref{sm-eq:nonzero-states}. The state with $n \bm k = n' \bm k'$ is nothing but the ground state itself, and it is therefore excluded from the sum in Eq.~\eqref{app:eq:force_constant_el}. The relevant expectation values can be recasted into ground-state averages:
\begin{equation}
\label{expect1}
    \left\langle \Omega \middle|  \gamma_{n''\vphantom{'}\bm k''\vphantom{'}}^\dagger \gamma_{n'''\bm k'''}^{\vphantom{\dagger}} \middle| \phi^{(n \bm k,\, n'\bm k')}\right\rangle = \delta_{n n''} \delta_{\bm k \bm k''} \delta_{n'n'''} \delta_{\bm k'\bm k'''}   \Bigl\langle \gamma^\dagger_{n\bm k\vphantom{'}} \gamma^{\vphantom{\dagger}}_{n'\bm k'}  \gamma^\dagger_{n'\bm k'} \gamma^{\vphantom{\dagger}}_{n\bm k\vphantom{'}} \Bigr\rangle_0\,;
\end{equation}
\begin{equation}
\label{expect2}
    \left\langle\phi^{(n \bm k,\, n'\bm k')} \middle|  \gamma_{n''\vphantom{'}\bm k''\vphantom{'}}^\dagger \gamma_{n'''\bm k'''}^{\vphantom{\dagger}} \middle| \Omega \right\rangle = \delta_{n n'''} \delta_{\bm k \bm k'''} \delta_{n'n''} \delta_{\bm k'\bm k''}   \Bigl\langle \gamma^\dagger_{n\bm k\vphantom{'}} \gamma^{\vphantom{\dagger}}_{n'\bm k'}  \gamma^\dagger_{n'\bm k'} \gamma^{\vphantom{\dagger}}_{n\bm k\vphantom{'}} \Bigr\rangle_0\,.
\end{equation}
The resulting two-body ground-state average is easily evaluated
\begin{equation}
    \begin{split}
        \Bigl\langle 
        \gamma^\dagger_{n\bm k\vphantom{'}} \gamma^{\vphantom{\dagger}}_{n'\bm k'}  \gamma^\dagger_{n'\bm k'} \gamma^{\vphantom{\dagger}}_{n\bm k\vphantom{'}} \Bigr\rangle_0 
        & = \Bigl\langle 
        \gamma^\dagger_{n\bm k\vphantom{'}} \left(1-  \gamma^\dagger_{n'\bm k'}\gamma^{\vphantom{\dagger}}_{n'\bm k'} \right)
        \gamma^{\vphantom{\dagger}}_{n\bm k\vphantom{'}} \Bigr\rangle_0 \\
        & = \Bigl\langle 
        \gamma^\dagger_{n\bm k\vphantom{'}} \gamma^{\vphantom{\dagger}}_{n\bm k} 
        \Bigr\rangle -\Bigl\langle \gamma^{\dagger}_{n\bm k\vphantom{'}} \gamma^{\vphantom{\dagger}}_{n\bm k\vphantom{'}}  \gamma^{\dagger}_{n'\bm k'} \gamma^{\vphantom{\dagger}}_{n'\bm k'} \Bigr\rangle_0 \\
        & = n_{n\bm k}\left(1-n_{n'\bm k'}\right)\\
        & = \Theta\left[\mu - E_n(\bm k)\right] \Theta\left[ E_{n'}(\bm k') -\mu\right]\,.
    \end{split}
\end{equation}
It is immediate to see that, under the restrictions in Eq.~\eqref{sm-eq:nonzero-states}, this counting factor is just unity. Moreover, the energy denominator appearing in Eq.~\eqref{app:eq:force_constant_el} is just the difference between the hole energy and the above-the-surface electron energy:
\begin{equation}
    \mathcal{E}_0 - \mathcal{E}_{n\bm k,\,n'\bm k'} =  E_{n}(\bm k) - E_{n'}(\bm k') ~.
\end{equation}
Plugging the expectation values in formula~\eqref{app:eq:force_constant_el}, we immediately arrive via Fourier transform to expressions~\eqref{sm-eq:dI-explicit} and~\eqref{sm-eq:dII-explicit} for the dynamical matrix:
\begin{equation}
\label{di}
    \begin{split}
       \bigl[{\cal D}^{\text{(el, 1)}}(\bm q)\bigr]_{\nu\nu'}^{ij} & = \frac{1}{\sqrt{M_{\nu} M_{\nu'}}} \sum_p e^{-i\bm q \cdot(\bm R_p + \bm \tau_{\nu\vphantom{'}} - \bm \tau_{\nu'})}\left\{\bigl[ {\cal C}^{\text{(el, 1)}}_{ij}\bigr]^{p,\,p'=0}_{\nu \nu'} + \bigl[ {\cal C}^{\text{(el, 1)}}_{ji}\bigr]^{p'=0, \,p}_{\nu' \nu}\right\} \\
        & = \frac{1}{\sqrt{M_{\nu} M_{\nu'}}}\,  \frac{2}{N} \sum_{n\bm k}^{\rm occ.} \sum_{n'\bm k'}^{\rm unocc.}
            \left\{\delta_{\bm k', \,\bm k+\bm q}\,\frac{\bigl[F_i(n\bm k, n'\bm k')\bigr]_{\nu} \bigl[F_j(n'\bm k', n\bm k)\bigr]_{\nu'}}{E_{n}(\bm k) - E_{n'}(\bm k')}\right. \\
            &\phantom{\frac{1}{\sqrt{M_{\nu} M_{\nu'}}} \sum_p  \frac{1}{N^2} }\qquad\quad +
            \left. \delta_{\bm k', \,\bm k-\bm q}\,\frac{\bigl[F_j(n\bm k, n'\bm k')\bigr]_{\nu'} \bigl[F_i(n'\bm k', n\bm k)\bigr]_{\nu}}{E_{n}(\bm k) - E_{n'}(\bm k')}\right\} \\
        & = \frac{1}{\sqrt{M_{\nu} M_{\nu'}}}\,  \frac{2}{N} \sum_{n}^{\rm occ.} \sum_{n'}^{\rm unocc.} \sum_{\bm k}^{\text{FBZ}}
            \left\{\frac{\bigl[F_i(n\bm k, n'\bm k+\bm q)\bigr]_{\nu} \bigl[F_j(n'\bm k+\bm q, n\bm k)\bigr]_{\nu'}}{E_{n}(\bm k) - E_{n'}(\bm k+\bm q)} \right\} + 
       {\rm H.c.}~,
    \end{split}
\end{equation}
\begin{equation}
\label{dii}
    \begin{split}
        \bigl[{\cal D}^{\text{(el, 2)}}(\bm q)\bigr]_{\nu\nu'}^{ij} & = \frac{1}{\sqrt{M_\nu M_{\nu'}}} \sum_p e^{-i\bm q \cdot \left(\bm R_p + \bm \tau_{\nu\vphantom{'}}-\bm \tau_{\nu'}\right)}\, \bigl[ {\cal C}^{{\text{(el, 2)}}}_{ij}\bigr]^{p,\, p'=0}_{\nu \nu'} \\
        & = \frac{1}{M_\nu} \frac{2}{N} \delta_{\nu\nu'}   \sum_{\Tilde{\nu}} \sum_{n}^{\rm occ.} \sum_{\bm k }^{\text{FBZ}} \left\{   M_{\nu\Tilde{\nu};\,ij}(\bm k) \bigl[P_n(\bm k)\bigr]_{\Tilde{\nu}\nu}  
             \right\} + \rm{H.c.} \\
            & -  \frac{1}{\sqrt{M_\nu M_{\nu'}}}\,\frac{2}{N} \sum_{n}^{\rm occ.} \sum_{\bm k}^{\text{FBZ}}  \left\{M_{\nu\nu';\,ij}(\bm k + \bm q) \bigl[P_n(\bm k)\bigr]_{\nu'\nu}\right\} + \rm{H.c.}~.
    \end{split}
\end{equation}

\section{Formal analogy with linear response theory}\label{section5}

The analogy with linear response theory is built as follows. The small-displacement expansion of the electronic Hamiltonian (Eq.~(1) in the main text) is
\begin{equation}
\label{sm-eq:hamiltonian-expansion}
        \hat{{\cal H}}_{\bm u} = \left. \hat{{\cal H}}_{\bm u}\right|_0 + \sum_{p,\nu,i}\left.\frac{\partial \hat{{\cal H}}_{\bm u}}{\partial u_{p\nu i}} \right|_{0}\,u_{p\nu i} + \frac{1}{2}\sum_{p,\nu,i} \sum_{p'\nu',j}\left.\frac{\partial^ 2  \hat{{\cal H}}_{\bm u}}{\partial u_{p\nu i}\partial u_{p'\nu'j}}\right|_{0} \,u_{p\nu i}\,u_{p'\nu' j}~,
\end{equation}
where we introduced the shorthand notation $ \hat{{\cal H}}_{\bm u}\equiv \hat{\cal H}_{\rm e}(\{\bm u_{p\nu}\})$, and the $|_0$ subscript once again denotes the evaluation on the mechanical equilibrium configuration. The most general Hamiltonian of an electron system electron which has been minimally coupled to a time-independent gauge field $\bm A(\bm r, t) = \bm A(\bm r)$ is
\begin{equation}
\label{sm-eq:minimal-coupling}
    \hat{\cal H}_{\bm A} = \left.\hat{\cal H}_{\bm A}\right|_0 + \sum_i\int{\rm d}^ D\bm r\,\left.\frac{\delta \hat{\cal H}_{\bm A}}{\delta A_i(\bm r)} \right|_{{0}}\,A_i(\bm r) + \frac{1}{2}\sum_{i,j}\iint{\rm d}^ D\bm r\,{\rm d}^ D\bm r'\,\left.\frac{\delta^ 2 \hat{\cal H}_{\bm A}}{\delta A_i(\bm r) \delta A_j(\bm r)} \right|_{0} \, A_i(\bm r)A_j(\bm r')~,
\end{equation}
where $\delta$ denotes a functional derivative, and the $|_0$ subscript means $\bm A = 0$. Considering a time-independent vector field parallels the Born-Oppenheimer approximation. The two expressions~\eqref{sm-eq:hamiltonian-expansion} and~\eqref{sm-eq:minimal-coupling} are formally equivalent under exchange of
\begin{equation}
\label{sm-eq:substitutions}
      \bm r \leftrightarrow \bm R_{p\nu}~, \qquad \sum_i \int {\rm d}^D \bm r \leftrightarrow \sum_i \sum_{p,\,\nu}~, \qquad
    \hat{\cal H}_{\bm A}  \leftrightarrow \hat{{\cal H}}_{\bm u}~,\qquad  \bm A(\bm r)\leftrightarrow \bm u_{p\nu}  ~,
\end{equation}
Identifying $\bm A(\bm r)$ with an electromagnetic vector potential, we can define the paramagnetic current operator $\hat{j}_i(\bm r)$ and the diamagnetic operator $\hat{\cal T}_{ij}(\bm r, \bm r')$ respectively as
\begin{equation}
\label{sm-eq:paramagnetic-current}
    \hat{j}_{i}(\bm r) \equiv \frac{c}{e}\left. \frac{\delta {\hat{\cal H}}_{\bm A}}{\delta A_i(\bm r)}\right|_{0}~,
\end{equation}
and
\begin{equation}
\label{sm-eq:diamagnetic-operator}
    \hat{\cal T}_{ij}(\bm r, \bm r') \equiv \frac{c^ 2}{e^ 2}\left.\frac{\delta^ 2 \hat{\cal H}_{\bm A}}{\delta A_i(\bm r) \delta A_j(\bm r')} \right|_{0}~.
\end{equation}
The {\it physical} current operator for the Hamiltonian in Eq.~\eqref{sm-eq:minimal-coupling} can be thus written as
\begin{equation}
\label{sm-eq:physical-current}
    \hat{J}_{i}(\bm r) \equiv \frac{c}{e}\frac{\delta {\hat{\cal H}}_{\bm A}}{\delta A_i(\bm r)} = \hat{j}_{p,i}(\bm r)+ \frac{e}{c}\sum_j\int {\rm d}^D \bm r'\,\hat{\cal T}_{ij}(\bm r, \bm r')\,A_j(\bm r') ~.
\end{equation}
By formal analogy, we can build the same operators for the theory described by the Hamiltonian~\eqref{sm-eq:hamiltonian-expansion}:
\begin{align}
    \hat{j}_{p\nu i} & \equiv\left.\frac{\partial \hat{{\cal H}}_{\bm u}}{\partial u_{p\nu i}} \right|_{0}~,\\[3pt]
    \hat{\cal T}_{p\nu i}^ {p'\nu'j} & \equiv \left.\frac{\partial^ 2  \hat{{\cal H}}_{\bm u}}{\partial u_{p\nu i}\partial u_{p'\nu'j}}\right|_{0} ~,\\[3pt]
    \hat{J}_{p\nu i} & \equiv \hat{j}_{p\nu i} + \sum_{p',\nu', j}  \hat{\cal T}_{p\nu i}^ {p'\nu'j} u_{p'\nu'j}~.
\end{align}
For each of the two theories, we can build a zero-temperature, static, paramagnetic current-current response function, which can be written as follows in the Lehmann representation :
\begin{align}
\label{sm-eq:response-function}
    \chi_{j_{i}\, j_{j}}^ {\vphantom{\dagger}}(\bm r, \bm r')& = \sum_{m\neq 0} \frac{\left \langle \phi^{(m )}_{\bm A = \bm 0} \middle| \hat{j}_{i}(\bm r)\middle| \phi^ {(0)}_{\bm A=\bm 0}\right\rangle \left \langle \phi^{(0 )}_{\bm A = \bm 0} \middle| \hat{j}_{j}(\bm r')\middle| \phi^ {(m)}_{\bm A = \bm 0}\right\rangle}{{\cal E}_0-{\cal E}_m} + {\rm H.c.}~,\\
\label{sm-eq:equivalent-response-function}
    \chi_{j_{p\nu i}\, j_{p'\nu'j}}^ {\vphantom{\dagger}} &= \sum_{m\neq 0} \frac{\left \langle \phi^{(m )}_{\{\bm 0\}} \middle| \hat{j}_{p\nu i}\middle| \phi^ {(0)}_{\{\bm 0\}}\right\rangle \left \langle \phi^{(0 )}_{\{\bm 0\}} \middle| \hat{j}_{p'\nu'j}\middle| \phi^ {(m)}_{\{\bm 0\}}\right\rangle}{{\cal E}_0-{\cal E}_m} + {\rm H.c.}~,
\end{align}
with the same notation for excited states of the Hamiltonian as in the main text. To obtain the electronic force constants, we start from Eq.~\eqref{sm-eq:hellmann-feynman}, which states the following:
\begin{equation}
   \bigl[{\cal C}^ {\rm (el)}\bigr]_{p\nu i}^ {p'\nu'j} = \left[\frac{\partial}{\partial u_{p'\nu' j}}\left\langle \phi^ {(0)}_{\{\bm u_{p\nu}\}}\middle |\frac{\partial\hat{\cal H}_{\bm u}}{\partial u_{p\nu i}}\middle|\phi^ {(0)}_{\{\bm u_{p\nu}\}}\right\rangle\right]_0 \equiv \left[\frac{\partial}{\partial u_{p'\nu' j}}\left\langle \hat{J}_{p\nu i}\right\rangle\right]_0~,
\end{equation}
where we introduced the shorthand notation $\langle\,\cdot\,\rangle \equiv \langle \phi^ {(0)}_{\{\bm u_{p\nu}\}}|\cdot|\phi^ {(0)}_{\{\bm u_{p\nu}\}} \rangle$. This equation parallels the definition of the physical current-current response function~\cite{GiulianiVignale_SM}, i.e.
\begin{equation}
   \chi^ J_{ij}(\bm r, \bm r') \equiv\left[\frac{\delta}{\delta A_j(\bm r')} \left\langle\phi^{(0)}_{\bm A}\middle| \frac{\delta \hat{\cal H}_{\bm A}}{\delta A_i(\bm r)} \middle|\phi^{(0)}_{\bm A}\right\rangle\right]_{0} 
   \equiv \left[\frac{\delta}{\delta A_j(\bm r')} \left\langle\hat{J}_i(\bm r)\right\rangle\right]_{0} ~ . 
\end{equation}
The expectation value of the physical current operator to first order in the gauge field amplitude is obtained as
\begin{equation}
\label{sm-eq:physical-currrent-expvalue}
    \left\langle\hat{J}_i(\bm r)\right\rangle = \frac{e}{c}\sum_l \int {\rm d}^ D \bm r' \left[\chi_{j_{i}\, j_{l}}^ {\vphantom{\dagger}}(\bm r, \bm r') + \frac{e}{c}\left\langle \hat{\cal T}_{il}(\bm r, \bm r')\right\rangle \right] A_l(\bm r') + {\cal O}(A^ 2)~,
\end{equation}
so we obtain
\begin{align}
\label{sm-eq:formal-analogy1}
   \chi^ J_{ij}(\bm r, \bm r') &=  \chi_{j_{i}\, j_{j}}^ {\vphantom{\dagger}}(\bm r, \bm r') + \frac{e}{c}\left\langle \hat{\cal T}_{ij}(\bm r, \bm r')\right\rangle  ~,\\
\label{sm-eq:formal-analogy2}
    \bigl[{\cal C}^ {\rm (el)}\bigr]_{p\nu i}^ {p'\nu'j} & =  \chi_{j_{p\nu i}\, j_{p'\nu'j}}^ {\vphantom{\dagger}} +\left\langle\hat{\cal T}_{p\nu i}^ {p'\nu'j}\right\rangle
\end{align}
We can therefore identify the linear contribution ${\cal C}^ {(\rm el,\, 1)}$ to the force constants as a paramagnetic current-current response function, through the formal substitution~\eqref{sm-eq:paramagnetic-current}. The second-order contribution ${\cal C}^ {\rm (el,\,2)}$ is therefore the parallel of the diamagnetic operator ground-state expectation value. 

Via this identification, the acoustic sum rule restricted to ${\cal D}^ {(\rm el)}(\bm q)$ can be inferred as follows. Since a static, uniform field $\bm A(\bm r, t) \equiv \bm A$ can be gauged away by a unitary transformation of the Hamiltonian, the system cannot respond with a physical current to such a potential. Therefore, imposing that the physical current expectation value (Eq.~\eqref{sm-eq:physical-currrent-expvalue}) vanishes at every order in $A$, i.e. 
\begin{equation}
   \sum_j\int {\rm d}^ D\bm r' \chi^J_{ij}(\bm r, \bm r') A_j(\bm r)=  \sum_j A_j \int {\rm d}^ D\bm r' \chi^J_{ij}(\bm r, \bm r') = 0\qquad \forall ~\bm A \in \mathds{R}^ D~,
\end{equation}
we find the real-space TRK sum rule~\cite{Sakurai2017_SM, Andolina2019_SM}:
\begin{equation}
    \int {\rm d}^ D\bm r' \chi^J_{ij}(\bm r, \bm r') =\int {\rm d}^ D\bm r' \left[\chi_{j_{i}\, j_{ j}}^ {\vphantom{\dagger}}(\bm r, \bm r') + \frac{e}{c}\left\langle \hat{\cal T}_{ij}(\bm r, \bm r')\right\rangle \right] = 0\qquad \forall \, \bm r\in \mathds{R}^D, i, j~.
\end{equation}
In Sect.~VIII, we prove the TRK sum rule explicitly for a 1D toy model. By formal analogy, we can apply the same procedure to the parallel theory described by the Hamiltonian~\eqref{sm-eq:hamiltonian-expansion}, for a static and uniform displacement ${\bm u}_{p\nu} \equiv {\bm u}~\forall~p,\nu$ of all the atoms. We impose
\begin{equation}
   \sum_{p',\nu',j}  \bigl[ {\cal C}^ {\rm (el)} \bigr]_{p\nu i}^ {p'\nu' j} u_{p'\nu'j}=  \sum_j u_{j} \sum_{p',\nu'}  \bigl[ {\cal C}^ {\rm (el)} \bigr]_{p\nu i}^ {p'\nu' j}=0\qquad \forall ~\bm u \in \mathds{R}^ D~,
\end{equation}
thus obtaining the real-space version of the acoustic sum rule (Eq.~(9) in the main text) as the counterpart of the TRK sum rule:
\begin{equation}
   \sum_{p',\nu'}  \bigl[ {\cal C}^ {\rm (el)} \bigr]_{p\nu i}^ {p'\nu' j} =  \sum_{p',\nu'}\left[\chi^{\vphantom{\dagger}}_{j_{p\nu i}\,j_{p'\nu'j}}+\left\langle\hat{\cal T}_{p\nu i}^ {p'\nu'j}\right\rangle\right] = 0\qquad \forall~p,\nu, i,j~.
\end{equation}

\section{Proof of the TRK sum rule in a simple case}\label{section6}

We will now explicitly prove the TRK sum rule for a simple 1D toy model, the so-called {\it extended Falikov-Kimball (EFK) model} (see e.g.~Ref.~\cite{Andolina2019_SM} and references therein to earlier work), which describes a 1D chain of identical atoms. EFK models are often used in the literature to describe excitonic insulators provided that interactions are added to the noninteracting Hamiltonian below (for a complete list of references see Ref.~\cite{Andolina2019_SM}). Each atom has an orbital degree of freedom $\alpha = \{s, p\}$, which may correspond to the $s$ and $p$ orbitals of a hydrogenoid atom. In the following, we will consider nearest neighbors (NN) interactions only. The first Brillouin zone (FBZ) for a monopartite 1D lattice of lattice spacing $a$ is just the ring $(-\pi/a, \pi/a]$.

We begin by explicitly writing the EFK Hamiltonian, including only NN terms, as defined in Ref.~\cite{Andolina2019_SM}:
\begin{equation}\label{EFK_model}
    \hat{\cal H} = \sum_{k \in {\rm FBZ}} 
    \begin{pmatrix}
    \hat{c}^ {\dagger}_{ k, s}, \hat{c}^ {\dagger}_{ k, p}
    \end{pmatrix}
    \begin{pmatrix}
        \varepsilon_s - 2 t_s \cos ka & 2i\tilde{t} \sin ka \\[3pt]
        -2i\tilde{t} \sin ka &  \varepsilon_p + 2 t_p \cos ka
    \end{pmatrix}
    \begin{pmatrix}
        \hat{c}_{k,s}\\[3pt]
        \hat{c}_{k, p}
    \end{pmatrix}
    \equiv 
    \sum_{k \in {\rm FBZ}}\sum_{\alpha, \,\beta} \hat{c}^ {\dagger}_{ k, \alpha} H_{\alpha\beta}(k) \hat{c}_{k, \beta}~,
\end{equation}
where $\hat{c}^ \dagger_{k, \alpha}$ creates an electron in a Bloch state with momentum $k$ and orbital flavor $\alpha\in \{s, p\}$. In Eq.~(\ref{EFK_model}) we introduce on-site energies $\varepsilon_\alpha$, intra-orbital hopping amplitudes $t_\alpha$, and an inter-orbital amplitude $\tilde{t} \equiv t_{sp} = t_{ps}$, which is real by gauge choice. To simplify the model, we shift the zero-point energy by setting $\delta \equiv (\varepsilon_s - \varepsilon_p)/2$, and assume  $t_s = t_p \equiv t$. The Hamiltonian can therefore be written in the following pseudospin form:
\begin{equation}
    \underline{H}(k) = 
    \begin{pmatrix}
        \delta - 2 t \cos ka & 2i\tilde{t} \sin ka \\[3pt]
         -2i\tilde{t} \sin ka & -\delta + 2 t \cos ka 
    \end{pmatrix}
    = \bm d_k \cdot  {\bm \sigma}~,
\end{equation}
where $\bm \sigma = (\sigma_x, \sigma_y, \sigma_z)$ is the usual vector of spin-$1/2$ Pauli matrices and $\bm d_k = (0, - 2 \tilde{t} \sin ka, \delta - 2t \cos k a)$.  

The energy bands of the noninteracting 1D EFK model can be readily obtained using the straightforward procedure outlined in Ref.~\cite{bernevig_book_2013_SM}:
\begin{equation}
    \varepsilon_{\pm,\, k} = \pm \varepsilon_k = \pm \sqrt{4\tilde t\,^ 2 \sin^ 2 ka + (\delta - 2 t \cos ka)^ 2}~,
\end{equation}
together with the Bloch eigenvectors
\begin{align}
    U_{+, k} &= \frac{1}{\sqrt{2 \varepsilon_k(\varepsilon_k+\delta - 2 t \cos ka)}} 
    \begin{pmatrix}
        \varepsilon_k+\delta - 2 t \cos ka\\[3pt]
        -2i\tilde t \sin ka
    \end{pmatrix}~,\\
     U_{-,k} &= \frac{1}{\sqrt{2 \varepsilon_k(\varepsilon_k+\delta - 2 t \cos ka)}} 
    \begin{pmatrix}
    -2i\tilde t \sin ka\\[3pt]
    \varepsilon_k+\delta - 2 t \cos ka
    \end{pmatrix}~.
\end{align}
We impose $\delta \neq 2 t$ to avoid accidental band crossing. The Bloch states of the system are defined by the contraction of the Bloch eigenvectors with the orbital basis operators, i.e.
\begin{equation}
    |\psi_{\pm, k}\rangle = \hat{\gamma}_{\pm,k}^ \dagger|0\rangle= \sum_{\alpha} \bigl[U_{\pm, k}\bigr]_{\alpha} \,\hat{c}^ \dagger_{k, \alpha} |{0}\rangle \equiv \sum_{\alpha}\bigl[U_{\pm, k}\bigr]_{\alpha} |k,\alpha\rangle~,
\end{equation}
where $\hat{\gamma}_{\pm,k}^ \dagger$ is the fermionic creation operator in the band basis.

The static $\omega = 0, \,q = 0$ paramagnetic current operator can be presented as
\begin{equation}
\label{eqn:current-operator}
    \hat{j}_p \equiv 
    \sum_{k \in {\rm FBZ}}\sum_{\alpha, \,\beta} \hat{c}^ {\dagger}_{ k, \alpha} j_{p;\,\alpha\beta}(k) \hat{c}_{k, \beta} =
    \frac{2a}{\hbar}
    \sum_{k \in {\rm FBZ}}
    \begin{pmatrix}
    \hat{c}^ {\dagger}_{ k, s}, \hat{c}^ {\dagger}_{ k, p}
    \end{pmatrix}
    \begin{pmatrix}
      t \sin ka & i\tilde{t} \cos ka \\[3pt]
        -i\tilde{t} \cos ka & -t\sin ka
    \end{pmatrix}
    \begin{pmatrix}
        \hat{c}_{k,s}\\[3pt]
        \hat{c}_{k, p}
    \end{pmatrix}~,
\end{equation}
where we have introduced the following first derivative of the Hamiltonian with respect to $k$:
\begin{equation}
    j_{p;\,\alpha\beta}(k) \equiv \frac{1}{\hbar}\frac{\partial H_{\alpha\beta}(k)}{\partial k_p}~.
\end{equation}
This current operator finds a more convenient representation in the band basis:
\begin{equation}
    \sum_{k\in \rm FBZ}\sum_{\alpha, \,\beta} \hat{c}^ {\dagger}_{ k, \alpha} j_{p;\,\alpha\beta}(k) \hat{c}_{k, \beta} 
    = \sum_{k\in \rm FBZ}\sum_{\alpha, \,\beta} \sum_{n, m = \pm} \hat{\gamma}^ {\dagger}_{ n, k} \bigl[U^ \dagger_{n,k}\bigr]_{\alpha}\,j_{p;\,\alpha\beta}(k) \,\bigl[U_{n, k}\bigr]_\beta\hat{\gamma}_{n, k}
    \equiv \sum_{k\in \rm FBZ} \sum_{n, m = \pm} \hat{\gamma}^ {\dagger}_{ n, k} \,j_{p;\, nm}(k)\,\hat{\gamma}_{n, k}~,
\end{equation}
with
\begin{equation}
    j_{p;\,nm}(k) = \sum_{\alpha, \,\beta}  \bigl[U^ \dagger_{n,k}\bigr]_{\alpha}\,j_{p;\,\alpha\beta}(k) \,\bigl[U_{m, k}^ {\vphantom{\dagger}}\bigr]_\beta~.
\end{equation}
The current vector is parallel to the Bravais lattice vector $\hat{\bm u}$, i.e.  $\hat{\bm j}_p = \hat{ j}_p\hat{\bm u}$, where $\hat{ j}_p$ is defined above in Eq.~\eqref{eqn:current-operator}. The total (i.e.~physical) macroscopic (i.e.~$q = 0$) current operator, which measures the linear response of the system to a static and uniform vector potential $\bm A_0 = A_0 \hat{\bm u}$, contains the above paramagnetic contribution but also a {\it diamagnetic} one. This can be easily seen by coupling the 1D EFK model to ${\bm A}_0$ on the lattice via the Peierls substitution. The linear response of the system to ${\bm A}_0$ requires to expand the Peierls-coupled EFK model up to the second order in $A_0$:
\begin{equation}
    \hat{J}_{\rm phys} = \hat{j}_p - \frac{e}{c} A_0 \frac{a^2}{\hbar^2} \hat{\cal T}~.
\end{equation}
Here we have introduced the ``kinetic'' operator:
\begin{equation}
\label{eqn:kinetic-operator}
    \hat{\cal T} \equiv 
    \sum_{k\in \rm FBZ}\sum_{\alpha, \,\beta} \hat{c}^ {\dagger}_{ k, \alpha} {\cal T}_{\alpha\beta}(k) \hat{c}_{k, \beta} =  
    2\sum_{k\in \rm FBZ}
    \begin{pmatrix}
    \hat{c}^ {\dagger}_{ k, s}, \hat{c}^ {\dagger}_{ k, p}
    \end{pmatrix}
    \begin{pmatrix}
      -t \cos ka & i\tilde{t} \sin ka \\[3pt]
        -i\tilde{t} \sin ka & t\cos ka
    \end{pmatrix}
    \begin{pmatrix}
        \hat{c}_{k,s}\\[3pt]
        \hat{c}_{k, p}
    \end{pmatrix}~,
\end{equation}
where we have introduced the following second derivative of the Hamiltonian with respect to $k$:
\begin{equation}
   {\cal T}_{\alpha \beta}(k) \equiv -\frac{1}{a^2}\frac{\partial^2 H_{\alpha\beta}(k)}{\partial k^2}~.
\end{equation}
As discussed, for example, in the Supplemental Material of Ref.~\cite{Andolina2019_SM}, gauge invariance ensures that the system cannot generate a finite physical current in response to a static and spatially uniform vector potential, i.e.,
\begin{equation}\label{eq:TRK_sumrule}
    \langle \hat{J}_{\rm phys}\rangle = \frac{e}{c}\chi_{j_p j_p}^ {\vphantom{\dagger}} A_0 -\frac{e}{c} A_0 \frac{a^2}{\hbar^2} \langle \hat{\cal T} \rangle = \frac{e}{c} \left(\chi_{j_p j_p}^ {\vphantom{\dagger}}  -\frac{a^2}{\hbar^2} \langle \hat{\cal T} \rangle \right)  A_0 = 0 \qquad \forall~A_0~,
\end{equation}
where $\chi_{j_p j_p}$ is the $\omega=0$ and $q\to 0$ limit of the paramagnetic current-current response function, explicitly given by:
\begin{equation}
\label{eq:chi-jpjp}
    \begin{split}
   \chi_{j_p j_p}^ {\vphantom{\dagger}} \equiv  \lim_{q \to 0} \chi_{j_{p;\,q} \,j_{p;\,-q}}^{\vphantom{\dagger}}(\omega = 0) &= \lim_{q \to 0} \frac{1}{L} \sum_{n,m} \sum_{k \in {\rm FBZ}} \frac{f_{n,k}-f_{m,k+q}}{\varepsilon_{n,k} -\varepsilon_{m,k+q}}\,\langle \psi_{n,k}| \hat{\bm j}_{p;\,q}\cdot \hat{\bm u} |\psi_{m,k+q}\rangle \langle \psi_{m,k+q}| \hat{\bm j}_{p;\,-q}\cdot \hat{\bm u}|\psi_{n,k}\rangle\\ 
   &= - \frac{1}{L} \sum_n \sum_{k \in {\rm FBZ}}\delta(\varepsilon_{n,k})\bigl|\langle \psi_{n,k}| \hat{ j}_p|\psi_{n,k}\rangle\bigr|^ 2 \nonumber\\
   &
   +\sum_{n\neq m}\frac{1}{L}\sum_{k \in {\rm FBZ}} \frac{f_{n,k}-f_{m,k}}{\varepsilon_{n,k} -\varepsilon_{m,k}}\bigl|\langle \psi_{n,k}| \hat{ j}_p |\psi_{m,k}\rangle\bigr|^ 2=  
   -\frac{2}{L}\sum_{k \in {\rm FBZ}} \frac{\bigl|\langle \psi_{+,k}| \hat{ j}_p |\psi_{-,k}\rangle\bigr|^ 2}{\varepsilon_{+,\,k} -\varepsilon_{-,\,k}}~,
   \end{split}
   \tag{VI15}
\end{equation}%
\setcounter{equation}{15}%
where $L = Na$ is the lattice length and $N$ is the number of unit cells in the lattice. We have also employed the zero-temperature limit of the Fermi-Dirac distribution, i.e., $f_{n,k} \to \Theta(\mu - \varepsilon_{n, k})$. The first term in the second line of Eq.~\eqref{eq:chi-jpjp} is zero when the chemical potential $\mu = 0$ is chosen to lie within the insulating gap. On the other hand, the zero-temperature expectation value of the kinetic operator is given by
\begin{equation}
      \langle \hat{\cal T} \rangle=\frac{1}{N} \sum_{n = \pm}\sum_{k \in {\rm FBZ}} \Theta(-\varepsilon_{n, k})\langle \psi_{n,k}| \hat{\cal T}|\psi_{n,k}\rangle=\frac{1}{N} \sum_{k \in {\rm FBZ}}\langle \psi_{-,k}| \hat{\cal T}|\psi_{-,k}\rangle~.
\end{equation}
The TRK sum rule (\ref{eq:TRK_sumrule}) at temperature $T = 0$ for the 1D EFK model can thus be written as:
\begin{equation}
    -\frac{2}{L}\sum_{k \in {\rm FBZ}}\frac{\bigl|\langle \psi_{+,k}| \hat{ j}_p |\psi_{-,k}\rangle\bigr|^ 2}{\varepsilon_{+,\,k} -\varepsilon_{-,\,k}} = \frac{a^ 2}{N\hbar^2} \sum_{k \in {\rm FBZ}}\langle \psi_{-,k}| \hat{\cal T}|\psi_{-,k}\rangle ~.
\end{equation}
We start by evaluating the relevant current operator matrix element:
\begin{equation}
    \begin{split}
    \langle \psi_{+,k}| \hat{ j}_p|\psi_{-,k}\rangle &= j_{p;\,+-}(k) = \sum_{\alpha,\,\beta}  \bigl[U^ \dagger_{+,k}\bigr]_{\alpha}\,j_{p;\,\alpha\beta}(k) \,\bigl[U_{-, k}\bigr]_\beta~.
    \end{split}
\end{equation}
The argument of the FBZ sum in Eq.~\eqref{eq:TRK_sumrule} is therefore
\begin{equation}
     \frac{\bigl|\langle \psi_{+,k}| \hat{ j}_p|\psi_{-,k}\rangle\bigr|^ 2}{ \varepsilon_{+,k} - \varepsilon_{-,k}}=\frac{2a^ 2\tilde t\,^ 2}{\hbar^ 2} \frac{ (\delta \cos ka - 2t)^ 2}{\varepsilon_k^{3}}~.
\end{equation}
We take the thermodynamic limit,
\begin{equation}
    \frac{1}{N} \sum_{k \in {\rm FBZ}} \to \frac{1}{2\pi} \int_{-\pi}^ {\pi} \mathrm{d}(ka) ~,
\end{equation}
and find 
\begin{equation}
    -\frac{\hbar^ 2}{a^ 2}\frac{2}{L}\sum_{k \in {\rm FBZ}}\frac{\bigl|\langle \psi_{+,k}| \hat{ j}_p|\psi_{-,k}\rangle\bigr|^ 2}{\varepsilon_{+,\,k} -\varepsilon_{-,\,k}}
    \to
    -\frac{2}{\pi} \int_{0}^ {\pi} \mathrm{d}(ka)\,\frac{ 2(\delta \cos ka - 2t)^ 2}{\tilde t\left(4 \sin^ 2 ka + (\delta - 2 t \cos ka)^ 2/\tilde t\,^ 2\right)^{3/2}}~,
\end{equation}
where the integration domain was reduced due to parity of the integrand. We proceed to explicitly evaluate the kinetic operator expectation value, i.e.
\begin{equation}
\begin{split}
    \langle \hat{\cal T} \rangle &
    = \frac{1}{N}\sum_{\alpha,\,\beta} \sum_{k \in \rm FBZ}\bigl[U^ \dagger_{-,k}\bigr]_{\alpha}\,{\cal T}_{\alpha\beta}(k) \,\bigl[U_{-, k}\bigr]_\beta
    \\&= \frac{1}{N} \sum_{k \in \rm FBZ} \frac{2[t \cos ka(\delta - 2 t \cos ka )- 2 \tilde t\,^ 2 \sin^ 2ka)]}{\tilde t \,\left(4 \sin^ 2 ka + (\delta - 2 t \cos ka)^ 2/\tilde t\,^ 2\right)^{1/2}}\\
    & \to \frac{2}{\pi} \int_{0}^ {\pi} \mathrm{d}(ka)  \, \frac{t \cos ka(\delta - 2 t \cos ka )- 2 \tilde t\,^ 2 \sin^ 2ka}{\tilde t \,\left(4 \sin^ 2 ka + (\delta - 2 t \cos ka)^ 2/\tilde t\,^ 2\right)^{1/2}}~.
    \end{split}
\end{equation}
The TRK sum rule 
\begin{equation}
    \chi_{j_p j_p}^ {\vphantom{\dagger}}  =\frac{a^2}{\hbar^2} \langle \hat{\cal T} \rangle 
\end{equation}
thus reduces to an identity between two integrals:
\begin{equation}
    \int_{0}^ {\pi} \mathrm{d}x\,\frac{ 2(\delta \cos x - 2t)^ 2}{\left(4 \sin^ 2 x + (\delta - 2 t \cos x)^ 2/\tilde t\,^ 2\right)^{3/2}}=\int_{0}^ {\pi} \mathrm{d} x \, \frac{2 \tilde t\,^ 2 \sin^ 2x-t \cos x(\delta - 2 t \cos x ) }{ \,\left(4 \sin^ 2 x
    + (\delta - 2 t \cos x)^ 2/\tilde t\,^ 2\right)^{1/2}}~.
\end{equation}
The correctness of the previous equality can be easily verified by evaluating the integral of the difference between the two integrand functions. Defining:
\begin{align}
    F(x) & \equiv \frac{ 2(\delta \cos x - 2t)^ 2}{\left(4 \sin^ 2 x + (\delta - 2 t \cos x)^ 2/\tilde t\,^ 2\right)^{3/2}}~,\\
    G(x)& \equiv \frac{2 \tilde t\,^ 2 \sin^ 2x-t \cos x(\delta - 2 t \cos x ) }{ \,\left(4 \sin^ 2 x
    + (\delta - 2 t \cos x)^ 2/\tilde t\,^ 2\right)^{1/2}}~,
\end{align}
we obtain
\begin{equation}
    \int_0^ {\pi} \mathrm{d}x\,\Bigl[F(x)- G(x)\Bigr] = \left.\frac{[2(t^ 2-\tilde t\,^ 2)\cos x-\delta t]}{\left[{\delta^ 2+2(t^ 2+\tilde t\,^ 2)-4\delta t \cos x+2(t^ 2-\tilde t\,^ 2)\cos 2x}\right]^ {1/2}} \,\sin x\,\right|_{0}^ \pi = 0~,
\end{equation}
thus proving the TRK sum rule for the noninteracting 1D EFK model.

\section{Isolation of the non-trivial geometric terms}\label{section7}

The Gaussian approximation introduced in Eq.~(11) of the main text allows us to decompose the ${\bm k}$-space hopping gradient $f_i(\bm k)$, effectively isolating non-trivial quantum geometric terms, as demonstrated in Eqs.~(12) and~(13) of the main text:
\begin{equation}
\label{sm-eq:m-decomposition}
    M_{ij}(\bm k) =  M^{\,\rm H}_{ij}(\bm k) + M^{\,\rm E}_{ij}(\bm k) +  M^{\,\rm E-g }_{ij}(\bm k) +  M^{\,\rm g }_{ij}(\bm k)~,
\end{equation}
where we introduced the following quantities:
\begin{align}
    \label{sm-eq:mh}
    M^{\,\rm H}_{ij}(\bm k) & \equiv \delta_{ij} \gamma h(\bm k)~,\\
    \label{sm-eq:me}
    M^{\,\rm E}_{ij}(\bm k)  &\equiv - \gamma^2 \sum_{n} \partial_i \partial_j E_n(\bm k) P_n(\bm k)~,\\
    \label{sm-eq:m-e-g}
    M^{\,\rm E-g}_{ij}(\bm k) &\equiv -\gamma^2 \sum_n \partial_i E_n(\bm k) \partial_j P_n(\bm k) + (i\leftrightarrow j)~,\\
    \label{sm-eq:m-g}
    M^{\,\rm g}_{ij}(\bm k) &\equiv - \gamma^2 \sum_{n} E_n(\bm k) \partial_i \partial_j P_n(\bm k)~.
\end{align}
Here, $M^ {\rm H}_{ij}(\bm k)$ is directly proportional to the Hamiltonian $h({\bm k})$ and $M^ {\rm E-g}_{ij}(\bm k)$ is a cross term, which depends on both energy dispersion and quantum geometry. Therefore, using Eqs.~\eqref{sm-eq:dI-explicit} and~\eqref{sm-eq:dII-explicit}, together with the decomposition $f_i(\bm k) =  f_i^{\rm E}(\bm k) + f_i^{\rm g}(\bm k)$ of the $\bm k$-space hopping gradient provided in the main text and the analogous decomposition for the Hessian tensor given in Eqs.~\eqref{sm-eq:m-decomposition}-\eqref{sm-eq:m-g}, we obtain the following explicit formulas for the geometric term~${\cal D}_{\rm g}(\bm q) \equiv {\cal D}_{\rm g}^ {\rm (1)}(\bm q) + {\cal D}_{\rm g}^ {\rm (2)}(\bm q)$:
\begin{equation}
\label{sm-eq:dgI-explicit}
    \begin{split}
         \bigl[{\cal D}^{\rm (1)}_{\rm g}(\bm q)\bigr]_{\nu i}^ {\nu'j} = \frac{1}{\sqrt{M_{\nu} M_{\nu'}}}\, \frac{2}{N} \sum_{n}^{\rm occ.} \sum_{n'}^{\rm unocc.} \sum_{\bm k}^{\text{FBZ}} &
            \left\{\frac{\bigl[F_{i}^{}(n\bm k, n'\bm k+\bm q)\bigr]_{\nu} \bigl[F_{j}^{}(n'\bm k+\bm q, n\bm k)\bigr]_{\nu'}}{E_{n}(\bm k) - E_{n'}(\bm k+\bm q)}\right.+\\
            & \left.- \frac{\bigl[F_{i}^{\rm E}(n\bm k, n'\bm k+\bm q)\bigr]_{\nu} \bigl[F_{j}^{\rm E}(n'\bm k+\bm q, n\bm k)\bigr]_{\nu'}}{E_{n}(\bm k) - E_{n'}(\bm k+\bm q)}  \right\} + {\rm H.c.}~,
    \end{split}
\end{equation}
\begin{equation}
\label{sm-eq:dgII-explicit}
    \begin{split}
         \bigl[{\cal D}^{\rm (2)}_{\rm g}(\bm q)\bigr]_{\nu i}^ {\nu'j}
         =\frac{1}{\sqrt{M_\nu M_{\nu'}}}\,\frac{2}{N} &\sum_{\Tilde{\nu}} \sum_{n}^{\rm occ.} \sum_{\bm k \Tilde{\bm k}}^{\text{FBZ}} \sum_{\alpha\alpha'}\left[\delta_{\nu\nu'}\delta_{\Tilde{\bm k},\,\bm 0} - \delta_{\Tilde{\nu} \nu'} \delta_{\Tilde{\bm k},\,\bm q}\right] \times \\
         & \times\left\{  \Bigl[\bigl[M^{\rm g}_{ij}(\bm k + \Tilde{\bm k})\bigr]_{\nu \alpha}^ {\Tilde{\nu}\alpha'}  +\bigl[M^ {\rm E-g}_{ij}(\bm k + \Tilde{\bm k})\bigr]_{\nu \alpha}^ {\Tilde{\nu}\alpha'} \Bigr]\bigl[P_n(\bm k)\bigr]_{\Tilde{\nu}\alpha'}^{\nu\alpha}
           \right\} +\rm{H.c.}  ~,
    \end{split}
\end{equation}
where we have introduced the following quantities: 
\begin{align}
\label{sm-eq:ftensor-E}
    \bigl[F^{\rm E}_i(n\bm k, n'\bm k')\bigr]_{\nu} & =  \sum_{\Tilde{\nu}} \sum_{\nu_1 \nu_2 } \sum_{\alpha\alpha'}\bigl[U^\dagger_{n}(\bm k)\bigr]_{\nu_1\alpha}\biggl\{ \delta_{\nu\nu_1} \delta_{\Tilde{\nu}\nu_2} \bigl[ f^{\rm E}_i(\bm k')\bigr]_{\nu\alpha}^ {\Tilde{\nu}\alpha'}- \delta_{\nu \nu_2} \delta_{\Tilde{\nu}\nu_1} \bigl[f^{\rm E}_i(\bm k)]_{\Tilde{\nu}\alpha}^ {\nu\alpha'} \biggr\}\bigl[U^{\vphantom{\dagger}}_{n'}(\bm k')\bigr]_{\nu_2\alpha'}~,\\
\label{sm-eq:ftensor-g}
    \bigl[F^{\rm g}_i(n\bm k, n'\bm k')\bigr]_{\nu} & =  \sum_{\Tilde{\nu}} \sum_{\nu_1 \nu_2 } \sum_{\alpha\alpha'}\bigl[U^\dagger_{n}(\bm k)\bigr]_{\nu_1\alpha}\biggl\{ \delta_{\nu\nu_1} \delta_{\Tilde{\nu}\nu_2} \bigl[ f^{\rm g}_i(\bm k')\bigr]_{\nu\alpha}^ {\Tilde{\nu}\alpha'}- \delta_{\nu \nu_2} \delta_{\Tilde{\nu}\nu_1} \bigl[f^{\rm g}_i(\bm k)]_{\Tilde{\nu}\alpha}^ {\nu\alpha'} \biggr\}\bigl[U^{\vphantom{\dagger}}_{n'}(\bm k')\bigr]_{\nu_2\alpha'}~,\\
\label{sm-ftensor}
    \bigl[F^{}_i(n\bm k, n'\bm k')\bigr]_{\nu} & = \bigl[F^{\rm E}_i(n\bm k, n'\bm k')\bigr]_{\nu} + \bigl[F^{\rm g}_i(n\bm k, n'\bm k')\bigr]_{\nu}
\end{align}
Notice here the hermitian conjugate is taking the conjugate and transposing the $\nu$ index.

We now show that also ${\cal D}_{\rm g}(\bm q)$ alone satisfies the acoustic sum rule. In fact, the sum rule can be shown to hold separately for ${\cal D}_{\rm g}^{(1)}$ and ${\cal D}_{\rm g}^{(2)}$, as a consequence of continuous translational invariance. First we have, starting from formula~\eqref{sm-eq:dgI-explicit} given below:
\begin{equation}
\label{sm-eq:sumrule-proof-1}
    \begin{split}
         \sum_{\nu'}\sqrt{M_{\nu'}}\bigl[{\cal D}^{\rm (1)}_{\rm g}(\Gamma)\bigr]_{\nu i}^ {\nu'j} = \frac{1}{\sqrt{M_{\nu}}}\, \frac{2}{N} \sum_{n}^{\rm occ.} \sum_{n'}^{\rm unocc.} \sum_{\bm k}^{\text{FBZ}} &
            \left\{\frac{\bigl[F_{i}^{}(n\bm k, n'\bm k)\bigr]_{\nu}\sum_{\nu'} \bigl[F_{j}^{}(n'\bm k, n\bm k)\bigr]_{\nu'}}{E_{n}(\bm k) - E_{n'}(\bm k)}\right.+\\
            & \left.-\frac{\bigl[F_{i}^{\rm E}(n\bm k, n'\bm k)\bigr]_{\nu} \sum_{\nu'}\bigl[F_{j}^{\rm E}(n'\bm k, n\bm k)\bigr]_{\nu'}}{E_{n}(\bm k) - E_{n'}(\bm k)} \right\}  + {\rm H.c.} ~.
    \end{split}
\end{equation}
Using expression~\eqref{sm-eq:ftensor-g}, we get
\begin{equation}
    \begin{split}
        \sum_{\nu}\bigl[F^{}_i(n\bm k, n'\bm k)\bigr]_{\nu} & =  \sum_{\nu \Tilde{\nu}} \sum_{\nu_1 \nu_2 } \sum_{\alpha\alpha'}\bigl[U^\dagger_{n}(\bm k)\bigr]_{\nu_1\alpha}\biggl\{ \delta_{\nu\nu_1} \delta_{\Tilde{\nu}\nu_2} \bigl[ f^{}_i(\bm k)\bigr]_{\nu\alpha}^ {\Tilde{\nu}\alpha'}- \delta_{\nu \nu_2} \delta_{\Tilde{\nu}\nu_1} \bigl[f^{}_i(\bm k)]_{\Tilde{\nu}\alpha}^ {\nu\alpha'} \biggr\}\bigl[U^{\vphantom{\dagger}}_{n'}(\bm k)\bigr]_{\nu_2\alpha'} \\
        & =  \sum_{\nu \Tilde{\nu}}  \sum_{\alpha\alpha'}\biggl\{ \bigl[U^\dagger_{n}(\bm k)\bigr]_{\nu\alpha}\bigl[ f^{}_i(\bm k)\bigr]_{\nu\alpha}^ {\Tilde{\nu}\alpha'}\bigl[U^{\vphantom{\dagger}}_{n'}(\bm k)\bigr]_{\Tilde\nu\alpha'}- \bigl[U^\dagger_{n}(\bm k)\bigr]_{\Tilde \nu\alpha}\bigl[f^{}_i(\bm k)]_{\Tilde{\nu}\alpha}^ {\nu\alpha'} \bigl[U^{\vphantom{\dagger}}_{n'}(\bm k)\bigr]_{\nu\alpha'}\biggr\}\\
        &= 0~,
    \end{split}
\end{equation}
and identically we can show that $\sum_{\nu}\bigl[F^{\rm E}_i(n\bm k, n'\bm k)\bigr]_{\nu} = 0$, thus proving $\sum_{\nu'}\bigl[{\cal D}^{\rm (1)}_{\rm g}(\bm 0)\bigr]_{\nu i}^ {\nu'j} = 0$. 

For the second-order term, the calculation is straightforward:
\begin{equation}
\label{sm-eq:sumrule-proof-2}
    \begin{split}
         \sum_{\nu'}\sqrt{M_{\nu'}}\bigl[{\cal D}^{\rm (2)}_{\rm g}(\Gamma)\bigr]_{\nu i}^ {\nu'j} & =
         \frac{1}{\sqrt{M_\nu}}\,\frac{2}{N} \sum_{\nu'\Tilde{\nu}} \sum_{n}^{\rm occ.} \sum_{\bm k \Tilde{\bm k}}^{\text{FBZ}} \sum_{\alpha\alpha'}\left[\delta_{\nu\nu'}\delta_{\Tilde{\bm k},\,\bm 0} - \delta_{\Tilde{\nu} \nu'} \delta_{\Tilde{\bm k},\,\bm 0}\right] \times \\
         & \qquad\qquad\times\left\{  \Bigl[\bigl[M^{\rm g}_{ij}(\bm k + \Tilde{\bm k})\bigr]_{\nu \alpha}^ {\Tilde{\nu}\alpha'}  +\bigl[M^ {\rm E-g}_{ij}(\bm k + \Tilde{\bm k})\bigr]_{\nu \alpha}^ {\Tilde{\nu}\alpha'} \Bigr]\bigl[P_n(\bm k)\bigr]_{\Tilde{\nu}\alpha'}^{\nu\alpha}
            \right\}  +\rm{H.c.}\\
        & =
         \frac{1}{\sqrt{M_\nu}}\,\frac{2}{N} \sum_{\nu'\Tilde{\nu}} \sum_{n}^{\rm occ.} \sum_{\bm k }^{\text{FBZ}} \sum_{\alpha\alpha'}\left[\delta_{\nu\nu'} - \delta_{\Tilde{\nu} \nu'} \right] \times \\
         & \qquad\qquad\times\left\{  \Bigl[\bigl[M^{\rm g}_{ij}(\bm k)\bigr]_{\nu \alpha}^ {\Tilde{\nu}\alpha'}  +\bigl[M^ {\rm E-g}_{ij}(\bm k)\bigr]_{\nu \alpha}^ {\Tilde{\nu}\alpha'} \Bigr]\bigl[P_n(\bm k)\bigr]_{\Tilde{\nu}\alpha'}^{\nu\alpha}
            \right\}  +\rm{H.c.}~.
    \end{split}
\end{equation}
Since the quantity inside the braces does not depend on $\nu'$, we immediately get $\sum_{\nu '}\left[\delta_{\nu\nu'} - \delta_{\Tilde{\nu} \nu'} \right] = 1-1 = 0$, thus proving the sum rule $\sum_{\nu'}\sqrt{M_{\nu'}}\bigl[{\cal D}^{\rm (2)}_{\rm g}(\Gamma)\bigr]_{\nu i}^ {\nu'j}$. Therefore, the entire geometric contribution satisfies
\begin{equation}
    \sum_{\nu'}\sqrt{M_{\nu'}}\bigl[{\cal D}_{\rm g}(\bm q =\Gamma )\bigr]_{\nu i}^ {\nu'j} = 0~,
\end{equation}
thus enforcing the presence of two massless acoustic modes for both ${\cal D}_{\rm g}(\bm q)$ and ${\cal D}_{\rm ng}(\bm q)$. Indeed, the frequencies of the two lowest-lying eigenmodes of ${\cal D}_{\rm ng}(\bm q)$ will be vanishing in the long-wavelength limit, i.e.
\begin{equation}
    \lim_{\bm q \to \Gamma}\tilde{\omega}_{\rm LA/TA}(\bm q) = 0~.
\end{equation}

\section{Further details on the case of graphene}\label{section8}

In order to obtain a complete description of electrons and phonons in graphene, as well as extract the relevant parameters of the hopping function within the Gaussian approximation, we have carried out \textit{ab initio} density functional theory (DFT) and density-functional perturbation theory (DFPT)~\cite{DFPT_Baroni_SM} calculations using Quantum ESPRESSO (QE)~\cite{QE1_SM, QE2_SM}. 

The necessary pseudopotentials were taken from the standard solid-state pseudopotential (SSSP) accuracy library~\cite{prandini2018precision_SM, DalCorso_SM}. The exchange-correlation potential was treated in the Generalized Gradient Approximation (GGA), as parametrized by the Perdew-Burke-Ernzerhof (PBE) formula~\cite{PBE_SM}, with the vdW-D2 correction proposed by Grimme~\cite{Grimme2006_SM}. For integrations over the FBZ, we employed a Methfessel-Paxton smearing function~\cite{MP_smearing_SM} of $10^{-2}~{\rm Ryd}$. A dense Monkhorst-Pack (MP)~\cite{MP_SM} ${\bm k}$-point grid with $ 96\times96 \times 1$ points is chosen to sample for self-consistent calculations of the charge density. The equilibrium lattice parameter of graphene is $a=2.467~{\rm \AA}$. We considered a simulation cell with approximately $22~{\rm \AA}$ of vacuum between periodic images along the $c$-direction. We also introduced a 2D Coulomb cutoff for a better description of phonons at small wave vectors~\cite{Sohier2017_SM}.

The next-nearest-neighbor (NNN) Born-von Karman model for in-plane phonon dispersion has four force constants $\alpha, \beta, \gamma, \delta$ as free parameters. 
These parameters are optimized via least absolute deviations fitting to DFPT results.

\begin{figure}[t]
    \begin{tabular}{cc}
         \begin{overpic}[width=0.5\columnwidth]{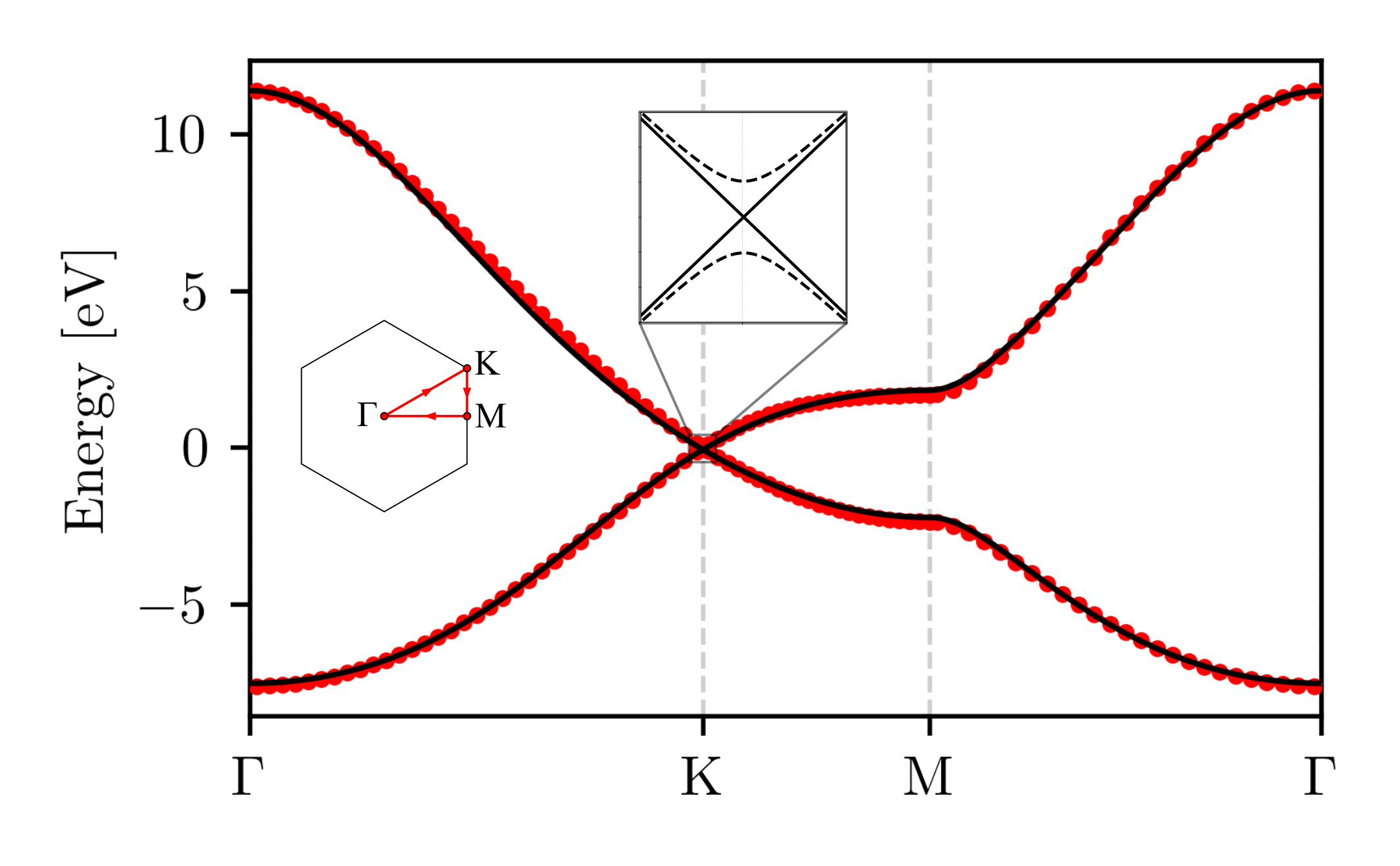}
             \put(10,140){\rm (a)}
         \end{overpic}&   
         \begin{overpic}[width=0.5\columnwidth]{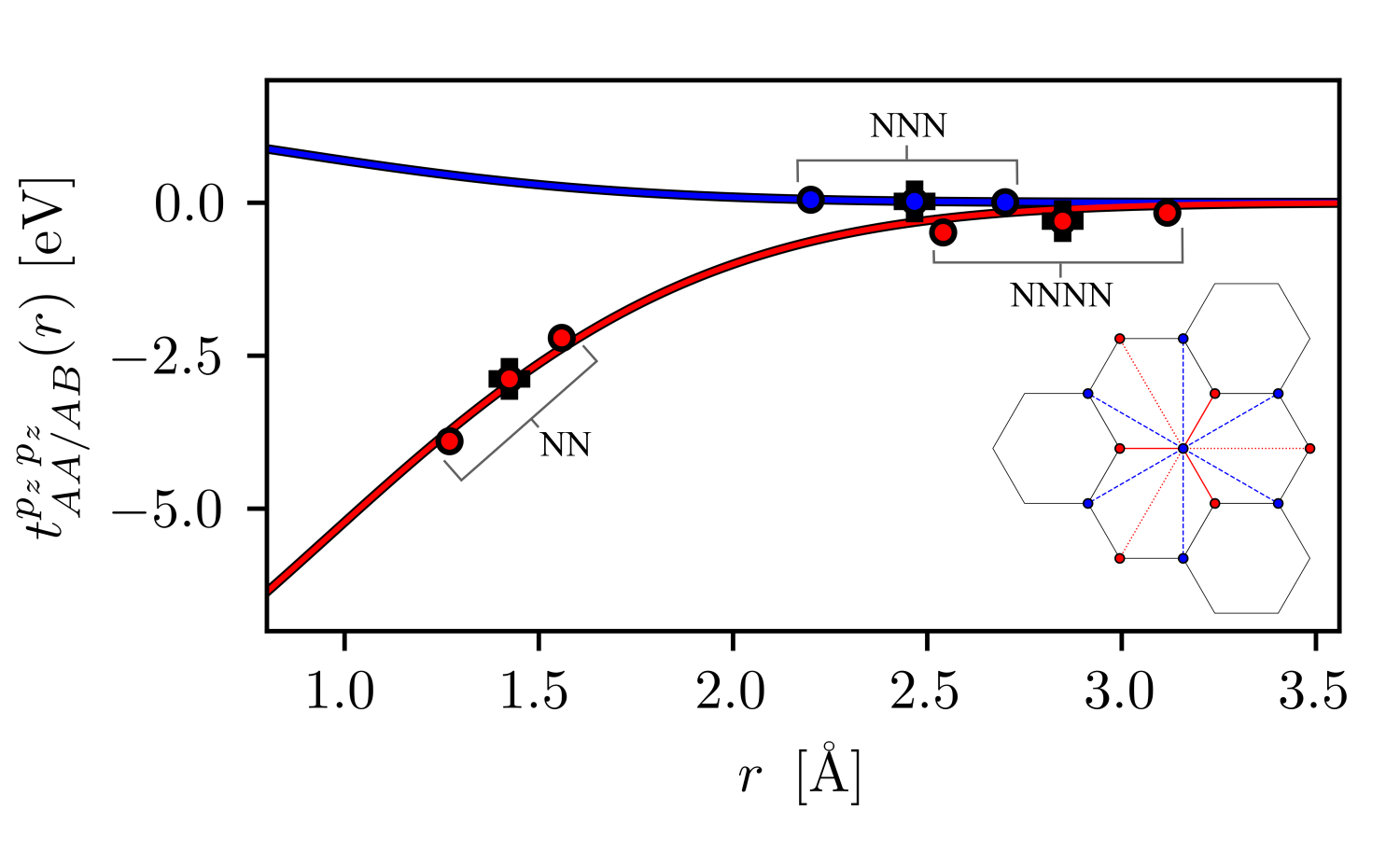}
             \put(10,140){\rm (b)}
         \end{overpic}\\
         \begin{overpic}[width=0.5\columnwidth]{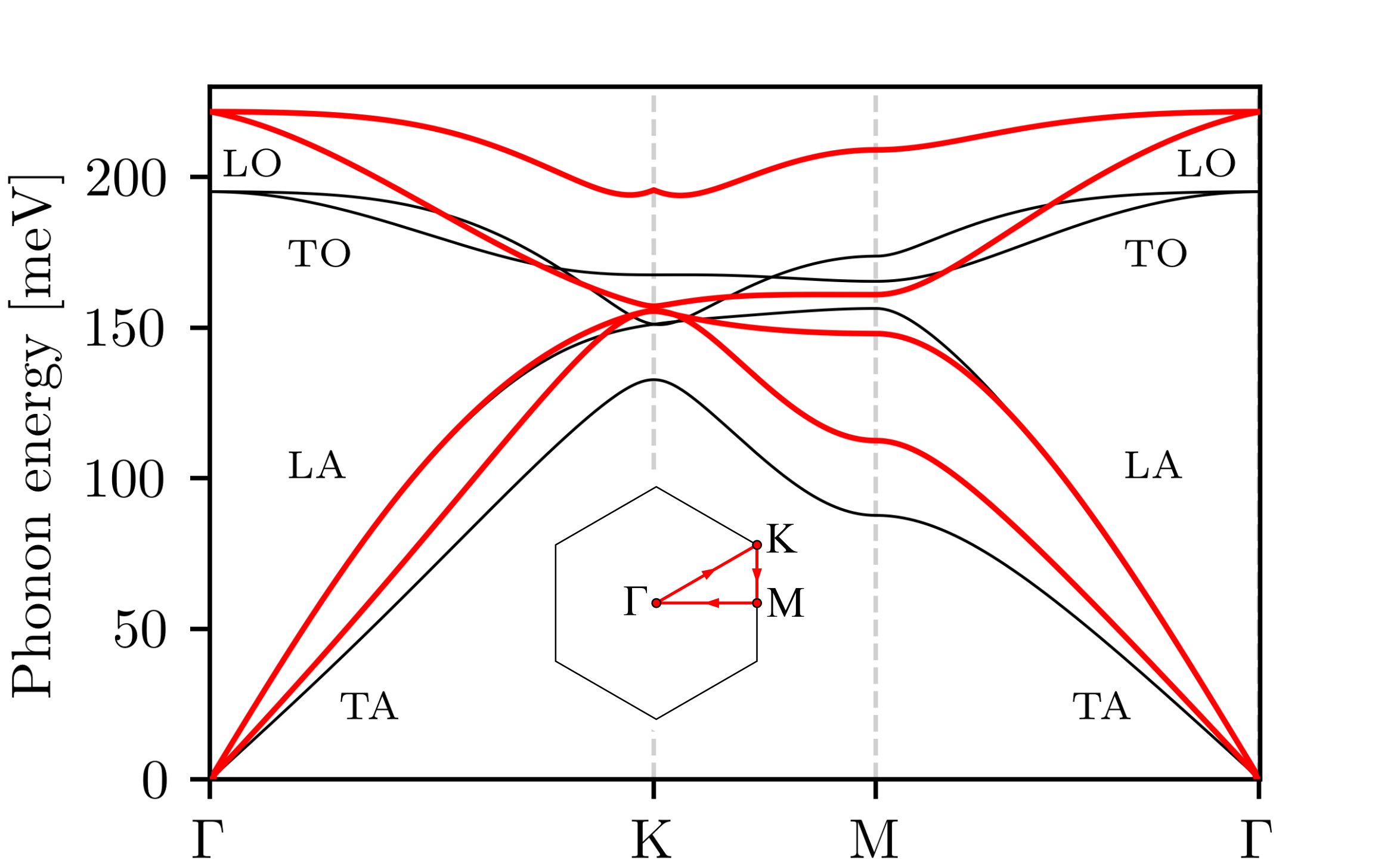}
             \put(10,160){\rm (c)}
         \end{overpic}&   
         \begin{overpic}[width=0.5\columnwidth]{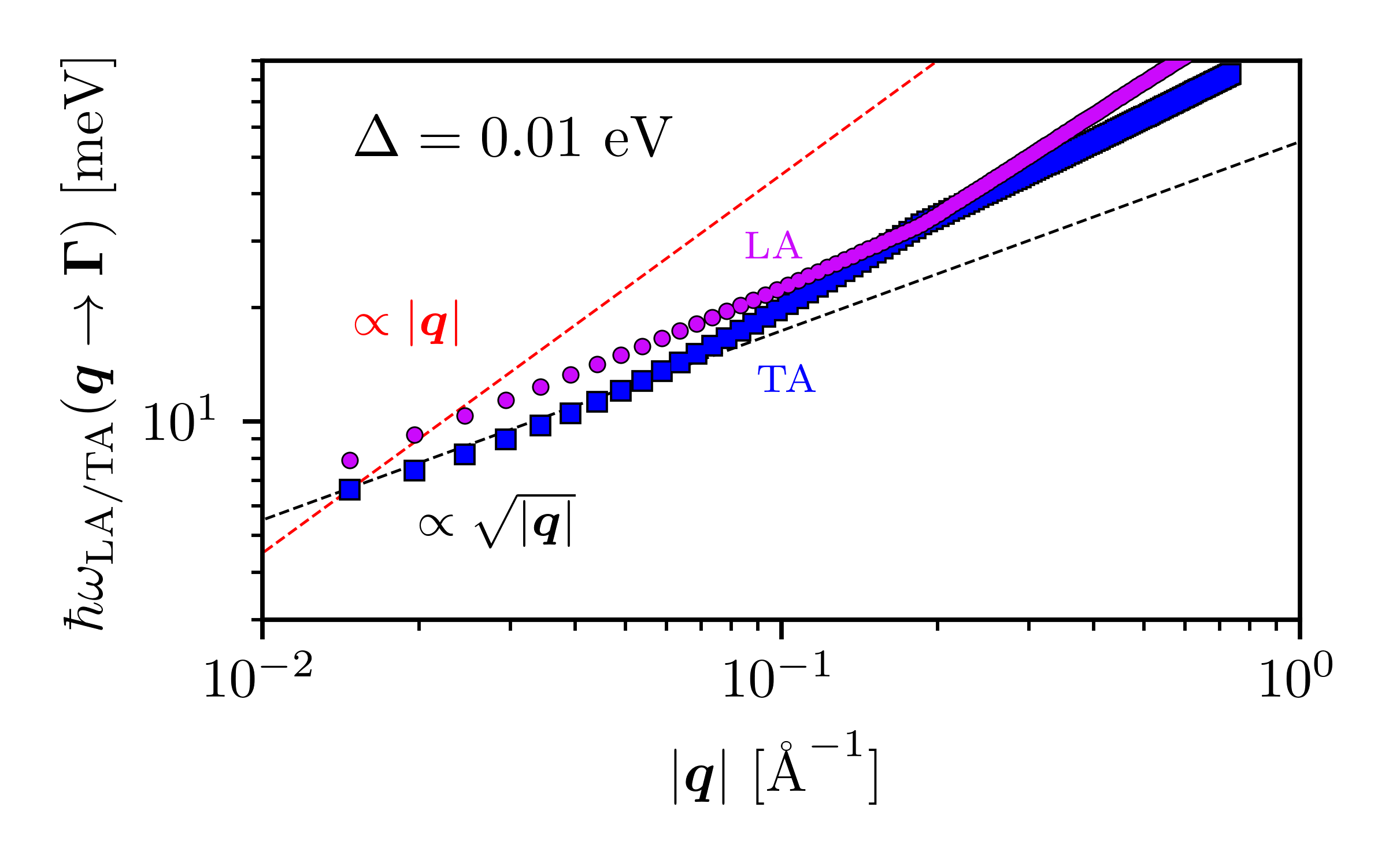}
             \put(10,160){\rm (d)}
        \end{overpic}    
    \end{tabular}
    \caption{(Color online) (a) Plot of the $\pi$ bands of graphene along the high-symmetry $\Gamma{\rm KM}\Gamma$ path in the first Brillouin zone. Red dots: Results of the {\it ab initio} DFT calculations. Black line: best fit based on our analytical overlap-inclusive NNNN model. Inset shows the low-energy electron dispersion near the Dirac point for $\Delta = 0$ (solid black line) and $\Delta = 10$ meV (dotted black line), corresponding respectively to massless and massive Dirac fermions. (b) Hopping integrals generating the graphene's $\pi$ bands, obtained via the strained-lattice method. Red circles: inter-sublattice ($AB$) hopping integrals. Blue dots: intra-sublattice ($AA$) hopping integrals. Both are obtained as optimal parameters from a least-square best-fit procedure on the \textit{ab initio} bands of strained lattices. Black crosses: hopping integrals in a relaxed graphene lattice. Red (blue) lines: Gaussian best fit to the inter- (intra-) sublattice hopping integral as a function of the inter-atomic distance $r$. The inset shows the hierarchy of nearest neighbors in a honeycomb lattice. (c) Black lines: Phonon dispersions in graphene along $\Gamma$KM$\Gamma$. Red lines: {eigenvalues of ${\cal D}(\bm q) - {\cal D}^ {{\rm (el},\, \pi )}(\bm q)$, where ${\cal D}^ {{\rm (el},\, \pi )}(\bm q)$ is the electronic contribution of the $\pi$ bands. The acoustic modes still behave linearly in $|\bm q|$ for $\bm q \to \bm \Gamma$ after the removal of the entire electronic contribution of the $\pi$ bands to the dynamical matrix, thus proving the non-analytic behavior of $\tilde{\omega}_{\rm LA/TA}(\bm q \to \bm \Gamma)$ to be of purely geometric nature.} (d) Behavior of the acoustic modes near $\Gamma$ for $|\bm q|\gtrsim q_{\rm thr}$, as illustrated in Fig.~2 of the main text; here, $\bm q$ is on the M$\to\Gamma$ portion of the high-symmetry contour.}
    \label{sm-fig:1}
\end{figure}

\begin{figure*}[t]
\centering
\begin{tabular}{cc}
\begin{overpic}[width=0.5\columnwidth]{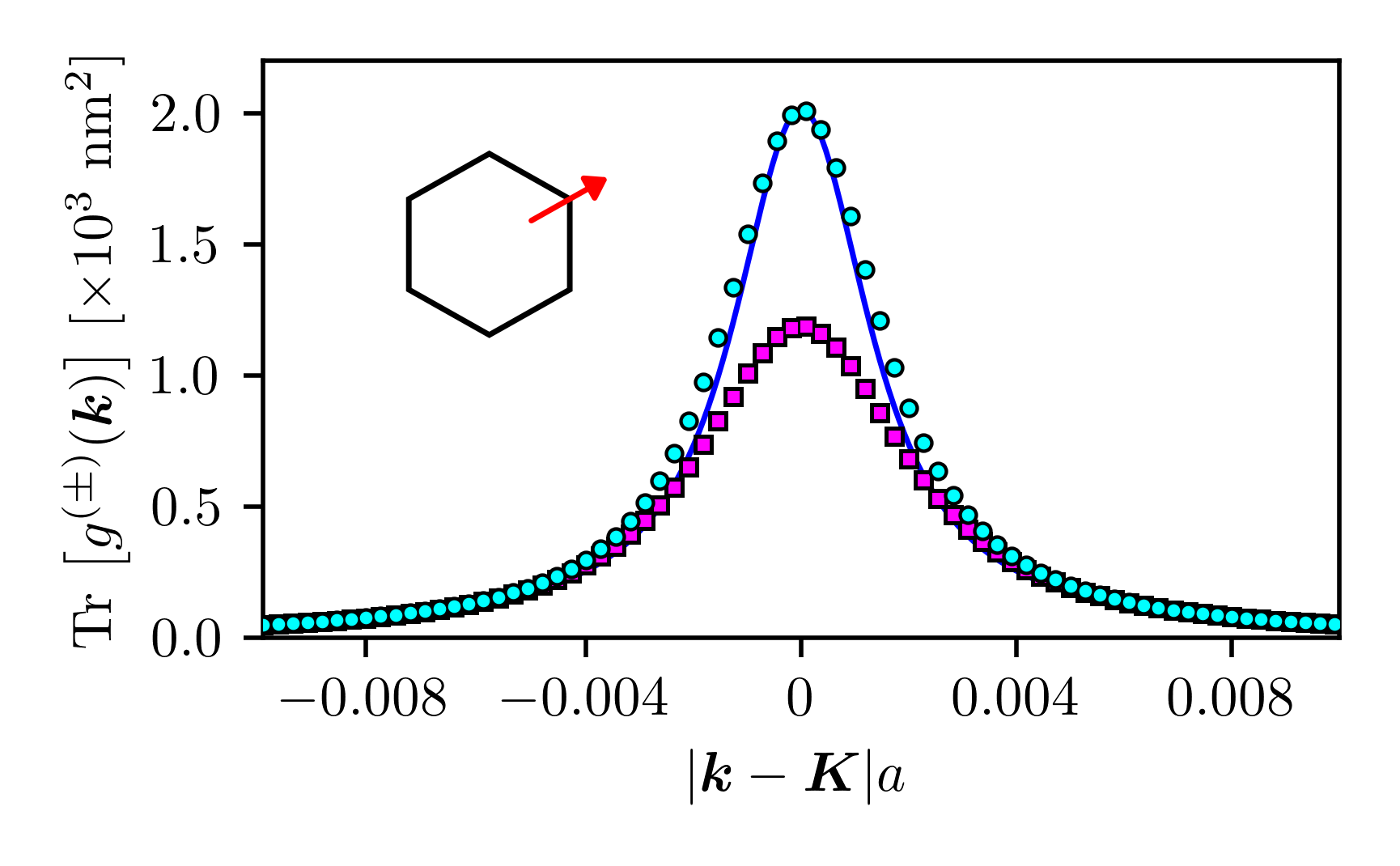}
\put(-4,147){\rm{(a)}}
\end{overpic} &
\begin{overpic}[width=0.5\columnwidth]{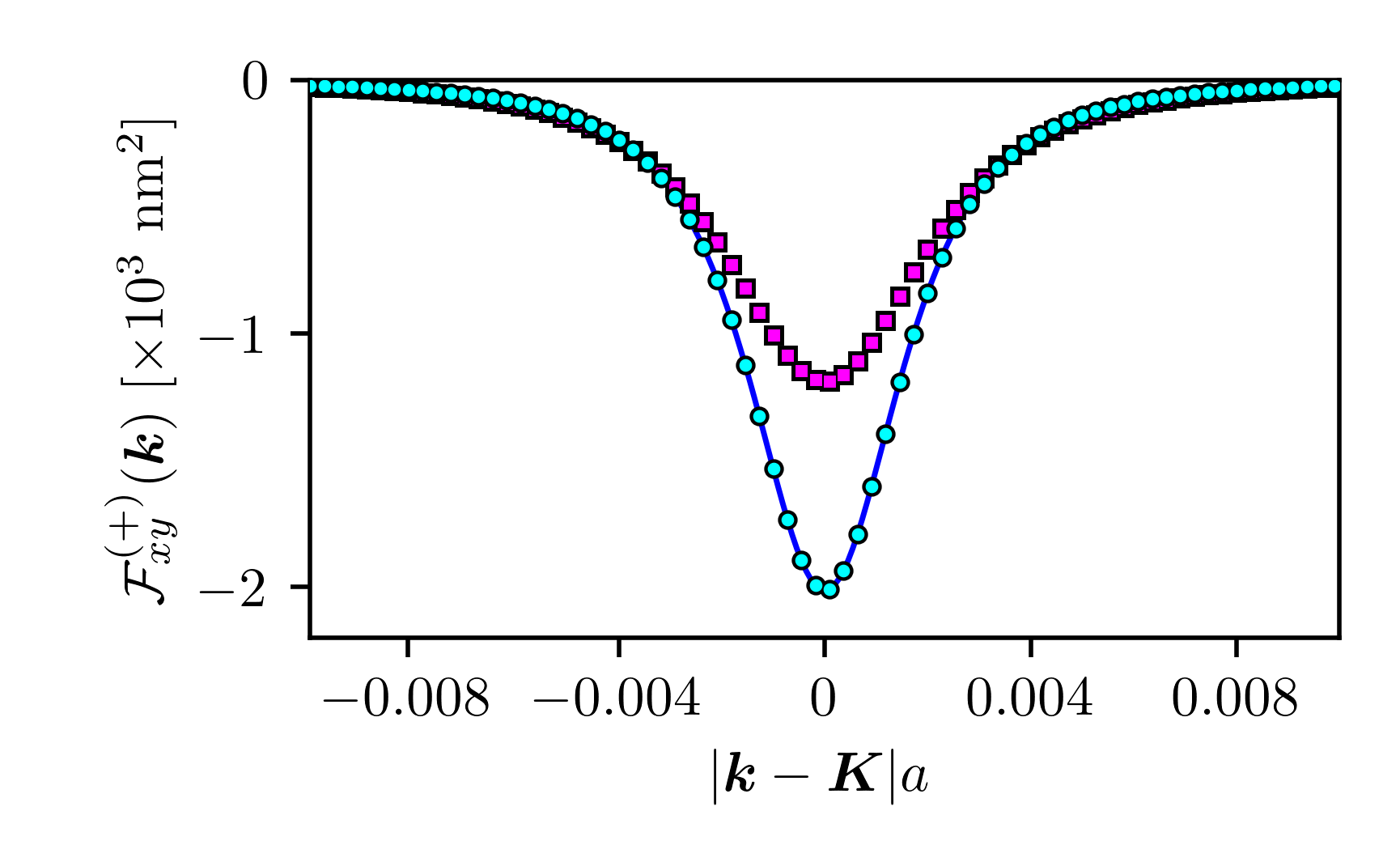}
\put(7,147){\rm{(b)}}
\end{overpic}
\\
\begin{overpic}[width=0.5\columnwidth]{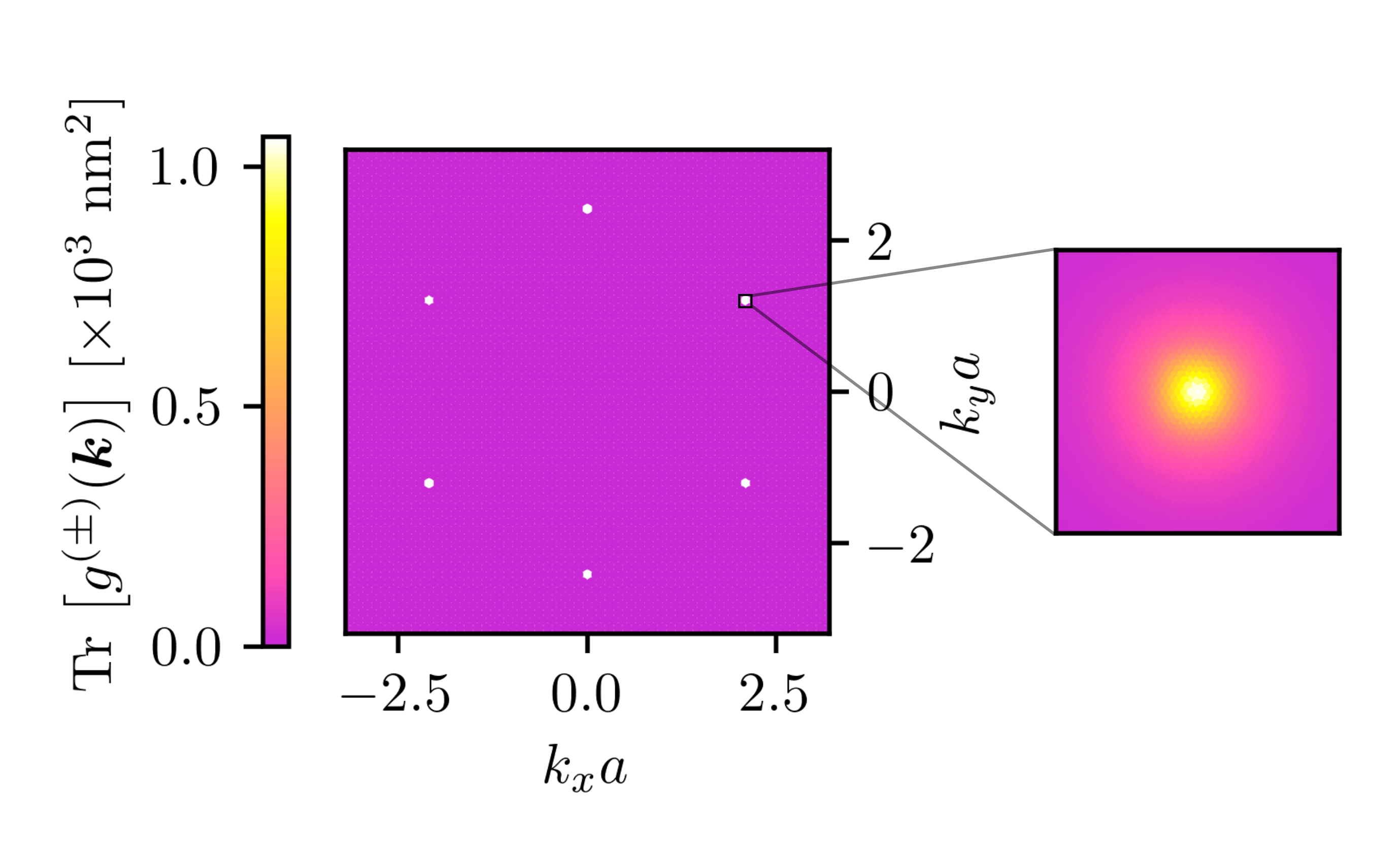}
\put(-4,140){\rm{(c)}}
\put(120, 100){$\bm{K}$}
\put(120, 57){$\bm{K ^ {\prime}}$}
\end{overpic} &
\begin{overpic}[width=0.5\columnwidth]{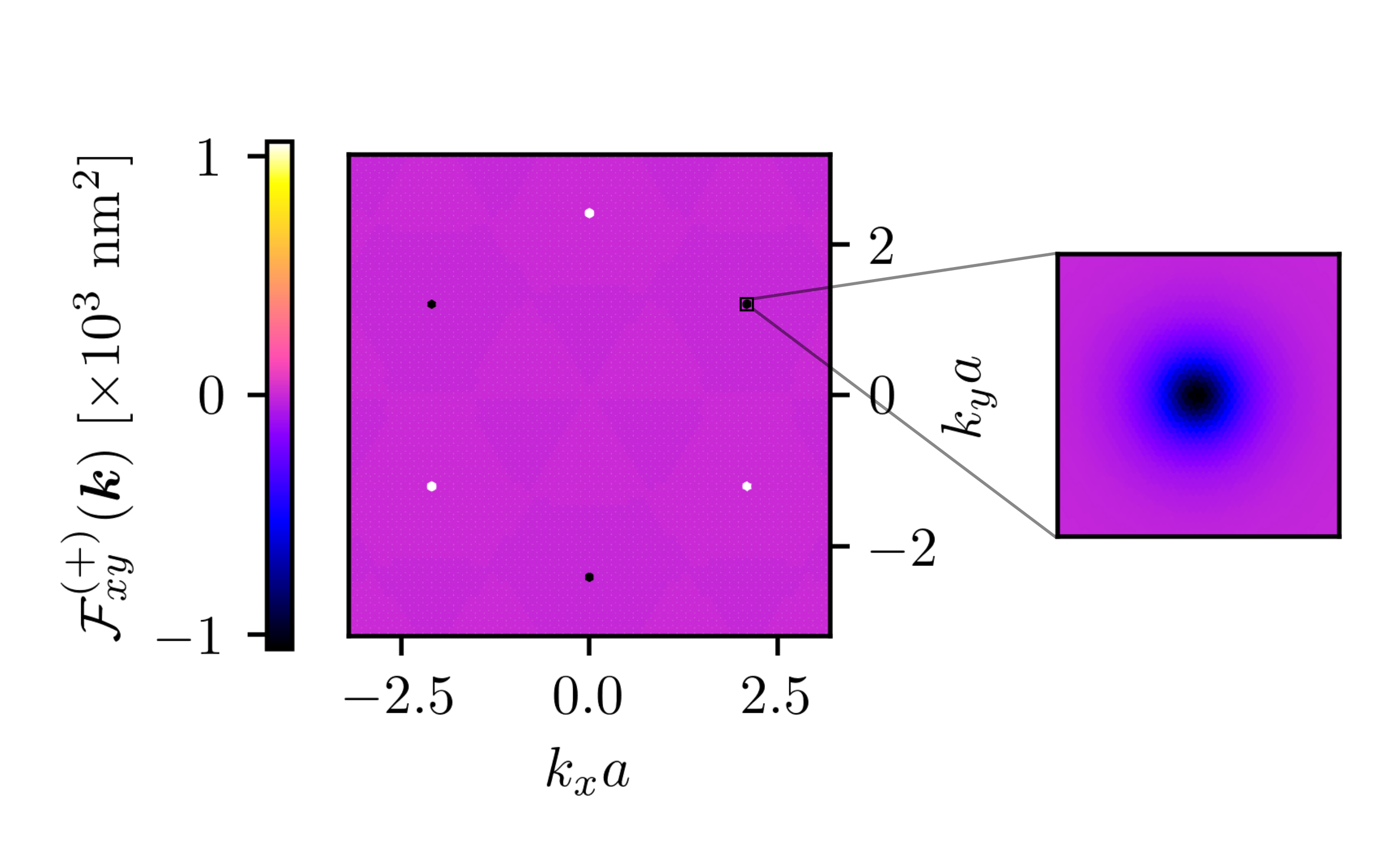}
\put(7,140){\rm{(d)}}
\put(120, 100){$\bm{K}$}
\put(120, 57){$\bm{K ^ {\prime}}$}
\end{overpic}
\end{tabular}
\caption{(Color online) Regularized QGT for gapped graphene. For better illustration, results in this figure have been calculated with a large gap of $\Delta = 20~{\rm meV}$. (a) Trace of the Fubini-Study metric $g^{(\pm)}_{ij}(\bm k)$--- which is the same for both conduction and valence bands due to the quasi-exact particle-hole symmetry in the vicinity of the $\bm K$ point---plotted along the path in the FBZ shown as a red arrow in the inset. Cyan circles: numerical results obtained within the NN tight-binding model. Magenta squares: numerical results obtained within the NNNN model. Solid blue line: analytical results for massive Dirac fermions. (b) The only non-zero component of the conduction-band Berry curvature, ${\cal F}_{xy}^ {(+)}(\bm k) \equiv -2\Im {\cal Q}_{xy}^ {(+)}(\bm k) = {\cal F}_{yx}^ {(-)}(\bm k)$, plotted along the same path as in panel (a). The valence-band curvature is identical apart from a sign. Legend is the same as in panel (a). (c)-(d) 2D color plots of $g^{(\pm)}_{ij}(\bm k)$ and ${\cal F}_{xy}^ {(+)}(\bm k)$ calculated within the NNNN model as functions of $\bm k$. The insets provide a zoom-in view of the $\bm K$ point, where the singularity discussed in the text has been cured.}
\label{sm-fig:2}
\end{figure*}

The electronic dispersion in graphene is described by a next-next-nearest-neighbor (NNNN) tight-binding model~\cite{Reich2002_SM}, which depends on seven parameters: three hopping integrals $t^ {(i)}$, three overlap integrals $s^ {(i)}$ with $i = 1,2,3$, and an on-site energy $\varepsilon_{p_z}$. The latter parameter can be set equal to zero due to gauge freedom. The remaining six physical parameters are determined via a least-squares procedure applied to the results of our DFT calculations. The results of this procedure are reported in Fig.~\ref{sm-fig:1}(a). Fixing the six above-mentioned parameters allows us to reconstruct the Bloch Hamiltonian $h(\bm k)$ on the entire FBZ, thus obtaining the Bloch eigenvectors $U_n(\bm k)$ and the electron bands $E_n(\bm k)$ with $\bm k \in \rm FBZ$. The reconstructed Hamiltonian is then regularized following the standard prescription~\cite{Katsnelson2012_SM}
\begin{equation}
    \label{sm-eq:regularized-hamiltonian}
    h(\bm k) \to h(\bm k) + 
    \begin{pmatrix}
        \Delta/2 & 0 \\
        0 & -\Delta/2
    \end{pmatrix}~,
\end{equation}
fixing a gap $\Delta$ at the Dirac point. The effect of the gap size is discussed in the main text. The regularized Fubini-Study metric and Berry curvature are shown in Fig.~\ref{sm-fig:2}(a)-(d). We note that the overlap matrix is close to unity, and the parameters $s^{(i)}$ can thus be set to zero without significantly affecting the numerical results.

The hopping functions in graphene are modeled within the Gaussian approximation (Eq.~(11) in the main text):
\begin{equation}
    \label{sm-eq:ga}
    t^ {p_z p_z}_{AB}(r) = t^ {p_z p_z}_{AB}(0) \exp \left(\frac{1}{2} \gamma^ {\vphantom{*}}_{AB} r^ 2\right)~, \qquad 
    t^ {p_z p_z}_{AA}(r) = t^ {p_z p_z}_{AA}(0) \exp \left(\frac{1}{2} \gamma^ {\vphantom{*}}_{AA} r^ 2\right)~.
\end{equation}
Clearly, NN- and NNNN-type hopping integrals are parametrized by the first of the two equations above, while NNN-type hopping integrals are parametrized by the second one. The spatial dependence of the hopping integrals is determined as follows: after calculating the hopping integrals $t^ {(1,2,3)}$ for a relaxed graphene lattice with lattice constant $a$, starting from the band structure as discussed above, the procedure is repeated by calculating electron bands on strained graphene lattices, i.e.~graphene lattices whose lattice spacing $a^*$ is imposed to be $a^ * \neq a$. This approach allows us to extract sufficient information about the spatial dependence of $t^ {(1,2,3)}(r)$ to perform a Gaussian interpolation, as shown in Fig.~\ref{sm-fig:1}(b). From this interpolation, the relevant parameters are estimated to be $\gamma_{AA}^{\vphantom{*}} = -1.37$~\AA$^ {-2}$ and $\gamma_{AB}^{\vphantom{*}} = -1.10$~\AA$^ {-2}$. In calculating the numeric result, for simplicity, a single effective parameter $\gamma_{\rm eff} = -1.18 $~\AA$^ {-2}$ was employed to parametrize both $t_{AA}(\bm r)$ and $t_{AB}(\bm r)$, changing the sign of the amplitude accordingly, with 
$t^ {p_z p_z}_{AB}(0) = -t^ {p_z p_z}_{AA}(0)=-9.462$ eV. 

Summations over electron wave vectors $\bm k \in {\rm FBZ}$ appearing in Eqs.~\eqref{sm-eq:dI-explicit}-\eqref{sm-eq:dII-explicit} have been carried out on a $N = 600 \times 600$-dense partition of the FBZ. The evaluation of ${\cal D}_{\rm g}(\bm q)$ at fixed $\bm q$ was repeated for each phonon wave vector $\bm q \in \Gamma {\rm KM} \Gamma$. 

Results for the phonon dispersion lines deprived of the electronic contribution (i.e. the eigenvalues of ${\cal D}(\bm q) - {\cal D}^{(\rm el)}(\bm q)$) are shown in Fig.~\ref{sm-fig:1}(c). The behavior of the phonon dispersion lines is still fully analytical and the dispersion near $\Gamma$ remains linear. Moreover, as shown by comparing Fig.~2(b) in the main text with Fig.~\ref{sm-fig:1}(d), the behavior of the acoustic modes for $\bm q \to \bm \Gamma$ is invariant under rotations.


\section{Analytical results for gapless and gapped Dirac fermions }\label{section9}

In this section, we consider the contribution of a single Dirac node of the graphene to $\mathcal{D}_{\rm g}(\bm q)$ in the long-wavelength limit. To this end, we consider the following effective Hamiltonian, which characterizes a single Dirac node in graphene 
\begin{equation}
    h({\bm k}) = 
    v_{\rm F} \begin{pmatrix}
        0 & k_x-ik_y \\ 
        k_x + ik_y & 0 
    \end{pmatrix}
\end{equation}
We restrict our analysis to electrons near the Dirac node with momenta satisfying  $|{\bm k}| < \Lambda$, where $\Lambda$ is an ultraviolet cutoff. The eigenvalues and eigenfucntions of the Dirac Hamiltonian are given by
 \begin{align}
 &\epsilon_{\pm, {\bm k}} = \pm v_{\rm F} |{\bm k} |~,\nonumber\\ 
  &    U_{+}({\bm k}) = \frac{1}{\sqrt{2}|{\bm k}|}
    \begin{pmatrix}
        |{\bm k}| \\ 
        k_x+ik_y 
    \end{pmatrix},\quad \quad 
    U_{-}({\bm k}) = \frac{1}{\sqrt{2}|{\bm k}|}
    \begin{pmatrix}
        ik_y-k_x\\ 
        |{\bm k}|  
    \end{pmatrix} ~.
 \end{align}
We now want to focus on the behavior of the acoustic modes. To this end, we project the force constant matrix onto the acoustic mode, which is characterized by the eigenvector  $[w^{\rm ac}]_\nu = \frac{1}{\sqrt{2}}(\delta_{\nu A}+ \delta_{\nu B})$. Since there are only two atoms with the same mass within the unit cell of the graphene, the vector $w^{\rm ac}$ takes the same value for all its components. The dynamical matrix elements after projection onto the acoustic modes are given by 
\begin{align}
&[\bar{\cal D}_{\rm g}(\bm q)]_{i}^j=\sum_{\nu\nu'}[w^{\rm ac}]_{\nu}^ * [{\cal D}_{\rm g}(\bm q)]_{\nu i}^{\nu' j}
[w^{\rm ac}]_{\nu'}~ \nonumber\\ 
&[\bar{\cal D}^{\rm (el)}(\bm q)]_{i}^j=\sum_{\nu\nu'}[w^{\rm ac}]_{\nu}^ * [{\cal D^{({\rm el})}}(\bm q)]_{\nu i}^{\nu' j}
[w^{\rm ac}]_{\nu'} ~,
\end{align}
where $i,j \in \{x,y\}$. $[\bar{\cal D}_{\rm g}(\bm q)]$ and $[\bar{\cal D}^{\rm (el)}(\bm q)]$ denote the geometric contribution and total electronic contribution respectively. $[\bar{\cal D}_{\rm g}(\bm q)]$ and $[\bar{\cal D}^{\rm (el)}(\bm q)]$ can be calculated directly via Eq.~\eqref{sm-eq:dgI-explicit} and Eq.~\eqref{sm-eq:dynamical-matrix}.

We now aim to evaluate the behaviors of $\bar{\cal D}_{\rm g}(\bm q)$ and $\bar{\cal D}^{\rm (el,1)}(\bm q)$ in the long-wavelength limit $\bm q \to \Gamma$. We first show that the total electronic contribution vanishes: $\bar{\cal D}^{\rm (el)}(\bm q) =0 $.
As in Eqs.~$\eqref{sm-eq:dI-explicit}$ and~\eqref{sm-eq:dII-explicit}, we introduce 
\begin{align}
    [\bar{\cal D}^{\rm (el,1)}(\bm q)]_i^ j  = \sum_{\nu\nu'}[w^{\rm ac}]_{\nu}^ * [{\cal D}^{\rm (el,1)}(\bm q)]_{\nu i}^{\nu' j}
[w^{\rm ac}]_{\nu'} ,\quad [\bar{\cal D}^{\rm (el,2)}(\bm q)]_{i}^ {j}  = \sum_{\nu\nu'}[w^{\rm ac}]_{\nu}^ *[{\cal D}^{\rm (el,2)}(\bm q)]_{\nu i}^{\nu' j}
[w^{\rm ac}]_{\nu'} 
\end{align}
where $\bar{\cal D}^{\rm (el)}(\bm q) = \bar{\cal D}^{\rm (el,1)}(\bm q)+\bar{\cal D}^{\rm (el,2)}(\bm q)$. 
In a more compact form,
\begin{align}
\label{eq:dyn_acc_proj}
   & [\bar{\cal D}^{\rm (el,1)}(\bm q)]^j_i = \frac{1}{M_{\rm C}N} \sum_{n}^{\rm occ.} \sum_{n'}^{\rm unocc.} \sum_{\bm k}^{\text{FBZ}}\frac{ \text{Tr}
    \bigg[ P_n({\bm k}) \cdot [f_{i}({\bm k+\bm{q}})-f_{i}({\bm k})] \cdot P_{n'}({\bm k+\bm{q}})\cdot  [f_{j}({\bm k})-f_{j}({\bm k+\bm{q}})]  \bigg] }{E_n(\bm{k}) -E_{n'}(\bm{k+q})}  +\text{H.c.}~,\nonumber\\ 
   & [\bar{\cal D}^{\rm (el,2)}(\bm q)]^j_i = \frac{1}{M_{\rm C}N} \sum_{n}^{\rm occ.} \sum_{\bm k}^{\text{FBZ}}
    \text{Tr}
    \bigg[\bigg( M_{ij}(\bm{k})-M_{ij}(\bm{k+q})\bigg)  \cdot P_n({\bm{k}})\bigg]  +\text{H.c.}~,
\end{align}
where $\cdot$ denotes the matrix product. 
For the case of a single Dirac node, we observe that
\begin{align}
        &f_i(\bm{k}) = i\gamma \partial_{k_i}h({\bm k}) = i\gamma \hbar v_{\rm F} \sigma_i~, \nonumber\\ 
        &M_{ij}(\bm{k}) = (\gamma \delta_{ij} - \gamma^2 \partial^2_{k_i,k_j}) h({\bm k}) = \gamma \delta_{ij} h(\bm{k})~.
\end{align}
From $f_i(\bm{k}) = i\gamma \hbar v_{\rm F} \sigma_i $, we find 
\begin{align}
    f_{i}({\bm k+\bm{q}})-f_{i}({\bm k}) = 0 ~,
\end{align}
thus concluding that $[{\cal D}_{\rm el,1}^{\rm ac}(\bm q)]=0$. The total electronic contribution to the dynamical matrix, projected on the acoustic mode, is vanishing:
\begin{align}
   [\bar{\cal D}^{\rm (el)}(\bm q)]_j^i =   &
   \frac{1}{M_{\rm C}N} \sum_{n}^{\rm occ.}\frac{1}{\Omega}  \int_{ |{\bm k}|<\Lambda} \delta_{ij}\gamma 
    \text{Tr}
    \bigg[\bigg( h(\bm k) - h(\bm{k+q})\bigg)  \cdot P_n({\bm{k}})\bigg]  \nonumber\\ 
    =& \delta_{ij}\gamma \frac{v_{\rm F}}{M_{\rm C}N} \sum_{n}^{\rm occ.}\frac{1}{\Omega}  \int_{ |{\bm k}|<\Lambda} 
    \text{Tr}
    \bigg[ (q_x \sigma^x +q_y\sigma^y) \cdot P_n({\bm{k}})\bigg]  \nonumber\\ 
    = & -\delta_{ij}\gamma \frac{v_{\rm F}}{M_{\rm C}N} \sum_{n}^{\rm occ.}\frac{1}{\Omega}  \int_{ |{\bm k}|<\Lambda} 
    \bigg[ q_x \sigma^x \frac{k_x}{ \sqrt{k_x^2+k_y^2}} +q_y \sigma^y
    \frac{k_y}{ \sqrt{k_x^2+k_y^2}} \bigg] = 0 ~.
\end{align}
We next discuss  the geometric contribution $[\bar{\cal D}_{\rm g}(\bm q)]$. We emphasize that $[\bar{\cal D}_{\rm g}(\bm q)]$ is a non-analytical function of $\bm q$, and thus a direct Taylor expansion in small $\bm q$ fails to describe the long-wavelength behavior. In fact, by trying to perform the latter, we obtain
\begin{align}
    & [\bar{\cal D}_{\rm g}(\bm q)]_{i}^j \approx \frac{-\gamma^2 v_{\rm F}}{2 M_{\rm C}  \Omega }
    \int_0^{\Lambda} dk \frac{1}{k^2}
    \begin{pmatrix}
        q_x^2 + 3q_y^2 & 2 q_x q_y \\ 
        2 q_x q_y & 3q_x^2 + q_y^2 
    \end{pmatrix} + O({\bm q}^3) 
\end{align}
where $\Lambda$ is a once again the ultraviolt cutoff, and $\Omega$ is the area of the Brillouin zone. In addition, we replace the momentum summation with an integral, as $\frac{1}{N}\sum_{{\bm k}} \rightarrow \frac{1}{\Omega} \int_{{\bm k} }$. We observe that the coefficient of the ${\bm q^2}$ term diverges due to the infrared (small $k$) divergence of the integral $\int_0^\Lambda 1/k^2 dk$.  
This divergence reflects the non-analytical properties of the $ [\bar{\cal D}_{\rm g}(\bm q)]_{i}^j $. To directly prove this non-analyticity, we focus on the specific case of $\bm q = (q_x,0)$ to simplify the problem. We first evaluate the $i=x,j=x$ component which gives 
\begin{align}
   & [\bar{\cal D}_{\rm g}(q_x,0)]_{x}^x = \frac{v_{\rm F} \gamma^2}{\Omega M_{\rm C}} 
    \int_{ |{\bm k}| < \Lambda} \nonumber\\ 
    &(-1)\bigg[ k_y^2 q_x \bigg(6 k_x^5+23 k_x^4 q_x+k_x^2 q_x \left(11 \sqrt{k_x^2+k_y^2} \sqrt{(k_x+q_x)^2+k_y^2}+28 k_y^2+27 q_x^2\right) \nonumber\\ 
    &+k_x
   \left(k_y^2 \left(6 \sqrt{k_x^2+k_y^2} \sqrt{(k_x+q_x)^2+k_y^2}+23 q_x^2\right)+6 q_x^2 \sqrt{k_x^2+k_y^2} \sqrt{(k_x+q_x)^2+k_y^2}+6 k_y^4+11
   q_x^4\right) \nonumber\\
   &+q_x \left(k_y^2 \left(3 \sqrt{k_x^2+k_y^2} \sqrt{(k_x+q_x)^2+k_y^2}+7 q_x^2\right)+q_x^2 \sqrt{k_x^2+k_y^2}
   \sqrt{(k_x+q_x)^2+k_y^2}+5 k_y^4+2 q_x^4\right)\nonumber\\
   &+k_x^3 \left(6 \sqrt{k_x^2+k_y^2} \sqrt{(k_x+q_x)^2+k_y^2}+12 k_y^2+35
   q_x^2\right)\bigg)\bigg]\nonumber\\ 
   &\bigg[ \left(k_x^2+k_y^2\right)^{3/2} \left((k_x+q_x)^2+k_y^2\right)^{5/2} \left(\sqrt{k_x^2+k_y^2}+\sqrt{(k_x+q_x)^2+k_y^2}\right)\bigg] ~.
\end{align}
We could then let 
\begin{align}
    &k_x \rightarrow p_x  q_x ~,\nonumber\\ 
    &k_y \rightarrow p_y q_x ~,
\end{align}
thus obtaining 
\begin{align}
   & [\bar{\cal D}_{\rm g}(q_x,0)]_{x}^x = \frac{v_{\rm F} \gamma^2|q_x| }{\Omega M_{\rm C}} C_{\Lambda/|q_x|} 
     \nonumber~,
\end{align}
with
\begin{align}
  &C_{\Lambda /|q_x|}  
   =\int_{ |{\bm p}| < \Lambda/|q_x|} \nonumber\\
   &(-1)\bigg[ p_y^2 \bigg(6 p_x^5+23 p_x^4 q_x+p_x^2  \left(11 \sqrt{p_x^2+p_y^2} \sqrt{(p_x+1)^2+p_y^2}+28 p_y^2+27 \right) \nonumber\\ 
    &+p_x
   \left(p_y^2 \left(6 \sqrt{p_x^2+p_y^2} \sqrt{(p_x+1)^2+p_y^2}+23\right)+6  \sqrt{p_x^2+p_y^2} \sqrt{(p_x+1)^2+p_y^2}+6 p_y^4+11
   \right) \nonumber\\
   &+ \left(p_y^2 \left(3 \sqrt{p_x^2+p_y^2} \sqrt{(p_x+1)^2+p_y^2}+7 \right)+\sqrt{p_x^2+p_y^2}
   \sqrt{(p_x+1)^2+p_y^2}+5 p_y^4+2 \right)\nonumber\\
   &+p_x^3 \left(6 \sqrt{p_x^2+p_y^2} \sqrt{(p_x+1)^2+p_y^2}+12 p_y^2+35
   \right)\bigg)\bigg]\nonumber\\ 
   &\bigg[ \left(p_x^2+p_y^2\right)^{3/2} \left((p_x+1)^2+p_y^2\right)^{5/2} \left(\sqrt{p_x^2+p_y^2}+\sqrt{(p_x+1)^2+p_y^2}\right)\bigg]~.
\end{align}
Here, we comment that $ [\bar{\cal D}_{\rm g}(q_x,0)]_{x}^x$ only depends on $|q_x|$ instead of $q_x$. In the $|q_x|\to 0$ limit, we find 
\begin{align}
    [\bar{\cal D}_{\rm g}(q_x,0)]_{x}^x = \frac{v_{\rm F} \gamma^2|q_x| }{\Omega M_{\rm C}} C_{\Lambda/|q_x|} \approx  \frac{v_{\rm F} \gamma^2|q_x| }{\Omega M_{\rm C}} C_{\Lambda/|q_x|\rightarrow \infty }~,
\end{align}
where we defined the real coefficient $C_{\Lambda/|q_x|\rightarrow \infty }$ by letting $\Lambda \to \infty$. This coefficient is evaluated numerically, yielding 
\begin{align}
    C_{\Lambda/ |q_x|\rightarrow \infty } = -5.08~.
\end{align}
Therefore, we conclude that 
\begin{align}
    [\bar{\cal D}_{\rm g}(q_x,0)]_{x}^x \approx -5.08\frac{v_{\rm F} \gamma^2|q_x| }{\Omega M_{\rm C}} ~,
\end{align}
which is indeed a non-analytic function of $q_x$, since it depends on the absolute value of $q_x$. 
Similarly, for the other three components of the dynamical matrix, we find 
\begin{align}
    &[\bar{\cal D}_{\rm g}(q_x,0)]_{x}^y \approx  \frac{v_{\rm F} \gamma^2|q_x| }{\Omega M_{\rm C}} \times 0 \approx 0 \nonumber\\ 
    &[\bar{\cal D}_{\rm g}(q_x,0)]_{y}^x \approx  \frac{v_{\rm F} \gamma^2|q_x| }{\Omega M_{\rm C}} \times 0 \approx 0 \nonumber\\ 
    &[\bar{\cal D}_{\rm g}(q_x,0)]_{y}^y \approx - 2.09 \frac{v_{\rm F} \gamma^2|q_x| }{\Omega M_{\rm C}} 
\end{align}
Therefore, we have proved the non-analytic nature of $\bar{\cal D}_{\rm g}(q_x,0)$. 

To further understand the behaviors of the Dirac node, we also consider the Hamiltonian of the gapped Dirac node 
\begin{align}
     h({\bm k}) \to 
    \begin{pmatrix}
        \Delta/2 &v_{\rm F} ( k_x-ik_y )\\ 
       v_{\rm F} ( k_x + ik_y) & -\Delta/2
    \end{pmatrix}
\end{align}
with the following eigenvalue and eigenvectors 
\begin{align}
 &\varepsilon_{\pm, {\bm k}} = \pm \sqrt{v_{\rm F}^2 |{\bm k}|^2 + \frac{\Delta^2}{4}} \nonumber\\ 
  &    U_{+}({\bm k}) = \frac{1}{\sqrt{2\varepsilon_{+,\bm{k}}(\varepsilon_{+,\bm{k}}+\frac{\Delta}{2}) }}
    \begin{pmatrix}
         \varepsilon_{\bm{k},+}+\frac{\Delta}{2}\\ 
        v_{\rm F}(k_x+ik_y)
    \end{pmatrix},\quad \quad 
    U_{-}({\bm k}) = \frac{1}{\sqrt{2\varepsilon_{\bm{k},+}(\varepsilon_{\bm{k},+}+\frac{\Delta}{2}) }}
    \begin{pmatrix}
        v_{\rm F}(ik_y-k_x)\\ 
      \varepsilon_{\bm{k},+}+\frac{\Delta}{2} 
    \end{pmatrix} 
 \end{align}
We now evaluate the electronic contributions to the dynamical matrix of the acoustic mode. We first consider $[\bar{\cal D}^{\rm (el)}(\bm q)]_{i}^j$. We notice that, for the gapped Dirac node, 
\begin{align}
         &f_i(\bm{k}) = i\gamma \partial_{k_i}h({\bm k}) = i\gamma v_{\rm F} \sigma_i \nonumber\\ 
        &M_{ij}(\bm{k}) = (\gamma \delta_{ij} - \gamma^2 \partial^2_{k_i,k_j}) h({\bm k}) = \delta_{ij} \gamma h(\bm k)
    \end{align}
    Then we observe 
    \begin{align}
    f_{i}({\bm k+\bm{q}})-f_{i}({\bm k}) = 0 
    \end{align}
    Therefore, from \eqref{eq:dyn_acc_proj}, we conclude that the total electronic contribution of the gapped Dirac node gives 
    \begin{align}
        [\bar{\cal D}^{\rm (el)}(\bm q)]_i^j
       &= \delta_{ij}\gamma^2 \frac{v_{\rm F}}{M_{\rm C}N} \sum_{n}^{\rm occ.}\frac{1}{\Omega}  \int_{ |{\bm k}|<\Lambda} 
    \text{Tr}
    \bigg[ (q_x \sigma^x +q_y\sigma^y) \cdot P_n({\bm{k}})\bigg]  \nonumber\\ 
    &=  -\delta_{ij}\gamma^2 \frac{v_{\rm F}}{M_{\rm C}N} \sum_{n}^{\rm occ.}\frac{1}{\Omega}  \int_{ |{\bm k}|<\Lambda} 
    \bigg[ q_x \sigma^x \frac{k_x}{ \sqrt{k_x^2+k_y^2 + \frac{\Delta^2}{4}}} +q_y \sigma^y
    \frac{k_y}{ \sqrt{k_x^2+k_y^2 + \frac{\Delta^2}{4} } } \bigg] = 0 
    \end{align}
    which also vanishes. We next evaluate the geometric contribution. Since the finite $\Delta$ removes the singular behaviors of the Dirac node, we can perform a direct expansion in powers of ${\bm q}$, which gives 
\begin{align}
    &[\bar{\cal D}_{\rm g}(\bm q)]_{i}^j \nonumber 
    =\frac{\gamma^2\pi v_{\rm F}^4}{4M_{\rm C}\Omega}
    \int_0^{\Lambda}kdk \frac{1}{[\Delta^2/4 + v_{\rm F}^2k^2]^{7/2}} 
    \nonumber~\times\\[3pt] 
    &
\begin{pmatrix} 
 -\left(\Delta^4 \left(2 q_x^2+q_y^2\right)\right)+\Delta^2 k^2 v_{\rm F}^2 \left(5 q_x^2+q_y^2\right)+k^4 v_{\rm F}^4 \left(q_x^2+3 q_y^2\right) & -q_x
   q_y \left(\Delta^4-4 \Delta^2 k^2 v_{\rm F}^2+2 k^4 v_{\rm F}^4\right) \\[3pt]
 -q_x q_y \left(\Delta^4-4 \Delta^2 k^2 v_{\rm F}^2+2 k^4 v_{\rm F}^4\right) & -\left(\Delta^4 \left(q_x^2+2 q_y^2\right)\right)+\Delta^2 k^2 v_{\rm F}^2
   \left(q_x^2+5 q_y^2\right)+k^4 v_{\rm F}^4 \left(3 q_x^2+q_y^2\right) 
\end{pmatrix}_{ij} 
\end{align}
We introduce the auxiliary variable $k= y\Delta/v_{\rm F}$, and find 
\begin{align}
    &[\bar{\cal D}_{\rm g}(\bm q)]_{i}^j \nonumber
    =\frac{\gamma^2\pi v_{\rm F}^2}{4M_{\rm C} \Delta^5\Omega}
    \int_0^{v_{\rm F}\Lambda/\Delta }ydy \frac{1}{[1/4 + y^2]^{7/2}} 
    \nonumber~\times\\[3pt]
    &
\begin{pmatrix} 
 -\left(\left(2 q_x^2+q_y^2\right)\right)+ y^2 \left(5 q_x^2+q_y^2\right)+ y^4\left(q_x^2+3 q_y^2\right) & -q_x
   q_y \left(1-4 y^2+2  y^4 \right) \\[3pt]
 -q_x q_y \left(1-4 y^4 +2  y^4\right) & -\left( \left(q_x^2+2 q_y^2\right)\right)+y^4
   \left(q_x^2+5 q_y^2\right)+ y^4 \left(3 q_x^2+q_y^2\right) 
\end{pmatrix}_{ij}
\end{align}
By performing the integral on $y$, we find 
\begin{align}
     &[\bar{\cal D}_{\rm g}(\bm q)]_{i}^j  \nonumber
     =\frac{8\gamma^2\pi v_{\rm F}^2}{15\Delta M_{\rm C} \Omega}~\times \\[3pt]
     &\begin{pmatrix}
         \frac{\left(1-5 \eta ^2 \left(9 \eta ^2+4\right)\right) q_y^2-3 \left(5 \eta ^2 \left(\eta ^2+2\right)-1\right) q_x^2}{\left(4 \eta ^2+1\right)^{5/2}}-3 q_x^2-q_y^2 & -\frac{2 \left(-15 \eta ^4+5 \eta
   ^2+\left(4 \eta ^2+1\right)^{5/2}-1\right) q_x q_y}{\left(4 \eta ^2+1\right)^{5/2}} \\
 -\frac{2 \left(-15 \eta ^4+5 \eta ^2+\left(4 \eta ^2+1\right)^{5/2}-1\right) q_x q_y}{\left(4 \eta ^2+1\right)^{5/2}} & \frac{\left(1-5 \eta ^2 \left(9 \eta ^2+4\right)\right) q_x^2-3 \left(5 \eta ^2 \left(\eta
   ^2+2\right)-1\right) q_y^2}{\left(4 \eta ^2+1\right)^{5/2}}-q_x^2-3 q_y^2 
     \end{pmatrix} + \text{H.c.}~,
\end{align}
where 
\begin{align}
    \eta = \frac{v_{\rm F}\Lambda}{\Delta}~.
\end{align}
To get a simple formula, we could let $\Lambda \rightarrow \infty$, which then gives 
\begin{align}
    &[\bar{\cal D}_{\rm g}(\bm q)]_{i}^j 
    \approx \frac{-8\gamma^2\pi v_{\rm F}^2}{15\Delta M_{\rm C}\Omega}
\begin{pmatrix}
    3q_x^2+q_y^2 & 2 q_x q_y \\[3pt]
    2 q_x q_y & 3q_y^2+q_x^2
\end{pmatrix}~,
\end{align}
We can see that the dynamical matrix of the acoustic mode is proportional to $1/\Delta$, which diverges as $\Delta\rightarrow 0$, reflecting the non-analytical behaviors of the geometric contribution for the case of a gapless Dirac node $(\Delta \rightarrow 0)$. 

\section{Quantum geometric velocity renormalization for the EFK Hamiltonian}\label{section10}

In this section, we consider the velocity renormalization from the EFK Hamiltonian defined in \ref{section6}.
The geometric terms in the dynamical matrix are
\begin{equation}
\begin{aligned}
  {\cal D}_g^{(1)}(q) =& \frac{1}{M} \frac{2}{N} \sum^{\text{FBZ}}_{{k}}
  \bigg\{
\frac{\left(
F(-,k,+,k+q)F(+,k+q,k)
-
F^E(-,k,+,k+q)F^E(+,k+q,k)
\right)}{E_{-}(k)-E_{+}(k+q)}
  \bigg\}+\text{H.c.}~,\\
  {\cal D}_g^{(2)}(q) =& \frac{1}{M} \frac{2}{N} \sum^{\text{FBZ}} _{{k} \tilde{k}}
  \sum_{\alpha } 
  \bigg[\delta_{\tilde{k},0}-\delta_{\tilde{k},k}
  \bigg]\bigg\{
  \left[[M^{\text{E-g}}({\widetilde{k}})]_{\alpha}^{\alpha^{\prime} }  + [M^{\rm g}({\widetilde{k}})]_{\alpha}^{\alpha^{\prime} }
  \right]
  \bigl[P_n( k)\bigr]_{\alpha'}^{\alpha}
  +\text{H.c.}\bigg\}~,\\
\end{aligned}
\label{}
\end{equation}
where $F(n,k,n^{\prime},k^{\prime}),F^E(n,k,n^{\prime},k^{\prime})$ is given by 
\begin{equation}
    \begin{aligned}
        F^{\rm E}(n k, n' k') & = 
        \sum_{\alpha\alpha'}
        \bigl[U^\dagger_{n}(\bm k)\bigr]_{\alpha}
        \biggl\{  \bigl[ f^{\rm E}( k')\bigr]_{\alpha}^{\alpha'}-  \bigl[f^{\rm E}(k)]_{\alpha}^{\alpha'} \biggr\}
        \bigl[U^{\vphantom{\dagger}}_{n'}(\bm k')\bigr]_{\alpha'}~,\\
        F^{}(n k, n' k') & = 
        \sum_{\alpha\alpha'}
        \bigl[U^\dagger_{n}(\bm k)\bigr]_{\alpha}
        \biggl\{  \bigl[ f^{}( k')\bigr]_{\alpha}^{\alpha'}-  \bigl[f^{}(k)]_{\alpha}^{\alpha'} \biggr\}
        \bigl[U^{\vphantom{\dagger}}_{n'}(\bm k')\bigr]_{\alpha'}~,\\
        % & = \Tr \left[ U^{\dagger}_{n}(k) \left(f^{E}(k^{\prime})-f^{E}(k)\right)U_{n^{\prime}}(k^{\prime})\right]
    \end{aligned}
    \label{}
    \end{equation}
The leading order term in $q$ of the dynamical matrix is thus given by 
\begin{equation}
\begin{aligned}
 {\cal D}_{\rm g}^{(1)}(q) = &\frac{1}{M}\frac{2}{N}\sum^{\text{FBZ}}_{{k}}
 \bigg\{
\frac{1}{E_{+}(k)-E_{-}(k)}
\Bigg(
\Tr \Bigg[P_{-} \Bigg(\partial_{k}f^{}(k)\Bigg) P_{+}\Bigg( \partial_{k}f^{}(k)\Bigg)\Bigg]+\\
&\qquad \qquad \qquad\qquad\qquad\qquad\,\,-
\Tr \Bigg[P_{-} \Bigg(\partial_{k}f^{\rm E}(k)\Bigg) P_{+}\Bigg( \partial_{k}f^{\rm E}(k)\Bigg)\Bigg]
\Bigg)\bigg\}q^2 +\text{H.c.} + O(q^4)\\
\\
 {\cal D}_{\rm g}^{(2)}(q) = &-\frac{1}{M}\frac{2}{N}\sum^{\text{FBZ}}_{{k}}
 \bigg\{
\Tr \Bigg[
    \bigg(M^{\text{E-g}}({\widetilde{k}})+ M^{\rm g}({\widetilde{k}})
    \bigg)
    P_n(k)\Bigg]
    \bigg\}q^2+\text{H.c.} + O(q^4)\\
\end{aligned}
\label{}
\end{equation}
where $M$ is the mass of each atom in the chain. The quantities $f^{\rm g}(\bm{k}),f^{\rm E}(\bm{k})$ are given by 
\begin{equation}
\begin{aligned}
    {f}^{\rm g}({k}) =& \frac{1}{2 \left(t^2-\tilde{t}^2\right) \cos (2 a k)-4 \delta  t \cos (a k)+\delta ^2+2 t^2+2 \tilde{t}^2}~\times
    \\
    &\left(
        \begin{array}{cc}
         -4 i a \gamma \tilde{t}^2 \sin (a k) (\delta  \cos (a k)-2 t) & -2 a \gamma \tilde{t} \left(\left(\delta ^2+4 t^2\right) \cos (a k)-\delta  t (\cos (2 a k)+3)\right) \\[3pt]
         2 a \gamma \tilde{t} \left(\left(\delta ^2+4 t^2\right) \cos (a k)-\delta  t (\cos (2 a k)+3)\right) & 4 i a \gamma \tilde{t}^2 \sin (a k) (\delta  \cos (a k)-2 t) \\
        \end{array}
        \right)~,\\
        {f}^{E}({k}) =& \frac{1}{2 \left(t^2-\tilde{t}^2\right) \cos (2 a k)-4 \delta  t \cos (a k)+\delta ^2+2 t^2+2 \tilde{t}^2}~\times
        \\
        &\left(
            \begin{array}{cc}
             2 i a \gamma \sin (a k) (\delta -2 t \cos (a k)) \left(\delta  t-2 \left(t^2-\tilde{t}^2\right) \cos (a k)\right) & -4 a \gamma \tilde{t} \sin ^2(a k) \left(\delta  t-2 \left(t^2-\tilde{t}^2\right) \cos (a k)\right) \\[3pt]
             4 a \gamma \sin ^2(a k) \left(\delta  t-2 \left(t^2-\tilde{t}^2\right) \cos (a k)\right) & -2 i a \gamma \sin (a k) (\delta -2 t \cos (a k)) \left(\delta  t-2 \left(t^2-\tilde{t}^2\right) \cos (a k)\right) \\
            \end{array}
            \right)~.
\end{aligned}
\label{}
\end{equation}
while $ M^{\text{E-g}}({k}) + M^{\rm g}({k})$ is given by
\begin{equation}
\begin{aligned}
  M^{\text{E-g}}({k}) + M^{\rm g}({k}) =&
  \frac{2 a^2 \gamma^2 \tilde{t}^2}{\left(-4 \delta t \cos (a k)+2 \left(t^2-\tilde{t}^2\right) \cos (2 a k)+\delta^2+2 t^2+2 \tilde{t}^2\right)^2}
\Bigg(\begin{array}{cc}
m_{11} & m_{12}\\
m_{12}^* & -m_{11}    
\end{array}\Bigg)~,\\[5pt]
m_{11} =& \delta \left(2 \cos (2 a k) \left(\delta^2+4 t^2+2 \tilde{t}^2\right)-3 \delta t \cos (3 a k)+t^2 \cos (4 a k)-\tilde{t}^2 \cos (4 a k)+15 t^2-3 \tilde{t}^2\right)~,\\
 &-t \left(9 \delta^2+16 t^2\right) \cos (a k)\\[5pt]
m_{12} =& \frac{1}{\tilde{t}}
\Bigg[i \sin (a k) \left(-6 \delta t \cos (a k) \left(\delta^2+t^2+3 \tilde{t}^2\right)+6 \delta^2 t^2 \cos (2 a k)-2 \delta t^3 \cos (3 a k)+2 \delta t \tilde{t}^2 \cos (3 a k)\right.\\
&\left.+\delta^4+6 \delta^2 t^2+4 \delta^2 \tilde{t}^2+16 t^2 \tilde{t}^2\right)\Bigg]~.
\end{aligned}
\label{}
\end{equation}
Incorporating the formulas above, we the geometric DM as
\begin{equation}
\begin{aligned}
    &{\cal D}_{\rm g}(q) = -\frac{1}{M}\frac{2}{N}\sum_{k}\frac{a^4 \gamma^2 \tilde{t}^2}{\left(-4 \delta t \cos (a k)+2 \left(t^2-\tilde{t}^2\right) \cos (2 a k)+\delta^2+2 t^2+2 \tilde{t}^2\right)^{7/2}}\\
    &\Bigg(-14 \delta^5 t \cos (3 a k)-11 \delta^4 t^2 \cos (4 a k)+20 \delta^4 \tilde{t}^2 \cos (4 a k)+92 \delta^3 t^3 \cos (3 a k)\\
    &+12 \delta^3 t^3 \cos (5 a k)-204 \delta^3 t \tilde{t}^2 \cos (3 a k)-12 \delta^3 t \tilde{t}^2 \cos (5 a k)+8 \delta^2 t^4 \cos (4 a k)\\
    &-88 \delta^2 t^2 \tilde{t}^2 \cos (4 a k)+80 \delta^2 \tilde{t}^4 \cos (4 a k)-2 \delta t \cos (a k) \left[61 \delta^4+4 \delta^2 \left(53 t^2+37 \tilde{t}^2\right)-8 \left(123 t^4-206 t^2 \tilde{t}^2+43 \tilde{t}^4\right)\right]\\
    &+\cos (2 a k) \left(13 \delta^6+72 \delta^4 \left(t^2+\tilde{t}^2\right)+16 \delta^2 \left(5 t^4+22 t^2 \tilde{t}^2+5 \tilde{t}^4\right)-384 \left(t^6-t^2 \tilde{t}^4\right)\right)-504 \delta t^5 \cos (3 a k)\\
    &-56 \delta t^5 \cos (5 a k)+1136 \delta t^3 \tilde{t}^2 \cos (3 a k)+112 \delta t^3 \tilde{t}^2 \cos (5 a k)-632 \delta t \tilde{t}^4 \cos (3 a k)-56 \delta t \tilde{t}^4 \cos (5 a k)+304 t^6 \cos (4 a k)\\
    &-608 t^4 \tilde{t}^2 \cos (4 a k)+304 t^2 \tilde{t}^4 \cos (4 a k)+3 \delta^6+339 \delta^4 t^2-28 \delta^4 \tilde{t}^2-728 \delta^2 t^4+1272 \delta^2 t^2 \tilde{t}^2-160 \delta^2 \tilde{t}^4-688 t^6+1632 t^4 \tilde{t}^2\\
    &-688 t^2 \tilde{t}^4 \Bigg)q^2 +O(q^4)~,
\end{aligned}
\label{}
\end{equation}
thus observing a non-zero and non-divergent contribution to the velocity of the acoustic mode.


\begin{thebibliography}{70}
\bibitem{note_on_indices}
Throughout this work we have taken the liberty to omit orbital and sublattice indices when referring to the projection matrix $[P_n(\bm k)]_{\nu\alpha}^ {\nu'\alpha'}$ and the Bloch eigenvectors $[U_n(\bm k)]_{\nu\alpha}$. These indices have instead been written down explicitly for the hopping matrix $[t(\bm r)]_{\nu\alpha}^ {\nu'\alpha'}$, the force-constant matrix $[{\cal C}]_{p \nu i}^ {p'\nu'j}$, and the DM $[{\cal D}]_{\nu i}^ {\nu'j}$. Other quantities defined in the following, such as the Bloch Hamiltonian $h(\bm k)$ and its derivatives $f_i(\bm k)$ and $M_{ij}(\bm k)$, trivially inherit their matrix structure from the hopping matrix $t(\bm r)$. Wherever the $\gamma$ symbol appears in front of a matrix, e.g. $\gamma t(\bm r)$, the product is intended to be element-wise, i.e.~$[\gamma t(\bm r)]_{\nu\alpha}^{\nu'\alpha'} = [\gamma]_{\nu\alpha}^{\nu'\alpha'} [t(\bm r)]_{\nu\alpha}^{\nu'\alpha'}$ and~$[\gamma^2 t(\bm r)]_{\nu\alpha}^{\nu'\alpha'} = \bigl([\gamma]_{\nu\alpha}^{\nu'\alpha'}\bigr)^2 [t(\bm r)]_{\nu\alpha}^{\nu'\alpha'}$. The same applies when $t(\bm r) \leftrightarrow h(\bm k)$.
\end{thebibliography}
\end{document}